\documentclass[longauth,traditabstract]{aa} 
\usepackage{amsmath}
\pdfoutput=1
\makeatletter
\def\input@path{{../../../../Repositories/LatexCommands/}{../../../../Repositories/AuthorLists/CosmologyProductPapers/AuthorLists/}{./}}
\makeatother

\def\setsymbol#1#2{\expandafter\def\csname #1\endcsname{#2}}
\def\getsymbol#1{\csname #1\endcsname}

\def\Planck{\textit{Planck}}

\def\all2013resultspapers{\nocite{planck2013-p01, planck2013-p02, planck2013-p02a, planck2013-p02d, planck2013-p02b, planck2013-p03, planck2013-p03c, planck2013-p03f, planck2013-p03d, planck2013-p03e, planck2013-p01a, planck2013-p06, planck2013-p03a, planck2013-pip88, planck2013-p08, planck2013-p11, planck2013-p12, planck2013-p13, planck2013-p14, planck2013-p15, planck2013-p05b, planck2013-p17, planck2013-p09, planck2013-p09a, planck2013-p20, planck2013-p19, planck2013-pipaberration, planck2013-p05, planck2013-p05a, planck2013-pip56, planck2013-p06b}}

\newbox\tablebox    \newdimen\tablewidth
\def\leaderfil{\leaders\hbox to 5pt{\hss.\hss}\hfil}
\def\endPlancktable{\tablewidth=\columnwidth 
    $$\hss\copy\tablebox\hss$$
    \vskip-\lastskip\vskip -2pt}
\def\endPlancktablewide{\tablewidth=\textwidth 
    $$\hss\copy\tablebox\hss$$
    \vskip-\lastskip\vskip -2pt}
\def\tablenote#1 #2\par{\begingroup \parindent=0.8em
    \abovedisplayshortskip=0pt\belowdisplayshortskip=0pt
    \noindent
    $$\hss\vbox{\hsize\tablewidth \hangindent=\parindent \hangafter=1 \noindent
    \hbox to \parindent{$^#1$\hss}\strut#2\strut\par}\hss$$
    \endgroup}
\def\doubleline{\vskip 3pt\hrule \vskip 1.5pt \hrule \vskip 5pt}

\def\L2{\ifmmode L_2\else $L_2$\fi}

\def\DeltaT{\ifmmode \Delta T\else $\Delta T$\fi}
\def\deltat{\ifmmode \Delta t\else $\Delta t$\fi}
\def\fknee{\ifmmode f_{\rm knee}\else $f_{\rm knee}$\fi}
\def\Fmax{\ifmmode F_{\rm max}\else $F_{\rm max}$\fi}
\def\solar{\ifmmode{\rm M}_{\mathord\odot}\else${\rm M}_{\mathord\odot}$\fi}
\def\Msolar{\ifmmode{\rm M}_{\mathord\odot}\else${\rm M}_{\mathord\odot}$\fi}
\def\Lsolar{\ifmmode{\rm L}_{\mathord\odot}\else${\rm L}_{\mathord\odot}$\fi}

\def\inv{\ifmmode^{-1}\else$^{-1}$\fi}
\def\mo{\ifmmode^{-1}\else$^{-1}$\fi}
\def\sup#1{\ifmmode ^{\rm #1}\else $^{\rm #1}$\fi}
\def\expo#1{\ifmmode \times 10^{#1}\else $\times 10^{#1}$\fi}
\def\,{\thinspace}
\def\lsim{\mathrel{\raise .4ex\hbox{\rlap{$<$}\lower 1.2ex\hbox{$\sim$}}}}
\def\gsim{\mathrel{\raise .4ex\hbox{\rlap{$>$}\lower 1.2ex\hbox{$\sim$}}}}

\def\simprop{\mathrel{\raise .4ex\hbox{\rlap{$\propto$}\lower 1.2ex\hbox{$\sim$}}}}
\def\deg{\ifmmode^\circ\else$^\circ$\fi}
\def\pdeg{\ifmmode $\setbox0=\hbox{$^{\circ}$}\rlap{\hskip.11\wd0 .}$^{\circ}
          \else \setbox0=\hbox{$^{\circ}$}\rlap{\hskip.11\wd0 .}$^{\circ}$\fi}
\def\arcs{\ifmmode {^{\scriptstyle\prime\prime}}
          \else $^{\scriptstyle\prime\prime}$\fi}
\def\arcm{\ifmmode {^{\scriptstyle\prime}}
          \else $^{\scriptstyle\prime}$\fi}
\newdimen\sa  \newdimen\sb
\def\parcs{\sa=.07em \sb=.03em
     \ifmmode \hbox{\rlap{.}}^{\scriptstyle\prime\kern -\sb\prime}\hbox{\kern -\sa}
     \else \rlap{.}$^{\scriptstyle\prime\kern -\sb\prime}$\kern -\sa\fi}
\def\parcm{\sa=.08em \sb=.03em
     \ifmmode \hbox{\rlap{.}\kern\sa}^{\scriptstyle\prime}\hbox{\kern-\sb}
     \else \rlap{.}\kern\sa$^{\scriptstyle\prime}$\kern-\sb\fi}
\def\ra[#1 #2 #3.#4]{#1\sup{h}#2\sup{m}#3\sup{s}\llap.#4}
\def\dec[#1 #2 #3.#4]{#1\deg#2\arcm#3\arcs\llap.#4}
\def\deco[#1 #2 #3]{#1\deg#2\arcm#3\arcs}
\def\rra[#1 #2]{#1\sup{h}#2\sup{m}}

\def\dots{\relax\ifmmode \ldots\else $\ldots$\fi}
\def\WHzsr{\ifmmode $W\,Hz\mo\,sr\mo$\else W\,Hz\mo\,sr\mo\fi}
\def\mHz{\ifmmode $\,mHz$\else \,mHz\fi}
\def\GHz{\ifmmode $\,GHz$\else \,GHz\fi}
\def\mKs{\ifmmode $\,mK\,s$^{1/2}\else \,mK\,s$^{1/2}$\fi}
\def\muKs{\ifmmode \,\mu$K\,s$^{1/2}\else \,$\mu$K\,s$^{1/2}$\fi}
\def\muKRJs{\ifmmode \,\mu$K$_{\rm RJ}$\,s$^{1/2}\else \,$\mu$K$_{\rm RJ}$\,s$^{1/2}$\fi}
\def\muKHz{\ifmmode \,\mu$K\,Hz$^{-1/2}\else \,$\mu$K\,Hz$^{-1/2}$\fi}
\def\MJysr{\ifmmode \,$MJy\,sr\mo$\else \,MJy\,sr\mo\fi}
\def\MJysrmK{\ifmmode \,$MJy\,sr\mo$\,mK$_{\rm CMB}\mo\else \,MJy\,sr\mo\,mK$_{\rm CMB}\mo$\fi}
\def\microns{\ifmmode \,\mu$m$\else \,$\mu$m\fi}

\def\muK{\ifmmode \,\mu$K$\else \,$\mu$\hbox{K}\fi}
\def\microK{\ifmmode \,\mu$K$\else \,$\mu$\hbox{K}\fi}
\def\muW{\ifmmode \,\mu$W$\else \,$\mu$\hbox{W}\fi}
\def\kms{\ifmmode $\,km\,s$^{-1}\else \,km\,s$^{-1}$\fi}
\def\kmsMpc{\ifmmode $\,\kms\,Mpc\mo$\else \,\kms\,Mpc\mo\fi}
\setsymbol{LFI:center:frequency:70GHz:units}{70.3\,GHz}
\setsymbol{LFI:center:frequency:44GHz:units}{44.1\,GHz}
\setsymbol{LFI:center:frequency:30GHz:units}{28.5\,GHz}

\setsymbol{LFI:center:frequency:70GHz}{70.3}
\setsymbol{LFI:center:frequency:44GHz}{44.1}
\setsymbol{LFI:center:frequency:30GHz}{28.5}

\setsymbol{LFI:center:frequency:LFI18:Rad:M:units}{71.7\GHz}
\setsymbol{LFI:center:frequency:LFI19:Rad:M:units}{67.5\GHz}
\setsymbol{LFI:center:frequency:LFI20:Rad:M:units}{69.2\GHz}
\setsymbol{LFI:center:frequency:LFI21:Rad:M:units}{70.4\GHz}
\setsymbol{LFI:center:frequency:LFI22:Rad:M:units}{71.5\GHz}
\setsymbol{LFI:center:frequency:LFI23:Rad:M:units}{70.8\GHz}
\setsymbol{LFI:center:frequency:LFI24:Rad:M:units}{44.4\GHz}
\setsymbol{LFI:center:frequency:LFI25:Rad:M:units}{44.0\GHz}
\setsymbol{LFI:center:frequency:LFI26:Rad:M:units}{43.9\GHz}
\setsymbol{LFI:center:frequency:LFI27:Rad:M:units}{28.3\GHz}
\setsymbol{LFI:center:frequency:LFI28:Rad:M:units}{28.8\GHz}
\setsymbol{LFI:center:frequency:LFI18:Rad:S:units}{70.1\GHz}
\setsymbol{LFI:center:frequency:LFI19:Rad:S:units}{69.6\GHz}
\setsymbol{LFI:center:frequency:LFI20:Rad:S:units}{69.5\GHz}
\setsymbol{LFI:center:frequency:LFI21:Rad:S:units}{69.5\GHz}
\setsymbol{LFI:center:frequency:LFI22:Rad:S:units}{72.8\GHz}
\setsymbol{LFI:center:frequency:LFI23:Rad:S:units}{71.3\GHz}
\setsymbol{LFI:center:frequency:LFI24:Rad:S:units}{44.1\GHz}
\setsymbol{LFI:center:frequency:LFI25:Rad:S:units}{44.1\GHz}
\setsymbol{LFI:center:frequency:LFI26:Rad:S:units}{44.1\GHz}
\setsymbol{LFI:center:frequency:LFI27:Rad:S:units}{28.5\GHz}
\setsymbol{LFI:center:frequency:LFI28:Rad:S:units}{28.2\GHz}

\setsymbol{LFI:center:frequency:LFI18:Rad:M}{71.7}
\setsymbol{LFI:center:frequency:LFI19:Rad:M}{67.5}
\setsymbol{LFI:center:frequency:LFI20:Rad:M}{69.2}
\setsymbol{LFI:center:frequency:LFI21:Rad:M}{70.4}
\setsymbol{LFI:center:frequency:LFI22:Rad:M}{71.5}
\setsymbol{LFI:center:frequency:LFI23:Rad:M}{70.8}
\setsymbol{LFI:center:frequency:LFI24:Rad:M}{44.4}
\setsymbol{LFI:center:frequency:LFI25:Rad:M}{44.0}
\setsymbol{LFI:center:frequency:LFI26:Rad:M}{43.9}
\setsymbol{LFI:center:frequency:LFI27:Rad:M}{28.3}
\setsymbol{LFI:center:frequency:LFI28:Rad:M}{28.8}
\setsymbol{LFI:center:frequency:LFI18:Rad:S}{70.1}
\setsymbol{LFI:center:frequency:LFI19:Rad:S}{69.6}
\setsymbol{LFI:center:frequency:LFI20:Rad:S}{69.5}
\setsymbol{LFI:center:frequency:LFI21:Rad:S}{69.5}
\setsymbol{LFI:center:frequency:LFI22:Rad:S}{72.8}
\setsymbol{LFI:center:frequency:LFI23:Rad:S}{71.3}
\setsymbol{LFI:center:frequency:LFI24:Rad:S}{44.1}
\setsymbol{LFI:center:frequency:LFI25:Rad:S}{44.1}
\setsymbol{LFI:center:frequency:LFI26:Rad:S}{44.1}
\setsymbol{LFI:center:frequency:LFI27:Rad:S}{28.5}
\setsymbol{LFI:center:frequency:LFI28:Rad:S}{28.2}

\setsymbol{LFI:white:noise:sensitivity:70GHz:units}{134.7\muKs}
\setsymbol{LFI:white:noise:sensitivity:44GHz:units}{164.7\muKs}
\setsymbol{LFI:white:noise:sensitivity:30GHz:units}{143.4\muKs}

\setsymbol{LFI:white:noise:sensitivity:70GHz}{134.7}
\setsymbol{LFI:white:noise:sensitivity:44GHz}{164.7}
\setsymbol{LFI:white:noise:sensitivity:30GHz}{143.4}

\setsymbol{LFI:white:noise:sensitivity:LFI18:Rad:M:units}{512.0\muKs}
\setsymbol{LFI:white:noise:sensitivity:LFI19:Rad:M:units}{581.4\muKs}
\setsymbol{LFI:white:noise:sensitivity:LFI20:Rad:M:units}{590.8\muKs}
\setsymbol{LFI:white:noise:sensitivity:LFI21:Rad:M:units}{455.2\muKs}
\setsymbol{LFI:white:noise:sensitivity:LFI22:Rad:M:units}{492.0\muKs}
\setsymbol{LFI:white:noise:sensitivity:LFI23:Rad:M:units}{507.7\muKs}
\setsymbol{LFI:white:noise:sensitivity:LFI24:Rad:M:units}{462.2\muKs}
\setsymbol{LFI:white:noise:sensitivity:LFI25:Rad:M:units}{413.6\muKs}
\setsymbol{LFI:white:noise:sensitivity:LFI26:Rad:M:units}{478.6\muKs}
\setsymbol{LFI:white:noise:sensitivity:LFI27:Rad:M:units}{277.7\muKs}
\setsymbol{LFI:white:noise:sensitivity:LFI28:Rad:M:units}{312.3\muKs}
\setsymbol{LFI:white:noise:sensitivity:LFI18:Rad:S:units}{465.7\muKs}
\setsymbol{LFI:white:noise:sensitivity:LFI19:Rad:S:units}{555.6\muKs}
\setsymbol{LFI:white:noise:sensitivity:LFI20:Rad:S:units}{623.2\muKs}
\setsymbol{LFI:white:noise:sensitivity:LFI21:Rad:S:units}{564.1\muKs}
\setsymbol{LFI:white:noise:sensitivity:LFI22:Rad:S:units}{534.4\muKs}
\setsymbol{LFI:white:noise:sensitivity:LFI23:Rad:S:units}{542.4\muKs}
\setsymbol{LFI:white:noise:sensitivity:LFI24:Rad:S:units}{399.2\muKs}
\setsymbol{LFI:white:noise:sensitivity:LFI25:Rad:S:units}{392.6\muKs}
\setsymbol{LFI:white:noise:sensitivity:LFI26:Rad:S:units}{418.6\muKs}
\setsymbol{LFI:white:noise:sensitivity:LFI27:Rad:S:units}{302.9\muKs}
\setsymbol{LFI:white:noise:sensitivity:LFI28:Rad:S:units}{285.3\muKs}

\setsymbol{LFI:white:noise:sensitivity:LFI18:Rad:M}{512.0}
\setsymbol{LFI:white:noise:sensitivity:LFI19:Rad:M}{581.4}
\setsymbol{LFI:white:noise:sensitivity:LFI20:Rad:M}{590.8}
\setsymbol{LFI:white:noise:sensitivity:LFI21:Rad:M}{455.2}
\setsymbol{LFI:white:noise:sensitivity:LFI22:Rad:M}{492.0}
\setsymbol{LFI:white:noise:sensitivity:LFI23:Rad:M}{507.7}
\setsymbol{LFI:white:noise:sensitivity:LFI24:Rad:M}{462.2}
\setsymbol{LFI:white:noise:sensitivity:LFI25:Rad:M}{413.6}
\setsymbol{LFI:white:noise:sensitivity:LFI26:Rad:M}{478.6}
\setsymbol{LFI:white:noise:sensitivity:LFI27:Rad:M}{277.7}
\setsymbol{LFI:white:noise:sensitivity:LFI28:Rad:M}{312.3}
\setsymbol{LFI:white:noise:sensitivity:LFI18:Rad:S}{465.7}
\setsymbol{LFI:white:noise:sensitivity:LFI19:Rad:S}{555.6}
\setsymbol{LFI:white:noise:sensitivity:LFI20:Rad:S}{623.2}
\setsymbol{LFI:white:noise:sensitivity:LFI21:Rad:S}{564.1}
\setsymbol{LFI:white:noise:sensitivity:LFI22:Rad:S}{534.4}
\setsymbol{LFI:white:noise:sensitivity:LFI23:Rad:S}{542.4}
\setsymbol{LFI:white:noise:sensitivity:LFI24:Rad:S}{399.2}
\setsymbol{LFI:white:noise:sensitivity:LFI25:Rad:S}{392.6}
\setsymbol{LFI:white:noise:sensitivity:LFI26:Rad:S}{418.6}
\setsymbol{LFI:white:noise:sensitivity:LFI27:Rad:S}{302.9}
\setsymbol{LFI:white:noise:sensitivity:LFI28:Rad:S}{285.3}

\setsymbol{LFI:knee:frequency:70GHz:units}{29.5\mHz}
\setsymbol{LFI:knee:frequency:44GHz:units}{56.2\mHz}
\setsymbol{LFI:knee:frequency:30GHz:units}{113.7\mHz}

\setsymbol{LFI:knee:frequency:70GHz}{29.5}
\setsymbol{LFI:knee:frequency:44GHz}{56.2}
\setsymbol{LFI:knee:frequency:30GHz}{113.7}

\setsymbol{LFI:knee:frequency:LFI18:Rad:M:units}{16.3\mHz}
\setsymbol{LFI:knee:frequency:LFI19:Rad:M:units}{15.1\mHz}
\setsymbol{LFI:knee:frequency:LFI20:Rad:M:units}{18.7\mHz}
\setsymbol{LFI:knee:frequency:LFI21:Rad:M:units}{37.2\mHz}
\setsymbol{LFI:knee:frequency:LFI22:Rad:M:units}{12.7\mHz}
\setsymbol{LFI:knee:frequency:LFI23:Rad:M:units}{34.6\mHz}
\setsymbol{LFI:knee:frequency:LFI24:Rad:M:units}{46.2\mHz}
\setsymbol{LFI:knee:frequency:LFI25:Rad:M:units}{24.9\mHz}
\setsymbol{LFI:knee:frequency:LFI26:Rad:M:units}{67.6\mHz}
\setsymbol{LFI:knee:frequency:LFI27:Rad:M:units}{187.4\mHz}
\setsymbol{LFI:knee:frequency:LFI28:Rad:M:units}{122.2\mHz}
\setsymbol{LFI:knee:frequency:LFI18:Rad:S:units}{17.7\mHz}
\setsymbol{LFI:knee:frequency:LFI19:Rad:S:units}{22.0\mHz}
\setsymbol{LFI:knee:frequency:LFI20:Rad:S:units}{8.7\mHz}
\setsymbol{LFI:knee:frequency:LFI21:Rad:S:units}{25.9\mHz}
\setsymbol{LFI:knee:frequency:LFI22:Rad:S:units}{15.8\mHz}
\setsymbol{LFI:knee:frequency:LFI23:Rad:S:units}{129.8\mHz}
\setsymbol{LFI:knee:frequency:LFI24:Rad:S:units}{100.9\mHz}
\setsymbol{LFI:knee:frequency:LFI25:Rad:S:units}{38.9\mHz}
\setsymbol{LFI:knee:frequency:LFI26:Rad:S:units}{58.9\mHz}
\setsymbol{LFI:knee:frequency:LFI27:Rad:S:units}{104.4\mHz}
\setsymbol{LFI:knee:frequency:LFI28:Rad:S:units}{40.7\mHz}

\setsymbol{LFI:knee:frequency:LFI18:Rad:M}{16.3}
\setsymbol{LFI:knee:frequency:LFI19:Rad:M}{15.1}
\setsymbol{LFI:knee:frequency:LFI20:Rad:M}{18.7}
\setsymbol{LFI:knee:frequency:LFI21:Rad:M}{37.2}
\setsymbol{LFI:knee:frequency:LFI22:Rad:M}{12.7}
\setsymbol{LFI:knee:frequency:LFI23:Rad:M}{34.6}
\setsymbol{LFI:knee:frequency:LFI24:Rad:M}{46.2}
\setsymbol{LFI:knee:frequency:LFI25:Rad:M}{24.9}
\setsymbol{LFI:knee:frequency:LFI26:Rad:M}{67.6}
\setsymbol{LFI:knee:frequency:LFI27:Rad:M}{187.4}
\setsymbol{LFI:knee:frequency:LFI28:Rad:M}{122.2}
\setsymbol{LFI:knee:frequency:LFI18:Rad:S}{17.7}
\setsymbol{LFI:knee:frequency:LFI19:Rad:S}{22.0}
\setsymbol{LFI:knee:frequency:LFI20:Rad:S}{8.7}
\setsymbol{LFI:knee:frequency:LFI21:Rad:S}{25.9}
\setsymbol{LFI:knee:frequency:LFI22:Rad:S}{15.8}
\setsymbol{LFI:knee:frequency:LFI23:Rad:S}{129.8}
\setsymbol{LFI:knee:frequency:LFI24:Rad:S}{100.9}
\setsymbol{LFI:knee:frequency:LFI25:Rad:S}{38.9}
\setsymbol{LFI:knee:frequency:LFI26:Rad:S}{58.9}
\setsymbol{LFI:knee:frequency:LFI27:Rad:S}{104.4}
\setsymbol{LFI:knee:frequency:LFI28:Rad:S}{40.7}

\setsymbol{LFI:slope:70GHz:units}{$-1.03$\mHz}
\setsymbol{LFI:slope:44GHz:units}{$-0.89$\mHz}
\setsymbol{LFI:slope:30GHz:units}{$-0.87$\mHz}

\setsymbol{LFI:slope:70GHz}{$-1.03$}
\setsymbol{LFI:slope:44GHz}{$-0.89$}
\setsymbol{LFI:slope:30GHz}{$-0.87$}

\setsymbol{LFI:slope:LFI18:Rad:M:units}{$-1.04$\mHz}
\setsymbol{LFI:slope:LFI19:Rad:M:units}{$-1.09$\mHz}
\setsymbol{LFI:slope:LFI20:Rad:M:units}{$-0.69$\mHz}
\setsymbol{LFI:slope:LFI21:Rad:M:units}{$-1.56$\mHz}
\setsymbol{LFI:slope:LFI22:Rad:M:units}{$-1.01$\mHz}
\setsymbol{LFI:slope:LFI23:Rad:M:units}{$-0.96$\mHz}
\setsymbol{LFI:slope:LFI24:Rad:M:units}{$-0.83$\mHz}
\setsymbol{LFI:slope:LFI25:Rad:M:units}{$-0.91$\mHz}
\setsymbol{LFI:slope:LFI26:Rad:M:units}{$-0.95$\mHz}
\setsymbol{LFI:slope:LFI27:Rad:M:units}{$-0.87$\mHz}
\setsymbol{LFI:slope:LFI28:Rad:M:units}{$-0.88$\mHz}
\setsymbol{LFI:slope:LFI18:Rad:S:units}{$-1.15$\mHz}
\setsymbol{LFI:slope:LFI19:Rad:S:units}{$-1.00$\mHz}
\setsymbol{LFI:slope:LFI20:Rad:S:units}{$-0.95$\mHz}
\setsymbol{LFI:slope:LFI21:Rad:S:units}{$-0.92$\mHz}
\setsymbol{LFI:slope:LFI22:Rad:S:units}{$-1.01$\mHz}
\setsymbol{LFI:slope:LFI23:Rad:S:units}{$-0.95$\mHz}
\setsymbol{LFI:slope:LFI24:Rad:S:units}{$-0.73$\mHz}
\setsymbol{LFI:slope:LFI25:Rad:S:units}{$-1.16$\mHz}
\setsymbol{LFI:slope:LFI26:Rad:S:units}{$-0.79$\mHz}
\setsymbol{LFI:slope:LFI27:Rad:S:units}{$-0.82$\mHz}
\setsymbol{LFI:slope:LFI28:Rad:S:units}{$-0.91$\mHz}

\setsymbol{LFI:slope:LFI18:Rad:M}{$-1.04$}
\setsymbol{LFI:slope:LFI19:Rad:M}{$-1.09$}
\setsymbol{LFI:slope:LFI20:Rad:M}{$-0.69$}
\setsymbol{LFI:slope:LFI21:Rad:M}{$-1.56$}
\setsymbol{LFI:slope:LFI22:Rad:M}{$-1.01$}
\setsymbol{LFI:slope:LFI23:Rad:M}{$-0.96$}
\setsymbol{LFI:slope:LFI24:Rad:M}{$-0.83$}
\setsymbol{LFI:slope:LFI25:Rad:M}{$-0.91$}
\setsymbol{LFI:slope:LFI26:Rad:M}{$-0.95$}
\setsymbol{LFI:slope:LFI27:Rad:M}{$-0.87$}
\setsymbol{LFI:slope:LFI28:Rad:M}{$-0.88$}
\setsymbol{LFI:slope:LFI18:Rad:S}{$-1.15$}
\setsymbol{LFI:slope:LFI19:Rad:S}{$-1.00$}
\setsymbol{LFI:slope:LFI20:Rad:S}{$-0.95$}
\setsymbol{LFI:slope:LFI21:Rad:S}{$-0.92$}
\setsymbol{LFI:slope:LFI22:Rad:S}{$-1.01$}
\setsymbol{LFI:slope:LFI23:Rad:S}{$-0.95$}
\setsymbol{LFI:slope:LFI24:Rad:S}{$-0.73$}
\setsymbol{LFI:slope:LFI25:Rad:S}{$-1.16$}
\setsymbol{LFI:slope:LFI26:Rad:S}{$-0.79$}
\setsymbol{LFI:slope:LFI27:Rad:S}{$-0.82$}
\setsymbol{LFI:slope:LFI28:Rad:S}{$-0.91$}

\setsymbol{LFI:FWHM:70GHz:units}{13\parcm01}
\setsymbol{LFI:FWHM:44GHz:units}{27\parcm92}
\setsymbol{LFI:FWHM:30GHz:units}{32\parcm65}

\setsymbol{LFI:FWHM:70GHz}{13.01}
\setsymbol{LFI:FWHM:44GHz}{27.92}
\setsymbol{LFI:FWHM:30GHz}{32.65}

\setsymbol{LFI:FWHM:LFI18:units}{13\parcm39}
\setsymbol{LFI:FWHM:LFI19:units}{13\parcm01}
\setsymbol{LFI:FWHM:LFI20:units}{12\parcm75}
\setsymbol{LFI:FWHM:LFI21:units}{12\parcm74}
\setsymbol{LFI:FWHM:LFI22:units}{12\parcm87}
\setsymbol{LFI:FWHM:LFI23:units}{13\parcm27}
\setsymbol{LFI:FWHM:LFI24:units}{22\parcm98}
\setsymbol{LFI:FWHM:LFI25:units}{30\parcm46}
\setsymbol{LFI:FWHM:LFI26:units}{30\parcm31}
\setsymbol{LFI:FWHM:LFI27:units}{32\parcm65}
\setsymbol{LFI:FWHM:LFI28:units}{32\parcm66}

\setsymbol{LFI:FWHM:LFI18}{13.39}
\setsymbol{LFI:FWHM:LFI19}{13.01}
\setsymbol{LFI:FWHM:LFI20}{12.75}
\setsymbol{LFI:FWHM:LFI21}{12.74}
\setsymbol{LFI:FWHM:LFI22}{12.87}
\setsymbol{LFI:FWHM:LFI23}{13.27}
\setsymbol{LFI:FWHM:LFI24}{22.98}
\setsymbol{LFI:FWHM:LFI25}{30.46}
\setsymbol{LFI:FWHM:LFI26}{30.31}
\setsymbol{LFI:FWHM:LFI27}{32.65}
\setsymbol{LFI:FWHM:LFI28}{32.66}

\setsymbol{LFI:FWHM:uncertainty:LFI18:units}{0.170\arcm}
\setsymbol{LFI:FWHM:uncertainty:LFI19:units}{0.174\arcm}
\setsymbol{LFI:FWHM:uncertainty:LFI20:units}{0.170\arcm}
\setsymbol{LFI:FWHM:uncertainty:LFI21:units}{0.156\arcm}
\setsymbol{LFI:FWHM:uncertainty:LFI22:units}{0.164\arcm}
\setsymbol{LFI:FWHM:uncertainty:LFI23:units}{0.171\arcm}
\setsymbol{LFI:FWHM:uncertainty:LFI24:units}{0.652\arcm}
\setsymbol{LFI:FWHM:uncertainty:LFI25:units}{1.075\arcm}
\setsymbol{LFI:FWHM:uncertainty:LFI26:units}{1.131\arcm}
\setsymbol{LFI:FWHM:uncertainty:LFI27:units}{1.266\arcm}
\setsymbol{LFI:FWHM:uncertainty:LFI28:units}{1.287\arcm}

\setsymbol{LFI:FWHM:uncertainty:LFI18}{0.170}
\setsymbol{LFI:FWHM:uncertainty:LFI19}{0.174}
\setsymbol{LFI:FWHM:uncertainty:LFI20}{0.170}
\setsymbol{LFI:FWHM:uncertainty:LFI21}{0.156}
\setsymbol{LFI:FWHM:uncertainty:LFI22}{0.164}
\setsymbol{LFI:FWHM:uncertainty:LFI23}{0.171}
\setsymbol{LFI:FWHM:uncertainty:LFI24}{0.652}
\setsymbol{LFI:FWHM:uncertainty:LFI25}{1.075}
\setsymbol{LFI:FWHM:uncertainty:LFI26}{1.131}
\setsymbol{LFI:FWHM:uncertainty:LFI27}{1.266}
\setsymbol{LFI:FWHM:uncertainty:LFI28}{1.287}

\setsymbol{HFI:center:frequency:100GHz:units}{100\,GHz}
\setsymbol{HFI:center:frequency:143GHz:units}{143\,GHz}
\setsymbol{HFI:center:frequency:217GHz:units}{217\,GHz}
\setsymbol{HFI:center:frequency:353GHz:units}{353\,GHz}
\setsymbol{HFI:center:frequency:545GHz:units}{545\,GHz}
\setsymbol{HFI:center:frequency:857GHz:units}{857\,GHz}

\setsymbol{HFI:center:frequency:100GHz}{100}
\setsymbol{HFI:center:frequency:143GHz}{143}
\setsymbol{HFI:center:frequency:217GHz}{217}
\setsymbol{HFI:center:frequency:353GHz}{353}
\setsymbol{HFI:center:frequency:545GHz}{545}
\setsymbol{HFI:center:frequency:857GHz}{857}

\setsymbol{HFI:Ndetectors:100GHz}{8}
\setsymbol{HFI:Ndetectors:143GHz}{11}
\setsymbol{HFI:Ndetectors:217GHz}{12}
\setsymbol{HFI:Ndetectors:353GHz}{12}
\setsymbol{HFI:Ndetectors:545GHz}{3}
\setsymbol{HFI:Ndetectors:857GHz}{4}

\setsymbol{HFI:FWHM:Maps:100GHz:units}{9\parcm88}
\setsymbol{HFI:FWHM:Maps:143GHz:units}{7\parcm18}
\setsymbol{HFI:FWHM:Maps:217GHz:units}{4\parcm87}
\setsymbol{HFI:FWHM:Maps:353GHz:units}{4\parcm65}
\setsymbol{HFI:FWHM:Maps:545GHz:units}{4\parcm72}
\setsymbol{HFI:FWHM:Maps:857GHz:units}{4\parcm39}
\setsymbol{HFI:FWHM:Maps:100GHz}{9.88}
\setsymbol{HFI:FWHM:Maps:143GHz}{7.18}
\setsymbol{HFI:FWHM:Maps:217GHz}{4.87}
\setsymbol{HFI:FWHM:Maps:353GHz}{4.65}
\setsymbol{HFI:FWHM:Maps:545GHz}{4.72}
\setsymbol{HFI:FWHM:Maps:857GHz}{4.39}

\setsymbol{HFI:beam:ellipticity:Maps:100GHz}{1.15}
\setsymbol{HFI:beam:ellipticity:Maps:143GHz}{1.01}
\setsymbol{HFI:beam:ellipticity:Maps:217GHz}{1.06}
\setsymbol{HFI:beam:ellipticity:Maps:353GHz}{1.05}
\setsymbol{HFI:beam:ellipticity:Maps:545GHz}{1.14}
\setsymbol{HFI:beam:ellipticity:Maps:857GHz}{1.19}

\setsymbol{HFI:FWHM:Mars:100GHz:units}{9\parcm37}
\setsymbol{HFI:FWHM:Mars:143GHz:units}{7\parcm04}
\setsymbol{HFI:FWHM:Mars:217GHz:units}{4\parcm68}
\setsymbol{HFI:FWHM:Mars:353GHz:units}{4\parcm43}
\setsymbol{HFI:FWHM:Mars:545GHz:units}{3\parcm80}
\setsymbol{HFI:FWHM:Mars:857GHz:units}{3\parcm67}

\setsymbol{HFI:FWHM:Mars:100GHz}{9.37}
\setsymbol{HFI:FWHM:Mars:143GHz}{7.04}
\setsymbol{HFI:FWHM:Mars:217GHz}{4.68}
\setsymbol{HFI:FWHM:Mars:353GHz}{4.43}
\setsymbol{HFI:FWHM:Mars:545GHz}{3.80}
\setsymbol{HFI:FWHM:Mars:857GHz}{3.67}

\setsymbol{HFI:beam:ellipticity:Mars:100GHz}{1.18}
\setsymbol{HFI:beam:ellipticity:Mars:143GHz}{1.03}
\setsymbol{HFI:beam:ellipticity:Mars:217GHz}{1.14}
\setsymbol{HFI:beam:ellipticity:Mars:353GHz}{1.09}
\setsymbol{HFI:beam:ellipticity:Mars:545GHz}{1.25}
\setsymbol{HFI:beam:ellipticity:Mars:857GHz}{1.03}

\setsymbol{HFI:CMB:relative:calibration:100GHz}{$\lsim 1\%$}
\setsymbol{HFI:CMB:relative:calibration:143GHz}{$\lsim 1\%$}
\setsymbol{HFI:CMB:relative:calibration:217GHz}{$\lsim 1\%$}
\setsymbol{HFI:CMB:relative:calibration:353GHz}{$\lsim 1\%$}
\setsymbol{HFI:CMB:relative:calibration:545GHz}{}
\setsymbol{HFI:CMB:relative:calibration:857GHz}{}

\setsymbol{HFI:CMB:absolute:calibration:100GHz}{$\lsim 2\%$}
\setsymbol{HFI:CMB:absolute:calibration:143GHz}{$\lsim 2\%$}
\setsymbol{HFI:CMB:absolute:calibration:217GHz}{$\lsim 2\%$}
\setsymbol{HFI:CMB:absolute:calibration:353GHz}{$\lsim 2\%$}
\setsymbol{HFI:CMB:absolute:calibration:545GHz}{}
\setsymbol{HFI:CMB:absolute:calibration:857GHz}{}

\setsymbol{HFI:FIRAS:gain:calibration:accuracy:statistical:100GHz}{}
\setsymbol{HFI:FIRAS:gain:calibration:accuracy:statistical:143GHz}{}
\setsymbol{HFI:FIRAS:gain:calibration:accuracy:statistical:217GHz}{}
\setsymbol{HFI:FIRAS:gain:calibration:accuracy:statistical:353GHz}{2.5\%}
\setsymbol{HFI:FIRAS:gain:calibration:accuracy:statistical:545GHz}{1\%}
\setsymbol{HFI:FIRAS:gain:calibration:accuracy:statistical:857GHz}{0.5\%}

\setsymbol{HFI:FIRAS:gain:calibration:accuracy:systematic:100GHz}{}
\setsymbol{HFI:FIRAS:gain:calibration:accuracy:systematic:143GHz}{}
\setsymbol{HFI:FIRAS:gain:calibration:accuracy:systematic:217GHz}{}
\setsymbol{HFI:FIRAS:gain:calibration:accuracy:systematic:353GHz}{}
\setsymbol{HFI:FIRAS:gain:calibration:accuracy:systematic:545GHz}{7\%}
\setsymbol{HFI:FIRAS:gain:calibration:accuracy:systematic:857GHz}{7\%}

\setsymbol{HFI:FIRAS:zero:point:accuracy:100GHz:units}{0.8\MJysr}
\setsymbol{HFI:FIRAS:zero:point:accuracy:143GHz:units}{}
\setsymbol{HFI:FIRAS:zero:point:accuracy:217GHz:units}{}
\setsymbol{HFI:FIRAS:zero:point:accuracy:353GHz:units}{1.4\MJysr}
\setsymbol{HFI:FIRAS:zero:point:accuracy:545GHz:units}{2.2\MJysr}
\setsymbol{HFI:FIRAS:zero:point:accuracy:857GHz:units}{1.7\MJysr}

\setsymbol{HFI:FIRAS:zero:point:accuracy:100GHz}{0.8}
\setsymbol{HFI:FIRAS:zero:point:accuracy:143GHz}{}
\setsymbol{HFI:FIRAS:zero:point:accuracy:217GHz}{}
\setsymbol{HFI:FIRAS:zero:point:accuracy:353GHz}{1.4}
\setsymbol{HFI:FIRAS:zero:point:accuracy:545GHz}{2.2}
\setsymbol{HFI:FIRAS:zero:point:accuracy:857GHz}{1.7}

\setsymbol{HFI:unit:conversion:100GHz:units}{0.2415\MJysrmK}
\setsymbol{HFI:unit:conversion:143GHz:units}{0.3694\MJysrmK}
\setsymbol{HFI:unit:conversion:217GHz:units}{0.4811\MJysrmK}
\setsymbol{HFI:unit:conversion:353GHz:units}{0.2883\MJysrmK}
\setsymbol{HFI:unit:conversion:545GHz:units}{0.05826\MJysrmK}
\setsymbol{HFI:unit:conversion:857GHz:units}{0.002238\MJysrmK}

\setsymbol{HFI:unit:conversion:100GHz}{0.2415}
\setsymbol{HFI:unit:conversion:143GHz}{0.3694}
\setsymbol{HFI:unit:conversion:217GHz}{0.4811}
\setsymbol{HFI:unit:conversion:353GHz}{0.2883}
\setsymbol{HFI:unit:conversion:545GHz}{0.05826}
\setsymbol{HFI:unit:conversion:857GHz}{0.002238}

\setsymbol{HFI:colour:correction:alpha=-2:V1.01:100GHz}{0.9893}
\setsymbol{HFI:colour:correction:alpha=-2:V1.01:143GHz}{0.9759}
\setsymbol{HFI:colour:correction:alpha=-2:V1.01:217GHz}{1.0007}
\setsymbol{HFI:colour:correction:alpha=-2:V1.01:353GHz}{1.0028}
\setsymbol{HFI:colour:correction:alpha=-2:V1.01:545GHz}{1.0019}
\setsymbol{HFI:colour:correction:alpha=-2:V1.01:857GHz}{0.9889}

\setsymbol{HFI:colour:correction:alpha=0:V1.01:100GHz}{1.0008}
\setsymbol{HFI:colour:correction:alpha=0:V1.01:143GHz}{1.0148}
\setsymbol{HFI:colour:correction:alpha=0:V1.01:217GHz}{0.9909}
\setsymbol{HFI:colour:correction:alpha=0:V1.01:353GHz}{0.9888}
\setsymbol{HFI:colour:correction:alpha=0:V1.01:545GHz}{0.9878}
\setsymbol{HFI:colour:correction:alpha=0:V1.01:857GHz}{1.0014}

\usepackage{natbib}
\bibpunct{(}{)}{;}{a}{}{,} %
\usepackage{graphicx}
\usepackage{txfonts}
\usepackage[table,usenames,dvipsnames]{xcolor}
\usepackage[breaklinks, colorlinks, citecolor=blue, colorlinks=true, linkcolor=blue]{hyperref}
\hypersetup{pdfauthor={Planck Collaboration},pdftitle={Planck 2013
    results.  VII. HFI time response and beams}}
\usepackage{tikz}
\usetikzlibrary{shapes.geometric,shapes.arrows,decorations.pathmorphing}
\usetikzlibrary{matrix,chains,scopes,positioning,arrows,fit}
\usepackage{fixltx2e} % should put figures in right order

\makeatletter
\tikzset{join/.code=\tikzset{after node path={%
\ifx\tikzchainprevious\pgfutil@empty\else(\tikzchainprevious)%
edge[every join]#1(\tikzchaincurrent)\fi}}}
\makeatother
\tikzset{>=stealth',every on chain/.append style={join},
  every join/.style={->}}

\tikzstyle{block} = [draw, fill=gray!30, circle, 
minimum height=0.2cm, minimum width=0.2cm]

\tikzstyle{beammap} = [draw, fill=gray!10, rectangle,scale=0.8,text width=2.0cm]

\tikzstyle{result} = [draw, fill=gray!30,circle,scale=0.8,text width=2.0cm]

\tikzstyle{lab} = [draw=none,fill=none,right]

\newcommand{\hatn}{\hat{n}}
\newcommand{\wslm}[3]{{}_{#1} w_{#2 #3}}
\newcommand{\threej}[6]{{\begm #1 & #2 & #3 \\ #4 & #5 & #6 \enm}}
\newcommand{\begm}{\begin{pmatrix}}
\newcommand{\enm}{\end{pmatrix}}
\newcommand{\muKarcmin}{\,\muK\,\arcmin}
\newcommand{\clo}{\mathcal{O}}

\begin{document}
%\input{../../../../Repositories/AuthorLists/CosmologyProductPapers/AuthorLists/AuthorList_P03c_HFI_Transfer_Function_and_Beams_authors_and_institutes.tex}
%This author list corresponds to \title{Author list for SVN P03c\_HFI\_Transfer\_Function\_and\_Beams, Proj. Ref. : HFI Time transfer functions and beams}
%Prepared by R. Leonardi (rleonardi@sciops.esa.int), ESAC/ESA
%This version is from Thu Sep 05 13:37:08 2013 CET
%\subtitle{There are 222 co-authors in this list}
\author{\small
Planck Collaboration:
P.~A.~R.~Ade\inst{82}
\and
N.~Aghanim\inst{57}
\and
C.~Armitage-Caplan\inst{86}
\and
M.~Arnaud\inst{69}
\and
M.~Ashdown\inst{66, 6}
\and
F.~Atrio-Barandela\inst{18}
\and
J.~Aumont\inst{57}
\and
C.~Baccigalupi\inst{81}
\and
A.~J.~Banday\inst{89, 10}
\and
R.~B.~Barreiro\inst{63}
% ...
}
\institute{\small
APC, AstroParticule et Cosmologie, Universit\'{e} Paris Diderot, CNRS/IN2P3, CEA/lrfu, Observatoire de Paris, Sorbonne Paris Cit\'{e}, 10, rue Alice Domon et L\'{e}onie Duquet, 75205 Paris Cedex 13, France\\
\and
Aalto University Mets\"{a}hovi Radio Observatory, Mets\"{a}hovintie 114, FIN-02540 Kylm\"{a}l\"{a}, Finland\\
\and
African Institute for Mathematical Sciences, 6-8 Melrose Road, Muizenberg, Cape Town, South Africa\\
\and
Agenzia Spaziale Italiana Science Data Center, Via del Politecnico snc, 00133, Roma, Italy\\
\and
Agenzia Spaziale Italiana, Viale Liegi 26, Roma, Italy\\
\and
Astrophysics Group, Cavendish Laboratory, University of Cambridge, J J Thomson Avenue, Cambridge CB3 0HE, U.K.\\
\and
Astrophysics \& Cosmology Research Unit, School of Mathematics, Statistics \& Computer Science, University of KwaZulu-Natal, Westville Campus, Private Bag X54001, Durban 4000, South Africa\\
\and
Atacama Large Millimeter/submillimeter Array, ALMA Santiago Central Offices, Alonso de Cordova 3107, Vitacura, Casilla 763 0355, Santiago, Chile\\
\and
CITA, University of Toronto, 60 St. George St., Toronto, ON M5S 3H8, Canada\\
\and
CNRS, IRAP, 9 Av. colonel Roche, BP 44346, F-31028 Toulouse cedex 4, France\\
\and
California Institute of Technology, Pasadena, California, U.S.A.\\
\and
Centre for Theoretical Cosmology, DAMTP, University of Cambridge, Wilberforce Road, Cambridge CB3 0WA, U.K.\\
\and
Centro de Estudios de F\'{i}sica del Cosmos de Arag\'{o}n (CEFCA), Plaza San Juan, 1, planta 2, E-44001, Teruel, Spain\\
\and
Computational Cosmology Center, Lawrence Berkeley National Laboratory, Berkeley, California, U.S.A.\\
\and
DSM/Irfu/SPP, CEA-Saclay, F-91191 Gif-sur-Yvette Cedex, France\\
\and
DTU Space, National Space Institute, Technical University of Denmark, Elektrovej 327, DK-2800 Kgs. Lyngby, Denmark\\
\and
D\'{e}partement de Physique Th\'{e}orique, Universit\'{e} de Gen\`{e}ve, 24, Quai E. Ansermet,1211 Gen\`{e}ve 4, Switzerland\\
\and
Departamento de F\'{\i}sica Fundamental, Facultad de Ciencias, Universidad de Salamanca, 37008 Salamanca, Spain\\
\and
Department of Astronomy and Astrophysics, University of Toronto, 50 Saint George Street, Toronto, Ontario, Canada\\
\and
Department of Astrophysics/IMAPP, Radboud University Nijmegen, P.O. Box 9010, 6500 GL Nijmegen, The Netherlands\\
\and
Department of Electrical Engineering and Computer Sciences, University of California, Berkeley, California, U.S.A.\\
\and
Department of Physics \& Astronomy, University of British Columbia, 6224 Agricultural Road, Vancouver, British Columbia, Canada\\
\and
Department of Physics and Astronomy, Dana and David Dornsife College of Letter, Arts and Sciences, University of Southern California, Los Angeles, CA 90089, U.S.A.\\
\and
Department of Physics and Astronomy, University College London, London WC1E 6BT, U.K.\\
\and
Department of Physics, Florida State University, Keen Physics Building, 77 Chieftan Way, Tallahassee, Florida, U.S.A.\\
\and
Department of Physics, Gustaf H\"{a}llstr\"{o}min katu 2a, University of Helsinki, Helsinki, Finland\\
\and
Department of Physics, Princeton University, Princeton, New Jersey, U.S.A.\\
\and
Department of Physics, University of California, One Shields Avenue, Davis, California, U.S.A.\\
\and
Department of Physics, University of California, Santa Barbara, California, U.S.A.\\
\and
Department of Physics, University of Illinois at Urbana-Champaign, 1110 West Green Street, Urbana, Illinois, U.S.A.\\
\and
Dipartimento di Fisica e Astronomia G. Galilei, Universit\`{a} degli Studi di Padova, via Marzolo 8, 35131 Padova, Italy\\
\and
Dipartimento di Fisica e Scienze della Terra, Universit\`{a} di Ferrara, Via Saragat 1, 44122 Ferrara, Italy\\
\and
Dipartimento di Fisica, Universit\`{a} La Sapienza, P. le A. Moro 2, Roma, Italy\\
\and
Dipartimento di Fisica, Universit\`{a} degli Studi di Milano, Via Celoria, 16, Milano, Italy\\
\and
Dipartimento di Fisica, Universit\`{a} degli Studi di Trieste, via A. Valerio 2, Trieste, Italy\\
\and
Dipartimento di Fisica, Universit\`{a} di Roma Tor Vergata, Via della Ricerca Scientifica, 1, Roma, Italy\\
\and
Discovery Center, Niels Bohr Institute, Blegdamsvej 17, Copenhagen, Denmark\\
\and
European Southern Observatory, ESO Vitacura, Alonso de Cordova 3107, Vitacura, Casilla 19001, Santiago, Chile\\
\and
European Space Agency, ESAC, Planck Science Office, Camino bajo del Castillo, s/n, Urbanizaci\'{o}n Villafranca del Castillo, Villanueva de la Ca\~{n}ada, Madrid, Spain\\
\and
European Space Agency, ESTEC, Keplerlaan 1, 2201 AZ Noordwijk, The Netherlands\\
\and
Helsinki Institute of Physics, Gustaf H\"{a}llstr\"{o}min katu 2, University of Helsinki, Helsinki, Finland\\
\and
INAF - Osservatorio Astrofisico di Catania, Via S. Sofia 78, Catania, Italy\\
\and
INAF - Osservatorio Astronomico di Padova, Vicolo dell'Osservatorio 5, Padova, Italy\\
\and
INAF - Osservatorio Astronomico di Roma, via di Frascati 33, Monte Porzio Catone, Italy\\
\and
INAF - Osservatorio Astronomico di Trieste, Via G.B. Tiepolo 11, Trieste, Italy\\
\and
INAF Istituto di Radioastronomia, Via P. Gobetti 101, 40129 Bologna, Italy\\
\and
INAF/IASF Bologna, Via Gobetti 101, Bologna, Italy\\
\and
INAF/IASF Milano, Via E. Bassini 15, Milano, Italy\\
\and
INFN, Sezione di Bologna, Via Irnerio 46, I-40126, Bologna, Italy\\
\and
INFN, Sezione di Roma 1, Universit\`{a} di Roma Sapienza, Piazzale Aldo Moro 2, 00185, Roma, Italy\\
\and
IPAG: Institut de Plan\'{e}tologie et d'Astrophysique de Grenoble, Universit\'{e} Joseph Fourier, Grenoble 1 / CNRS-INSU, UMR 5274, Grenoble, F-38041, France\\
\and
IUCAA, Post Bag 4, Ganeshkhind, Pune University Campus, Pune 411 007, India\\
\and
Imperial College London, Astrophysics group, Blackett Laboratory, Prince Consort Road, London, SW7 2AZ, U.K.\\
\and
Infrared Processing and Analysis Center, California Institute of Technology, Pasadena, CA 91125, U.S.A.\\
\and
Institut N\'{e}el, CNRS, Universit\'{e} Joseph Fourier Grenoble I, 25 rue des Martyrs, Grenoble, France\\
\and
Institut Universitaire de France, 103, bd Saint-Michel, 75005, Paris, France\\
\and
Institut d'Astrophysique Spatiale, CNRS (UMR8617) Universit\'{e} Paris-Sud 11, B\^{a}timent 121, Orsay, France\\
\and
Institut d'Astrophysique de Paris, CNRS (UMR7095), 98 bis Boulevard Arago, F-75014, Paris, France\\
\and
Institute for Space Sciences, Bucharest-Magurale, Romania\\
\and
Institute of Astronomy and Astrophysics, Academia Sinica, Taipei, Taiwan\\
\and
Institute of Astronomy, University of Cambridge, Madingley Road, Cambridge CB3 0HA, U.K.\\
\and
Institute of Theoretical Astrophysics, University of Oslo, Blindern, Oslo, Norway\\
\and
Instituto de F\'{\i}sica de Cantabria (CSIC-Universidad de Cantabria), Avda. de los Castros s/n, Santander, Spain\\
\and
Jet Propulsion Laboratory, California Institute of Technology, 4800 Oak Grove Drive, Pasadena, California, U.S.A.\\
\and
Jodrell Bank Centre for Astrophysics, Alan Turing Building, School of Physics and Astronomy, The University of Manchester, Oxford Road, Manchester, M13 9PL, U.K.\\
\and
Kavli Institute for Cosmology Cambridge, Madingley Road, Cambridge, CB3 0HA, U.K.\\
\and
LAL, Universit\'{e} Paris-Sud, CNRS/IN2P3, Orsay, France\\
\and
LERMA, CNRS, Observatoire de Paris, 61 Avenue de l'Observatoire, Paris, France\\
\and
Laboratoire AIM, IRFU/Service d'Astrophysique - CEA/DSM - CNRS - Universit\'{e} Paris Diderot, B\^{a}t. 709, CEA-Saclay, F-91191 Gif-sur-Yvette Cedex, France\\
\and
Laboratoire Traitement et Communication de l'Information, CNRS (UMR 5141) and T\'{e}l\'{e}com ParisTech, 46 rue Barrault F-75634 Paris Cedex 13, France\\
\and
Laboratoire de Physique Subatomique et de Cosmologie, Universit\'{e} Joseph Fourier Grenoble I, CNRS/IN2P3, Institut National Polytechnique de Grenoble, 53 rue des Martyrs, 38026 Grenoble cedex, France\\
\and
Laboratoire de Physique Th\'{e}orique, Universit\'{e} Paris-Sud 11 \& CNRS, B\^{a}timent 210, 91405 Orsay, France\\
\and
Lawrence Berkeley National Laboratory, Berkeley, California, U.S.A.\\
\and
Max-Planck-Institut f\"{u}r Astrophysik, Karl-Schwarzschild-Str. 1, 85741 Garching, Germany\\
\and
McGill Physics, Ernest Rutherford Physics Building, McGill University, 3600 rue University, Montr\'{e}al, QC, H3A 2T8, Canada\\
\and
National University of Ireland, Department of Experimental Physics, Maynooth, Co. Kildare, Ireland\\
\and
Niels Bohr Institute, Blegdamsvej 17, Copenhagen, Denmark\\
\and
Observational Cosmology, Mail Stop 367-17, California Institute of Technology, Pasadena, CA, 91125, U.S.A.\\
\and
Optical Science Laboratory, University College London, Gower Street, London, U.K.\\
\and
SB-ITP-LPPC, EPFL, CH-1015, Lausanne, Switzerland\\
\and
SISSA, Astrophysics Sector, via Bonomea 265, 34136, Trieste, Italy\\
\and
School of Physics and Astronomy, Cardiff University, Queens Buildings, The Parade, Cardiff, CF24 3AA, U.K.\\
\and
Space Sciences Laboratory, University of California, Berkeley, California, U.S.A.\\
\and
Special Astrophysical Observatory, Russian Academy of Sciences, Nizhnij Arkhyz, Zelenchukskiy region, Karachai-Cherkessian Republic, 369167, Russia\\
\and
Stanford University, Dept of Physics, Varian Physics Bldg, 382 Via Pueblo Mall, Stanford, California, U.S.A.\\
\and
Sub-Department of Astrophysics, University of Oxford, Keble Road, Oxford OX1 3RH, U.K.\\
\and
Theory Division, PH-TH, CERN, CH-1211, Geneva 23, Switzerland\\
\and
UPMC Univ Paris 06, UMR7095, 98 bis Boulevard Arago, F-75014, Paris, France\\
\and
Universit\'{e} de Toulouse, UPS-OMP, IRAP, F-31028 Toulouse cedex 4, France\\
% ...
}

 \title{\textit{Planck} 2013 results.  VII. HFI time response and beams}

% \date{Preprint online version: March 21, 2013}
\date{Preprint online version as accepted: December 9, 2013}

 \abstract{This paper characterizes the \textit{effective beams},
   the \textit{effective beam window functions} and the associated
   errors for the \Planck\ High Frequency Instrument (HFI) detectors.
   The effective beam is the angular response including the effect of
   the optics, detectors, data processing and the scan strategy.  The
   window function is the representation of this beam in the harmonic
   domain which is required to recover an unbiased measurement of the
   Cosmic microwave background (CMB) angular power spectrum. The HFI
   is a scanning instrument and its effective beams are the
   convolution of : (a) the optical response of the telescope and
   feeds;
%%(a) the collection of photons by the optical
%%beams, 
%%(b) the time response transfer functions of the bolometers and
%%associated readout electronics, 
(b) the processing of the time-ordered data and deconvolution of the
   bolometric and electronic transfer function; and (c) the merging of
   several surveys to produce maps. The time response transfer
   functions are measured using observations of Jupiter and Saturn and
   by minimizing survey difference residuals.  The \textit{scanning
     beam} is the post-deconvolution angular response of the
   instrument, and is characterized with observations of Mars.  The
   main beam solid angles are determined to better than 0.5\,\% at each
   HFI frequency band.  Observations of Jupiter and Saturn limit near
   sidelobes (within 5\deg) to about 0.1\,\% of the total solid angle.  Time
   response residuals remain as long tails in the scanning beams, but
   contribute less than 0.1\% of the total solid angle.  
%The
%   \textit{effective beams} of the instrument take into account the
%   averaging effects of \Planck's scan and the azimuthally asymmetric
%   scanning beams. 
   The bias and uncertainty in the beam products are estimated using
   ensembles of simulated planet observations that include the impact
   of instrumental noise and known systematic effects.  The
   correlation structure of these ensembles is well-described by five
   error eigenmodes that are sub-dominant to sample variance and
   instrumental noise in the harmonic domain.  A suite of consistency tests
   provide confidence that the error model represents a sufficient
   description of the data. The total error in the effective beam
   window functions is below 1\,\% at 100\,GHz up to multipole $\ell \sim
   1500$, and below 0.5\,\% at 143 and 217\,GHz up to $\ell \sim 2000$. }

 \keywords{Cosmology: observations -- Cosmic background radiation --
   Surveys -- Space vehicles: instruments -- Instrumentation:
   detectors}

%cosmology: observations
%cosmic background radiation
%space vehicles: instruments
%instrumentation: detectors
%ISM: general
%galaxies: active
%surveys
%methods: data analysis

\authorrunning{Planck Collaboration}
\titlerunning{Planck 2013 results.  VII. HFI time response and beams}
   \maketitle

\section{Introduction}

This paper, one of a set associated with the 2013 release
of data from the \Planck\footnote{\Planck\ is a project of the
  European Space Agency -- ESA -- with instruments provided by two
  scientific Consortia funded by ESA member states (in particular the
  lead countries: France and Italy) with contributions from NASA
  (USA), and telescope reflectors provided in a collaboration  between ESA and a scientific Consortium led and funded by Denmark.}\ mission \citep{planck2013-p01}, describes
the impact of the optical system, detector response, analogue and
digital filtering and the scan strategy on the determination of the
High Frequency Instrument (HFI) angular power spectra.  An accurate understanding of this
response, and the corresponding errors, is necessary in order to 
extract astrophysical and cosmological information from CMB data
\citep{hill2009,nolta2009,huffenberger2010}.

Bolometers, such as those used in the HFI 
on \Planck, are phonon-mediated thermal detectors with finite
response time to changes in the absorbed optical power.  The observations
are affected by attenuation and the phase shift of the signal resulting
from a detector's thermal response, as well as the analogue and digital
filtering in the associated electronics. 

In the small signal regime (appropriate for CMB fluctuations and 
Galactic emission), the receiver response can be well approximated
by a complex Fourier domain transfer function, termed the \emph{time
  response}.  The time-ordered data (also referred to as time-ordered
information, or TOI) are approximately deconvolved by the time
response function prior to calibration and mapmaking (for recent
examples of CMB observations with similar semiconductor bolometers see
\cite{archeops2005,boomerang2006}).

Ideally, the deconvolved TOIs represent the true sky signal convolved
with the optical response of the telescope (or \textit{physical beam})
and filtered by the TOI processing.  The combination of time
domain processing and physical beam convolution is, in practice,
degenerate, due to the nature of the scan strategy; the
\Planck\ spacecraft scans at $1\,$rpm, with variations less than
$0.1\,\%$ \citep{planck2011-1.1}.  The beams reconstructed from the
deconvolved planetary observations are referred to as the
\textit{scanning beam}.

These deconvolved data are then projected into a pixelized map, as
discussed in Sect. 6. of \cite{planck2013-p03}.  To a good
approximation, the effect of the mapmaking algorithm is to average the
beam over the observed locations in a given pixel. This average is
referred to as the \textit{effective beam}, which will vary from pixel
to pixel across the sky. The mapmaking procedure implicitly ignores any
smearing of the input TOI; no attempt is made to deconvolve the
optical beam and any remaining electronic time response. Thus, any
further use of the resulting maps must take into account the effective
beam.

To obtain an unbiased estimate of the angular power spectrum of the
CMB, one must determine the impact of this effective beam pattern on a
measurement of an isotropic Gaussian random signal in multipole ($\ell$)
space. The filtering effect is well approximated by a multiplicative
\emph{effective beam window function}, derived by coupling the scan
history with the scanning beam profile, which is used to relate the
angular power spectrum of the map to that of the underlying
sky \citep{Hivon2002}.

The solar and orbital motion of \Planck\ with respect to the surface
of last scattering provides a 60-second periodic signal in the time
ordered data that is used as a primary calibrator
\citep{planck2013-p02b,planck2013-p03f}.  This normalizes 
the window function at a multipole $\ell = 1$; the effective beam
window function is required to transfer this calibration to smaller
angular scales.

These successive products (the scanning beam, the effective beam and
the effective beam window function) must be accompanied by an account
of their errors, which are characterized through ensembles of
dedicated simulations of the planetary observations.  The error on the
effective beam window function is found to be sub-dominant to other
errors in the cosmological parameter analysis \citep{planck2013-p11}.

The scanning beam is thus measured with on-orbit
planetary data, coupling the response of the optical system to the
deconvolved time response function and additional filters in the
TOI processing. As shown in the following, the main effect of residual
deconvolution errors is a bias in the beam window function of order
$10^{-4}$ at $\ell > 100$, due to a residual tail in the beam along the
scan direction.  

The scanning beam can be further separated into the
following components;
\begin{enumerate}
\item The main beam, which is defined as extending to 30\arcm\ from the
  beam centroid. 
\item The near sidelobes, which extend between 30\arcm\ and 5\deg.
  These are typically features below $-30$\,dB, and include the optical
  effects of phase errors, consisting of both random and periodic surface
  errors and residuals due to the imperfect deconvolution of the time
  response.
\item The far sidelobes, which extend beyond 5\deg. These features are
  dominated by spillover: power coupling from the sky to the
  feed antennas directly, or via a single reflection around the
  mirrors and baffles. The minimum in the beam response between the
  main beam and direct spillover of the feed over the top of the
  secondary mirror falls at roughly 5\deg, making such a division
  natural \citep{tauber2010b}.
\end{enumerate}

This paper describes the main beam and the near sidelobes,
the resulting effective beam patterns on the sky and the effective
beam window function used for the measurement of angular power
spectra, along with their errors.  The effects of the far sidelobes
are mainly discussed in \cite{planck2013-p03} and
\cite{planck2013-pip88}, and their effects on calibration described in
\cite{planck2013-p03e}.  
A companion paper \citep{planck2013-p02d} computes the effective beams
and window functions for \Planck's Low Frequency Instrument
(LFI). Despite using very different methods which depend more strongly
on optical modelling, the two instruments produce compatible power
spectra, providing a cross-check of both approaches
\citep{planck2013-p01a}. 

The signal-to-noise ratio of Jupiter observations with HFI allows the
measurement of near sidelobes to a noise floor of $-45(-55)$\,dB 
  relative to the forward gain at 100
(857)\,GHz.  The HFI analysis does not rely on a physical optics model
to constrain the behaviour of the beam in this regime.

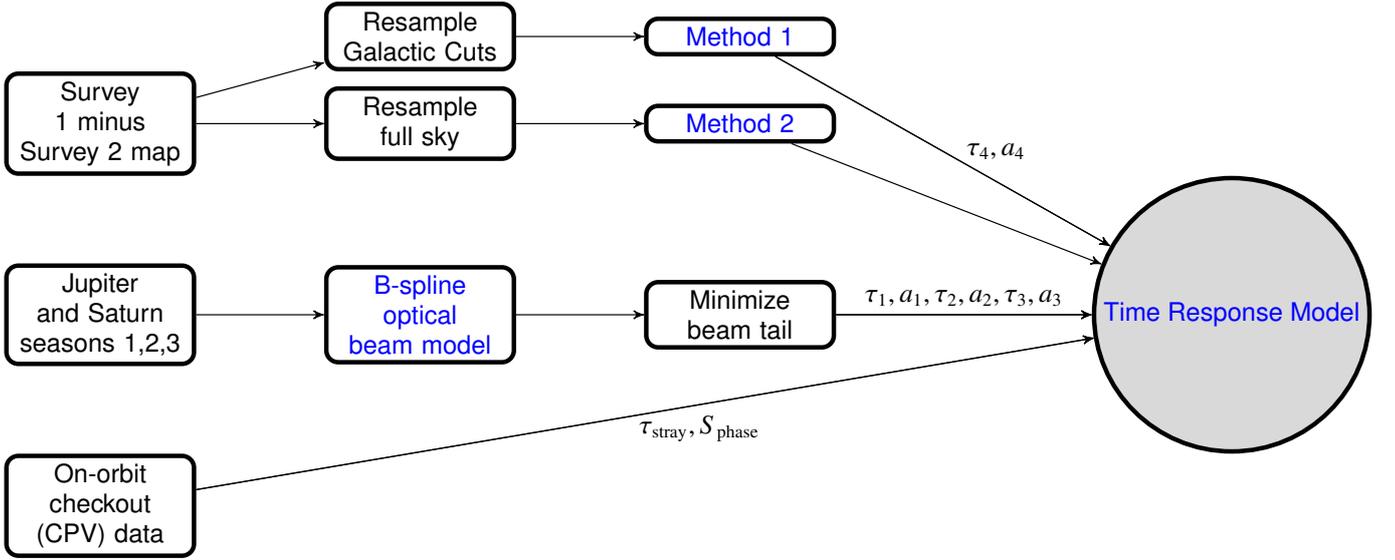
\begin{figure*}[!ht]
    \resizebox{\textwidth}{!}{
      \sffamily\begin{tikzpicture}
        \matrix (m) [matrix of nodes, 
ampersand replacement=\&,
        column sep=1.5cm,
        row sep=0mm,
        nodes={draw, % General options for all nodes
          line width=1.5pt,
          anchor=center, 
          text centered,
          rounded corners,
          minimum width=8mm, minimum height=1mm
        }, 
        % Define styles for some special nodes
        txt/.style={text width=2.0cm,anchor=center},
        empty/.style={draw=none}
        ]
        {
          % m-1-1 empty
          \& % m-1-2 
          |[txt]| {Resample Galactic Cuts}
          \& % m-1-3 
          |[txt]| {\hyperref[sec:method1]{Method 1}}
          \& % m-1-4 empty
          \& % m-1-5 empty
          \\
          % m-2-1 
          |[txt]|{Survey 1 minus Survey 2 map}
          \&% m-2-2
          |[txt]|{Resample full sky}
          \&% m-2-3
          |[txt]|{\hyperref[sec:method2]{Method 2}}
          \& % m-2-4 empty
          \& % m-2-5 
          \\
          % m-3-1 
          |[txt]|{Jupiter and Saturn seasons 1,2,3}
          \&% m-3-2
          |[txt]|{\hyperref[sec:bsplineopticalbeam]{B-spline optical beam model}}
          \&% m-3-3
          |[txt]|{Minimize beam tail}
          \&% m-3-4
          \&% m-3-5
          |[block]|{\hyperref[sec:electronicsmodel]{Time Response Model}}
          \\
          % m-4-1 
          |[txt]|{On-orbit checkout (CPV) data} 
          \&% m-4-2
          \&% m-4-3
          \& % m-4-4 empty
          \& % m-4-5 empty
          \\
        };  % End of matrix
        
        { [start chain] \chainin (m-2-1);
          {[start branch=A] \chainin (m-2-2);
            \chainin (m-2-3);
            \chainin (m-3-5);
          }
          {[start branch=B] \chainin (m-1-2);
            \chainin (m-1-3);
            \chainin (m-3-5)  [join={node[right,xshift=5pt] {$\tau_4,a_4$}}];  
          }
        }
        
        { [start chain] \chainin (m-3-1);
          \chainin(m-3-2);
          \chainin(m-3-3);
          \chainin(m-3-5) [join={node[above] {$\tau_1,a_1,\tau_2,a_2,\tau_3,a_3$}}]; 
          
        }
        
        { [start chain] \chainin(m-4-1);
          \chainin(m-3-5) [join={node[right,xshift=-5pt,yshift=-5pt]
            {$\tau_{\rm stray},S_{\rm phase}$}}];
        }
        
      \end{tikzpicture}
      
    }
\caption{\label{fig:FlowOfTimeResponse} Data flow for fitting time response model
  parameters.}
\end{figure*}

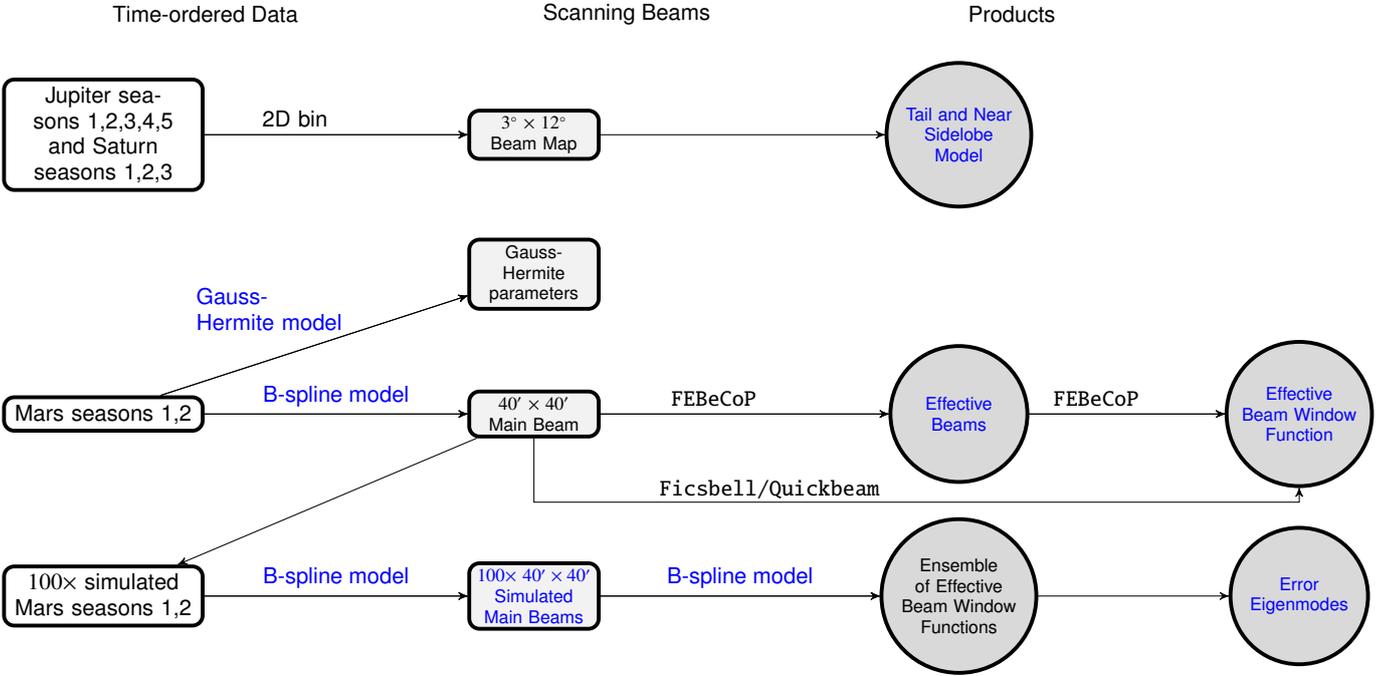
\begin{figure*}[!ht]
    \resizebox{\textwidth}{!}{
      \sffamily\begin{tikzpicture}
        \matrix (m) [matrix of nodes, 
        ampersand replacement=\&,
        column sep=2.2cm,
        row sep=0.4cm,
        nodes={draw, % General options for all nodes
          line width=1.5pt,
          anchor=center, 
          text centered,
          rounded corners,
          minimum width=8mm, minimum height=2.5mm
        }, 
        % Define styles for some special nodes
        txt/.style={text width=2.5cm,anchor=center},
        empty/.style={draw=none}
        ]
  {
  % m-1-1 
   |[lab]| {Time-ordered Data} 
  \& % m-1-2 
   |[lab]| {Scanning Beams}
  \& % m-1-3 empty
   |[lab]| {Products}
  \& % m-1-4 empty
  \\
  % m-2-1 
   |[txt]| {Jupiter seasons 1,2,3,4,5 and Saturn seasons 1,2,3} 
  \& % m-2-2 
   |[beammap]| {$3^{\circ} \times 12^{\circ}$ Beam Map}
  \& % m-2-3 
   |[result]| {\hyperref[fig:JupiterStack]{Tail and Near Sidelobe Model}}
  \& % m-2-4 empty
  \\
% m-3-1 
\&% m-3-2
  |[beammap]|{Gauss-Hermite parameters}
\&% m-3-3
\&% m-3-4
\\
  % m-4-1 
  |[txt]|{Mars seasons 1,2}
\&% m-4-2
   |[beammap]|{$40\arcm \times 40\arcm$ Main Beam}
\&% m-4-3
    |[result]| {\hyperref[sec:effectivebeams]{Effective Beams}}
  \& % m-4-4
    |[result]| {\hyperref[fig:FebecopWindowFunctions]{Effective Beam Window Function}}
  \\
  % m-5-1 
  |[txt]|{$100\times$ simulated Mars seasons 1,2} 
 \&% m-5-2
 |[beammap]|{\hyperref[sec:simulatedplanets]{$100\times$ $40\arcm \times 40\arcm$ Simulated Main Beams}}
 \&% m-5-3
|[result]| {Ensemble of Effective Beam Window Functions}
  \& % m-5-4
|[result]| {\hyperref[sec:erroreigenmodes]{Error Eigenmodes}}
  \\
};  % End of matrix

{ [start chain] \chainin (m-2-1);
  \chainin (m-2-2) [join={node[above,text width=2.0cm] {2D bin}}];
  \chainin (m-2-3);
}

{ [start chain] \chainin (m-4-1);
  {[start branch=A] \chainin (m-4-2) [join={node[above]      {\hyperref[sec:griddedbeamdescription]{B-spline model}}}];
    {[start branch=A1]
    \chainin (m-4-3) [join={node[above,text width=2.0cm] {{\tt FEBeCoP}}}];
    \chainin (m-4-4) [join={node[above,text width=2.0cm] {{\tt FEBeCoP}}}];
  }
% {[start branch=A2]
%\chainin (m-3-4) [join={node[below,text width=2.0cm] {{\tt Ficsbell/Quickbeam}}}];
%}
}
  {[start branch=B] \chainin (m-3-2) [join={node[above,yshift=0.1cm,xshift=-0.6cm,text width=2.0cm] {\hyperref[sec:gausshermite]{Gauss-Hermite model}}}];
  }
}

{ [start chain] \chainin (m-5-1);
  \chainin (m-5-2) [join={node[above]
  {\hyperlink{BSAppendix}{B-spline model}}}];
\chainin (m-5-3) [join={node[above]  {\hyperref[sec:griddedbeamdescription]{B-spline model}}}];

 \chainin (m-5-4); 
}

{ [start chain] \chainin (m-4-2);
  \chainin (m-5-1);
}

%{ [start chain] \chainin (m-6-1);
 % \chainin (m-6-2) [join={node[above] {\hyperref[sec:griddedbeamdescription]{B-spline model}}}];
%
 % \chainin (m-6-3) [join={node[above] {{\tt Ficsbell/Quickbeam}}}];
%  \chainin (m-6-4);
%}

% now add a squared off line around the effective beam node
\draw[->] (m-4-2.south) %-|,-|,->
|- ++(0,-25pt) node[right,text 
width=2.0cm,yshift=5pt,xshift=1.6cm]{{\tt Ficsbell/Quickbeam}} -|
(m-4-4.south);
%\tikzstyle{line} = [draw, -latex']
%\path [line] (m-3-2) -> (m-4-4);

      \end{tikzpicture}
      
    }
\caption{\label{fig:FlowOfBeams} Data flow for determining the HFI beams.}
\end{figure*}

On-orbit measurements of the \Planck\ HFI scanning beam and temporal
transfer function have been previously reported in
\cite{planck2011-1.5} for the Early Release Compact Source Catalog
(ERCSC; \cite{planck2011-1.10}) and early science from \Planck. Section~\ref{sec:timeresponse} presents an improved model for
the time response and explains how its parameters are derived using
on-orbit data, and how it is deconvolved from the data.  Figure~\ref{fig:FlowOfTimeResponse} provides a flowchart of the
determination of the time response parameters. Figure~\ref{fig:FlowOfBeams}
is the flowchart of the steps that lead from planet measurements to
effective beams, effective beam window functions and assessment of
uncertainties. 
Section~\ref{sec:scanningbeams} describes how the scanning beams are
reconstructed from planet observations.
Section~\ref{sec:jupiterresidual} specifically describes the effects of long time
scale residuals in the data due to imperfect deconvolution of the time
response.  Section~\ref{sec:effectivebeams} describes the propagation
of the scanning beams to effective beams and effective beam window
functions using the scanning history of \Planck.
Section~\ref{sec:errors} describes the techniques used to propagate
statistical and systematic errors and to check the consistency of the
beams and window functions.  Section~\ref{sec:totalerrors} describes
the final error budget and the eigenmode decomposition of the errors
in the effective beam window function.  

\section{Time response}
\label{sec:timeresponse}

In \Planck's early data release, a 10-parameter model TF10 
describes the time response of the bolometer/electronics system
\citep{planck2011-1.5}.  This section describes an improved model of the
time response based on the HFI readout electronics schematics  which
more accurately reproduces phase shifts in the system close to the 
Nyquist frequency.  The improved model also provides more degrees of
freedom for the bolometer's thermal response in order to describe more
accurately the low frequency response of the bolometer.

\subsection{Model}

The new model is named LFER4 (Low Frequency Excess Response with four
time constants) and consists of an analytic model of the HFI readout
electronics \citep{lamarre2010} and four thermal time
constants and associated amplitudes for the bolometer:

\begin{equation}
{\rm TF}(\omega) = F(\omega) H'(\omega;S_{\rm phase},\tau_{\rm stray}),
\end{equation}
where ${\rm TF} (\omega)$ represents the full time response as a function of angular
frequency $\omega$.   The time response of the bolometer alone is
modelled by 
\begin{equation}
F(\omega) = \sum_{i=1,4} \frac{a_i}{1 + i\omega\tau_i},
\end{equation}
and $H'(\omega;S_{\rm phase},\tau_{\rm stray})$ is the
analytic model of the electronics transfer function (whose detailed equations and parameters are given
in Appendix~\ref{sec:electronicsmodel}) with two parameters.

The overall normalization of the transfer function is forced to be 1.0 at
the signal frequency of the dipole, leaving a total of 9 free
parameters for each bolometer.  

The sum of single-pole low-pass filters represents a lumped-element
thermal model with four elements.  The thermal model underlying the
temporal transfer function is described elsewhere
\citep{spencer2013}; this work adopts an empirical approach to correcting
the data. 

The two parameters of $H'$ mainly affect the high
frequency portion of the time response. $S_{\rm phase}$ represents the
phase difference between the bias and the lock-in summation, and is
fixed in the model as a readout electronics setting.  The second
parameter $\tau_{\rm stray}$ is the time constant of the bolometer
resistance and the parasitic capacitance of the wiring and is measured
independently during the checkout and performance verification
(CPV) phase of the mission prior to the sky survey.  All resistance and
capacitance values of the readout electronics chain are fixed at values
from the circuit diagram. 

The in-flight data are used to determine the remaining seven free
parameters.  Low frequency parameters are constrained by minimizing the
difference between the first and second survey maps
(Sect.~\ref{sec:galaxyresidualfit}), while planetary observations are used to constrain those 
parameters governing the high frequency portion of the time response
(Sect.~\ref{sec:bsplineopticalbeam}). 

The fastest thermal time constants in the LFER4 model roughly
correspond to the time constants measured during pre-launch tests of the
bolometers \citep{holmes2008,pajot2010}. 
The slower time constants contribute lower frequency response at the
several percent level.  The time constants in the LFER4 model are not exactly
identical to those measured on glitches \citep{planck2013-p03e} due
to additional filtering applied by the deglitching. A future
publication \citep{spencer2013} will relate the
time constants and amplitudes to the thermal properties of the
bolometer and module. 

\subsection{Fitting slow time constants with galaxy residuals}
\label{sec:galaxyresidualfit}
\begin{figure}[!ht]
\centerline{\includegraphics[width=1.0\columnwidth]{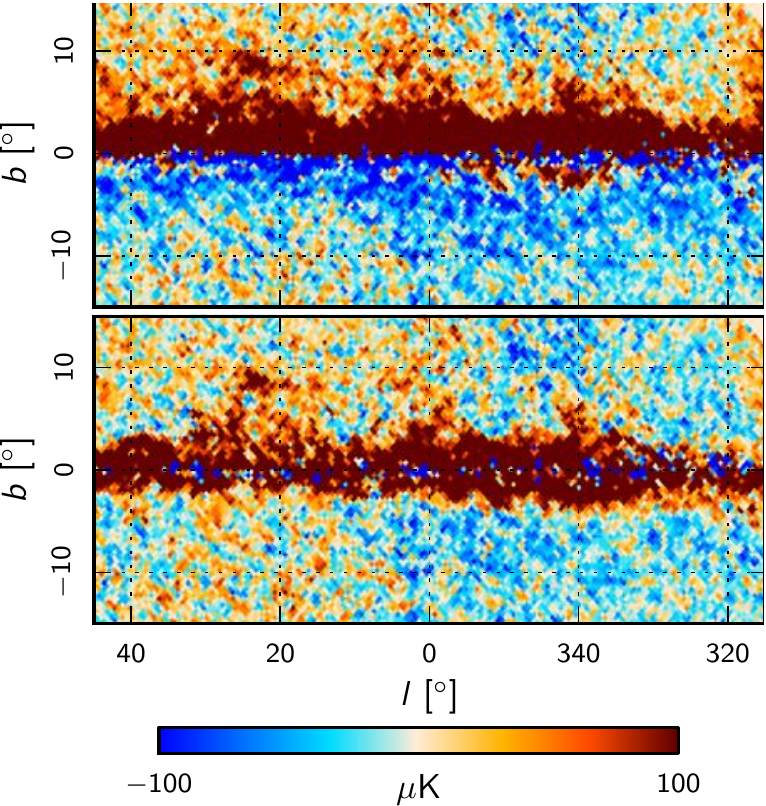}}
\caption{\label{fig:SurveydifferenceExample} Survey 2 minus
  Survey 1 residual close to the Galactic centre before (upper) and
  after (lower) fitting and deconvolving the low frequency part of the time
  response for bolometer 353-2. Remaining residuals are dominated
    by gain difference between the surveys due to ADC nonlinearity
    \citep{planck2013-p03f} and artifacts of the different scanning
    directions (beam asymmetry) and pixel coverage survey to survey.}
\end{figure}
 As \Planck\ scans across the Galactic plane,
the low frequency time response of HFI spreads Galactic signal away
from the plane.  Surveys 1 and 2 consist of roughly six months of data
each and cover nearly the same sky, scanned at almost opposite angles.
The difference between maps made with data from the two individual
surveys highlights the effect since the Galactic power is spread in
different directions in the two surveys, creating symmetric positive
and negative residuals in the difference map.  The LFER4 parameters
are varied to minimize the difference between these surveys.  For most
bolometers, the fit is limited to the slowest time constant and its
associated amplitude, and in others the fit is extended to the two
slowest time constants and associated amplitudes.

Other systematic effects can confuse the measurement of LFER4
parameters by creating a similar positive/negative residual pattern in
the survey difference.  The philosophy employed here is to minimize the survey
residuals fitting only for LFER4 parameters, but to use simulations
and data selections to test the dependency of the results with other
systematics.

\subsubsection{Survey difference method 1}
\label{sec:method1}
Two techniques are used to perform the fit. The first method is based
on map re-sampling in the time domain, using the pointing to generate
synthetic TOI.  The synthetic TOI of each survey is
compared with the synthetic TOI of both surveys combined. Before
the production of these synthetic TOI, the maps are smoothed to
30\arcm.  Given the fact that in consecutive surveys the scan
direction is nearly opposite, the survey 2 TOI is very similar
to the {\it time-reversal} of the survey 1 TOI. This symmetry is
assumed to be exact, ensuring that time reversal is equivalent to
taking the complex conjugate in the frequency domain. Since the sky
signal has been convolved by the true transfer function, ${\rm TF}_{\rm
  true}(\omega)$, and deconvolved by the estimated -- not fully
correct -- transfer function ${\rm TF}_{0}(\omega)$, the single survey
re-sampled time-stream is
\begin{equation} 
d_1(t)=\mathcal F^{-1}({\rm TF}_{\rm true}(\omega)/{\rm TF}_{0}(\omega) \mathcal 
F(s(t')),
\end{equation} 
where $\mathcal F$ is the Fourier transform operator, $s$ is the true sky map 
observed at time $t'$, and $d_1$ is the synthetic TOI. 
This can be written as, 
\begin{equation}
d_1(t)=\mathcal F^{-1}(\delta {\rm TF} (\omega) \mathcal F(s(t')), 
\end{equation}
having defined
$\delta {\rm TF} (\omega)={\rm TF}_{\rm true}(\omega) /{\rm TF}_{0}(\omega) $ 
as the corrective factor which should be applied to the estimated transfer 
function. The synthetic TOI of survey 2, $d_2$, is similarly defined.
For the TOI of both surveys combined, the average of the two 
maps is used. 

Using the time reversal  property, the synthetic TOI $d(t)$ of the full sky map (survey 1 and survey 2 
combined) may be written as
$$
d(t)= \mathcal F^{-1}\left( \frac{\delta {\rm TF} (\omega)+\delta {\rm
      TF}^*(\omega) }2  \mathcal 
F(s(t')) \right).
$$
In the Fourier domain, the ratio of the TOIs is
$$
R(\omega)=\frac{\mathcal F (d_1(t)) }
{\mathcal F (d(t)) }
=
\frac
{\delta {\rm TF} (\omega)}
{\Re(\delta {\rm TF} (\omega))}
=
1 + i\frac{\Im(\delta {\rm TF} (\omega))}
{\Re(\delta {\rm TF} (\omega))}.
$$
Since the real part of the transfer function at low frequency is close
to 1, the imaginary part of the ratio of the synthetic beams is 
equal to the imaginary part of the corrective factor, $\delta {\rm TF} (\omega)$:
$$
\Im(R) = \Im(\delta {\rm TF} (\omega)).
$$
The LFER4 model of the transfer function is fitted to this measure
of the imaginary part of the correction to infer the parameters 
of the true transfer function, in the low frequency regime, typically below 
a few Hz. 

\subsubsection{Survey difference method 2}
\label{sec:method2}
The second method looks at one-dimensional slices through the Galactic
plane for each survey independently.   A sky signal slice is obtained by
  resampling a sky map for a single survey and a single bolometer. 
The slices are taken along the scan direction and the sky signal 
  is averaged over $5\deg$ in longitude.    Only the sky region close to
the Galactic plane is considered (10\deg above
and below the Galactic plane and longitudes between $-40\deg$ and
$60\deg$).  The slice is convolved with the ratio of the
  transfer function used to create the map  and a new  LFER4 transfer
  function with trial parameter values. New sky survey maps are
obtained by re-projecting 
the slices into pixelized maps. As for the previous method, parameters
of the low-frequency part of the LFER4 transfer functions are chosen so
that they minimize the residuals in the difference between surveys 1
and 2.

In practice, the first method is used for the 100--353\,GHz bolometers,
and at 545\,GHz and 857\,GHz the second method is used, being better
adapted to the maps having the most structure in the Galactic
difference residuals.  Figure~\ref{fig:SurveydifferenceExample} shows
an example of the residual remaining in a
{\tt HEALPix}\footnote{\url{http://healpix.sf.net}} map \citep{gorski2005}
of the survey difference, after fitting the long time response, showing
reduced asymmetry in the Galactic plane.

\subsubsection{Survey difference systematics}

In the survey difference solution for the time response, any
systematic effect that creates a residual in the survey difference can
be confused with a time response effect, in particular affecting the
low frequency time response.  This section identifies a
number of residual-producing systematics and quantifies the resulting
bias in the transfer function. These residual-producing systematics include 
far sidelobes, zodiacal light, pointing offset
uncertainty, gain drifts, main beam asymmetry and polarized sky signal.  

\begin{figure}[!ht]
\centerline{\includegraphics[width=1.0\columnwidth]{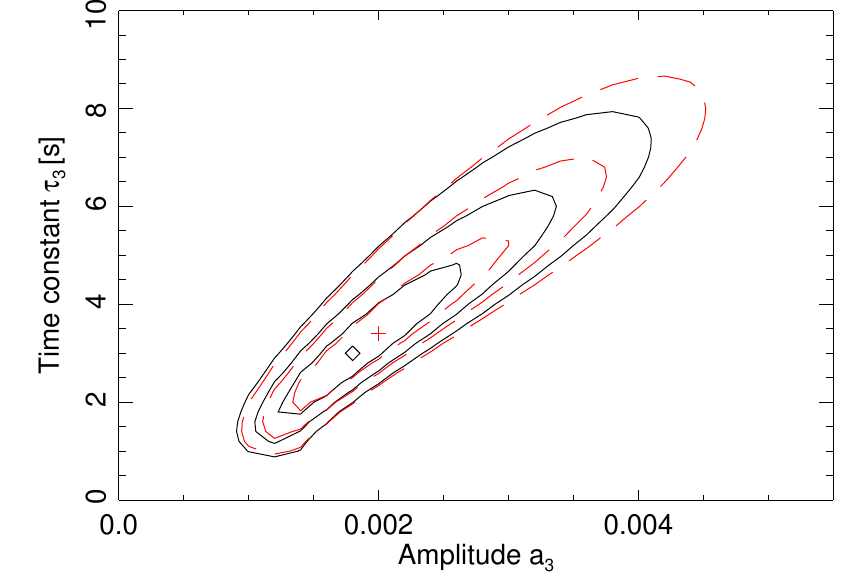}}
\caption{\label{fig:Surveydifferencebias} 68\,\%, 95\,\%, and 99\,\%
  likelihood contours for the long time constant $\tau_3$ and
  associated amplitude $a_3$ for a 545\,GHz bolometer (545-4) with
  (black) and without (red) zodiacal emission and far sidelobe removal.  The
  square and cross indicate the maximum likelihood values.}
\end{figure}

%\subsubsection{Far Sidelobes and Zodiacal Light}
 Due to the very high signal-to-noise ratio of Galactic signal at
 sub-millimetre wavelengths, far sidelobe pickup of the Galactic plane is
 detected in the 545\,GHz and 857\,GHz channels. A physical optics model
 of the far sidelobe pickup is used to estimate the signal from the
 Galactic centre.  The optical depth of the zodiacal dust cloud along
 the line of sight varies as \Planck\ orbits the Sun, leaving a
 residual in the survey difference \citep{planck2013-pip88}. A model
 of the zodiacal dust emission is subtracted in the reconstruction of
 the time response.  The reconstruction of the time response is then
 repeated without subtracting the models; these do
 not significantly affect the result (Fig.~\ref{fig:Surveydifferencebias}).

As a probe of the effect of far sidelobes on the time constant
determination, the pipeline is run on a survey difference map obtained
from the sidelobe model alone for each of the 100, 143, 217, and
353\,GHz channels.  The method did not find a long time constant, as
the sidelobe effect on the survey map is very different from the time
constant effect.

A systematic offset in the pixel pointing creates a residual in the
survey difference.  The pointing solution reduces the
pointing error to a few arcseconds RMS in both the co-scan and cross-scan
direction.   With the 6\deg\,s$^{-1}$ scanning speed, this error corresponds to 
frequencies greater than 1\,kHz, far from the range affected 
by the long time constant.

Gain variability can also bias the estimation; due to nonlinearity in
the analogue to digital converters (ADC), the HFI responsivity to the sky
signal varies at a few tenths of a 
percent throughout the mission.  As a probe of this effect,
gain-corrected data for the 100, 143 and 217\,GHz bolometers are used
to 
reconstruct the long time response.  This has a negligible
impact on the fitted parameters.  Residual gain errors tend to
leave monopolar residuals that are not coupled to the long time
constants in the fitting procedure.

\begin{figure}[!ht]
\centerline{\includegraphics[width=1.05\columnwidth]{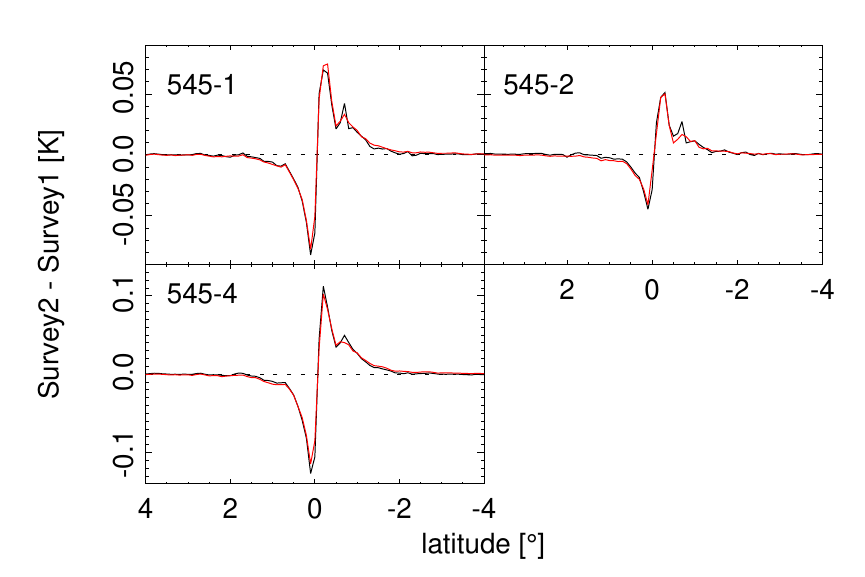}}
\caption{\label{fig:SurveydifferencevsFFP6}  Survey 1 minus
  survey 2  residual for the 545\,GHz bolometers, averaged  from
  Galactic longitude 0 through 20\deg.  The black curves show the
  \Planck\ data, and red is a simulation.}
\end{figure}

Some residual is expected in the survey difference maps because the
asymmetric beam scans the sky at different angles in the two surveys.
This is especially an issue at 545\,GHz where the beam is substantially
asymmetric. To quantify the amplitude, simulated survey difference maps
are generated using a realistic asymmetric beam model convolved with
the Planck Sky Model \citep{delabrouille2012}; the residuals in the Galactic plane at 545\,GHz
are dominated by the main beam asymmetry (see
Fig.~\ref{fig:SurveydifferencevsFFP6}).

Polarization sensitive bolometers (PSBs) show an additional residual in the
survey difference maps because of polarized sky signal observed at
slightly different crossing angles in the two surveys.  For the PSBs,
the low frequency time response is therefore determined using
different levels of polarization masking. These studies do not suggest
the presence of any significant level of bias from polarization, but
only higher noise with wider masking.  As an additional check on the
contribution of residual polarization to the low frequency response,
the survey difference of the sum of the two arms of each PSB
pair is input to the pipeline.  This sum is not sensitive to
polarization, and no bias is found in the determination of the time
response.
 
\subsection{Fitting fast time constants with planet crossings}
\label{sec:bsplineopticalbeam}
Filtering of the TOIs and errors in the deconvolution kernel results in ringing
along the scan direction.  The planets Mars, Jupiter and Saturn
are high signal-to-noise sources that can be used to minimize this
smearing by adjusting the parameters of the LFER4 model.  See
Sect.~\ref{sec:planetdata} for a description of the planet data. 

\begin{figure}[!ht]
\centerline{
\includegraphics[width=1.0\columnwidth]{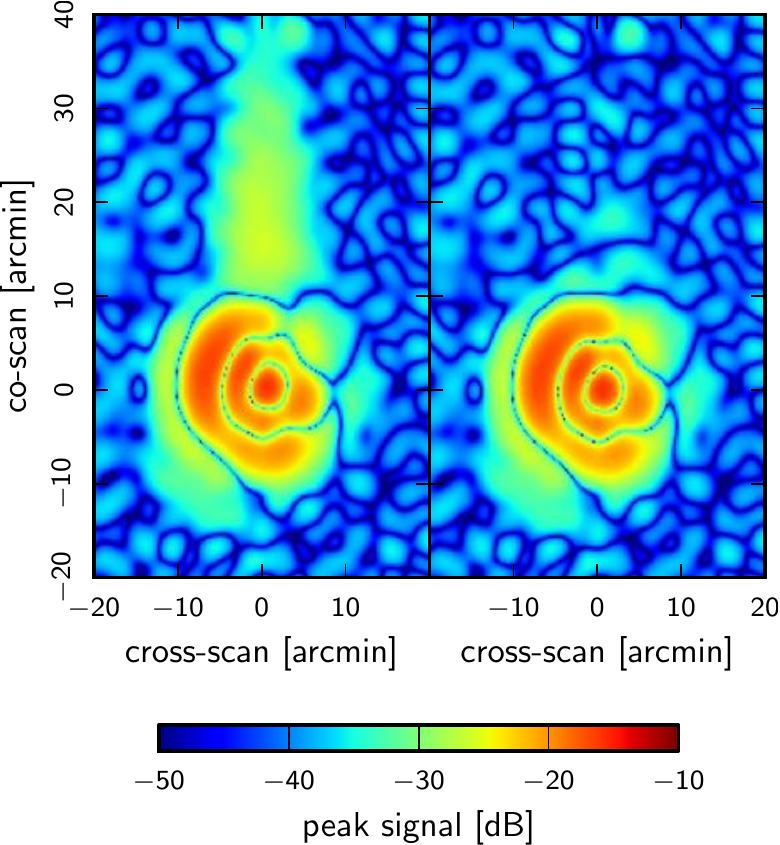}
}
\caption{\label{fig:BeamCompactnessExample} Gridded Jupiter data for
  bolometer 143-3b before and after fitting for LFER4 parameters. The
  best fit Gaussian is subtracted from each plot to emphasize
  residuals. Residuals in the main beam show the deviation  from a
  Gaussian shape, captured in the representations 
  of the scanning beam, as described in
  Appendices~\ref{sec:gausshermite} and
  \ref{sec:griddedbeamdescription}.} 
\end{figure}

In solving for the high frequency portion of the time response, the
beam profile is forced to be compact.  The optical beam is modelled as
a spline function on a two-dimensional grid (see
Appendix~\ref{sec:griddedbeamdescription} for details) and the LFER4
parameters 
are fit by forcing conditions on the resulting beam shape
(Fig.~\ref{fig:BeamCompactnessExample}). 

The planet data are first deconvolved with a time response model
derived from pre-launch data, to recover an initial estimate of the
beam profile.  Jupiter is used for the 100, 143 and 217\,GHz channels, 
while Saturn is used for higher frequencies (see
Sect.~\ref{sec:jupiterresidual} for a discussion of the nonlinearity of
Jupiter at higher frequencies).

Rather than deconvolve the planet data, the model parameters are
determined in the forward sense. Since the beam is decomposed into
B-spline functions, this basis is convolved with the temporal transfer
function to retrieve the coefficients for each basis function using
the planet data. These coefficients are applied to the original
deconvolved B-spline functions to recover an estimate of the optical
beam.  The convolution is made in the Fourier domain by re-sampling
the B-spline function onto a timeline with a sample separation
corresponding to 4\parcs5. The typical knot separation length of the
basis function is between 1\arcm\ and 2\arcm.

In the Fourier domain, the convolution of the temporal transfer
function with the planet signal is
\begin{equation}
\label{eq:tffit1}
\mathcal{PC}(\omega) = \mathcal{B}^{0}(\omega) {\rm TF}^{0}(\omega),
\end{equation}
where $\mathcal{B}^{0}(\omega)$ is the Fourier transform of the slice
through the peak planet signal in the scan direction $b^{0}(t)$, which
is obtained by de-convolving planet data using a fiducial estimate,
${\rm TF}^{0}(\omega)$ of the transfer function. The slice $b^{0}(t)$ is
then symmetrized about the origin (defined by the location of the
  maximum  of the B-spline representation)to obtain 
$b^{{\rm sym}}(t)$, and its Fourier transform $\mathcal{B}^{{\rm sym}}(\omega)$.
This operation aims to recover a beam that, by construction, is
symmetric, within the limits allowed by the model of the temporal
transfer function,
\begin{equation}
\label{eq:tffit2}
\mathcal{PC}(\omega) = \mathcal{B}^{\rm{sym}}(\omega) {\rm TF}^{*}(\omega).
\end{equation}
Here $\mathcal{B}^{\rm{sym}}(\omega)$ is entirely defined by
$\mathcal{PC}(\omega)$ and ${\rm TF}^{0}(\omega)$ and the new
estimate of the time response ${\rm TF}^{*}(\omega)$ is derived
from Eq.~\ref{eq:tffit2}.  This function is parameterized in terms
of the three shortest time constants ($\tau_1$, $\tau_2$, $\tau_3$)
and their associated amplitudes ($a_1$,$a_2$,$a_3$).

\subsection{Stationarity of the time response}

The time response of each HFI detector/readout channel is a function
of the cryogenic temperature of the bolometers and the
ambient-temperature components of the readout electronics.  Both
cryogenic and ambient temperatures change throughout the mission as the
Galactic particle flux and the \Planck\ spacecraft solar
distance are modulated.  The seasonal consistency of the scanning
beam sets an upper limit on changes in the time response through the
mission, shown below in Sect.~\ref{sec:seasonalconsistency}.

\subsection{Deconvolution of the data}
The time response transfer function is deconvolved from the data and
not included as part of the scanning beam, because the low frequency
response of the bolometer would give an extended scanning beam, 
stretching many degrees from the main lobe along the scan direction.

Since the time response amplitude decreases as a function of
frequency, the deconvolution operation increases the noise at high
frequency to an unacceptably high fraction of the RMS.  During the
deconvolution stage of the TOI processing,
a phaseless low-pass filter is applied in order to suppress the high
frequency noise and keep pixel aliasing at a manageable level.

In the early data release, a finite impulse response low-pass
filter was used for this purpose \citep{planck2011-1.7}.  In the 2013
cosmology data release, the low-pass filter is further tuned for the
100, 143, 217 and 353\,GHz channels to reduce filter ripple produced by
the lowpass filters used in the early-release data.  The filter is now implemented in the
Fourier domain, with a kernel consisting of the product of a Gaussian
and a squared cosine,
\begin{equation}\label{eqn:deconvolution}
K(f) = K_1 (f) K_2 (f),
\end{equation}
where
\begin{equation}
%K_1 (f) = \exp\left( - 0.5 (f/f_{\rm Gauss})^2\right) \\
K_1 (f) = e^{- \frac{1}{2} (f/f_{\rm Gauss})^2} \\
\end{equation}
and
\begin{equation}
K_2 (f) = 
\begin{cases}
1 & \text{if } f<f_c, \\
\cos^2 \left( \frac{\pi}{2} \frac{f-f_c}{f_{{\rm max}} -f_c} \right) & \text{if } f_c < f
< f_{\rm{max}}, \\
0 & \text{if } f>f_{\rm{max}}.
\end{cases}
\end{equation}
Here $f_{\rm{max}} = f_c + k (f_{\rm{samp}}/2 - f_c)$ and $f_{\rm{samp}}$ is the
sampling frequency of the data.  The parameters of the filter are the
same for all bolometers in the bands 100--353\,GHz: $f_{\rm{Gauss}}=$ 65\,Hz;
$f_c = $ 80\,Hz; and $k $ = 0.9.  To first order, this filter widens
the scanning beam along the scan in an equivalent way to convolving
the optical beam with a Gaussian with full-width at half-maximum
(FWHM) of 2\parcm07.  The filter introduces some rippling along the 
scan direction at the $-40$\,dB level at 217 and 353\,GHz, where the beams
allow harmonic signal content close to the filter edge.  The rippling
is captured by the B-spline beam representation (see
Fig.~\ref{fig:BeamNearSidelobeExamples}).

The 3-point finite impulse response filter is still used for the
545 and 857\,GHz channels\footnote{The data are convolved with the
  kernel $[0.25,0.5,0.25]$ in the time domain.}.  This extends the
scanning beams along the scan direction more than the Gaussian-cosine
Fourier filter (the 545/857\,GHz filter time scale corresponds to
3\parcm0 on the sky), but has the advantage of reducing rippling and
signal aliased from above the Nyquist frequency.

The deconvolution kernel multiplied by the data in the Fourier domain
is the product of the lowpass filter with the inverse of the
bolometer/electronics time response, 
\begin{equation}
D(f) = K(f) / {\rm TF} (f).
\end{equation}

\begin{figure}[!ht]

\centerline{\includegraphics[width=1.0\columnwidth]{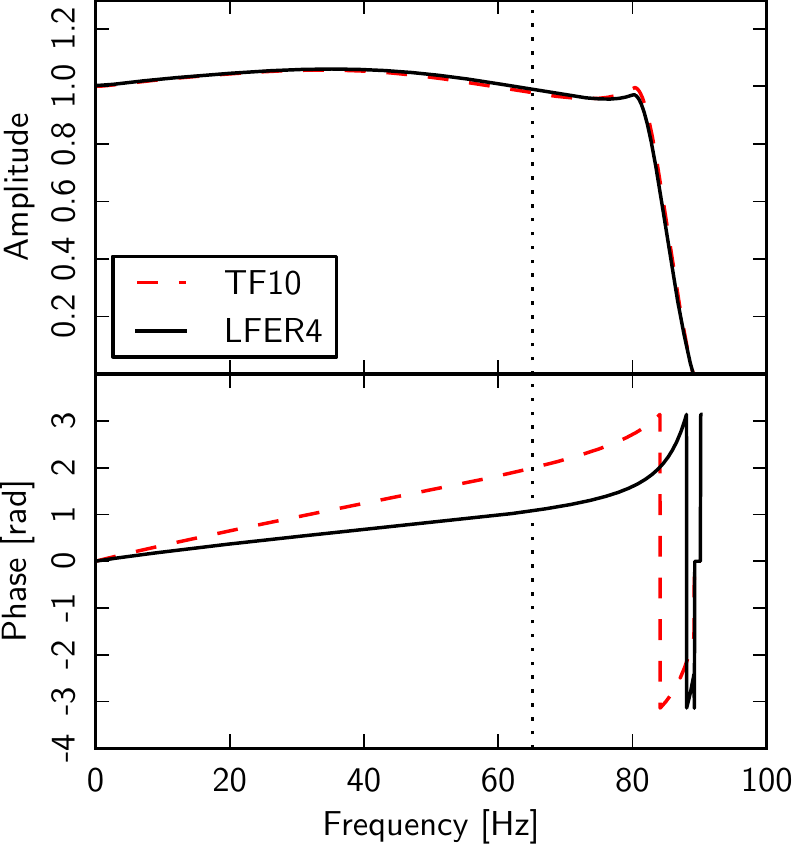}}
\caption{\label{fig:LFER4vsTF10} The phase and amplitude as a function
  of signal frequency of the deconvolution function of bolometer
  217-1.  The solid black curve is the LFER4 model, while the dashed
  red curve shows the TF10 model used in the earlier \Planck\ papers. The
  vertical dotted line marks the signal frequency corresponding to the 
  half power point of the average effective beam.}
\end{figure}

Figure~\ref{fig:LFER4vsTF10} shows a comparison of the deconvolution
functions resulting from the LFER4 model and 
from the TF10 model used in the ERCSC data.
The differences between the two models appear mainly in the 
phase at high frequency, mostly above the signal frequency
corresponding to the beam size.  Although the phase of LFER4 is a more
accurate description of the system, in practice replacing LFER4 with
TF10 had little effect on the data, because of the lowpass filter
applied at the time of deconvolution and the empirical determination
of an overall sample offset in the pointing reconstruction.

In the HFI data processing \citep{planck2013-p03}, data chunks of
length $2^{19} \left(\approx 5 \times 10^5\right)$ samples are Fourier
transformed at a time, overlapping by half 
with the subsequent chunk to avoid edge effects.

\section{Scanning beams}
\label{sec:scanningbeams}

The filtering of the TOI and the accuracy of the
deconvolution kernel affect the angular response of the HFI detectors.
An accurate estimate of the \emph{scanning beam}, resulting from the
filtering of the physical (optical) beam by these time domain
convolutions, is required to relate the angular power spectra of the
maps to that of the underlying sky.  As described in \cite{planck2011-1.5} and
\cite{planck2011-1.7}, the HFI scanning beam profiles are measured using
the planetary observations.  HFI uses two flat sky representations
of the two-dimensional scanning beams, one using Gauss-Hermite
polynomials, and another using B-spline functions.

Three selections of planetary data are used to derive estimates of different aspects of the beam:
\begin{enumerate}
\item the first two observing seasons of Mars (main beam, and window functions);
\item all available seasons of Mars [3], Jupiter [5] and Saturn [4]
  (near sidelobe); and
\item all five Jupiter observations (residual time response).
\end{enumerate}

The effective beam window functions used in the CMB analysis are
ultimately derived from the first of these.  In each case, the
statistical properties of the beam representations and the choice of
planetary data are studied using ensembles of simulated planet
observations (Sect.~\ref{sec:errors}).

While the signal-to-noise ratio of the Jupiter and Saturn data is 
significantly greater than the Mars data, at this stage of
the analysis a reconstruction bias results in the main beams
recovered from {\it simulated} Jupiter and Saturn observations that
is {\it not} present in the simulations of Mars.
Additionally, the
nonlinear response of some HFI detectors to the Jupiter
signal \citep{planck2011-1.5} makes the normalization of the planet
peak response uncertain at the few percent level (see
Sect.~\ref{sec:jupiterresidual}).  Therefore the main beam model is
established using Mars data, while observations of Jupiter and Saturn are used
only to estimate the near sidelobes and residuals in the deconvolution of the
transfer function.

The B-spline representation of the joint Mars observations is used as
input for the calculation of the effective beam and the effective beam
window function. Simulations have shown the B-spline representation
to better capture the features outside of the main lobe,
due to the necessarily finite order of the Gauss-Hermite
decomposition. However, the
Gauss-Hermite model is used for other systematics checks, including
the consistency of the planets and observing seasons.

Because the Jupiter and Saturn data allow measurement of the beam
response below $-45$\,dB from the peak, there is no need to rely on a
model of the telescope to determine the main beam or near sidelobe
structure.

\subsection{Planetary data handling}
\label{sec:planetdata}
The JPL Horizons package\footnote{\url{http://ssd.jpl.nasa.gov/?horizons}}
\citep{giorgini1996} is programmed with \Planck's orbit to calculate the
positions of the planets.  Table~\ref{table:PlanetSeasons} shows the
dates when the planets were within 2\deg\ of the centre of the focal
plane. By the end of HFI operations Mars was observed three times,
Saturn four times, and Jupiter five. 

The planets Jupiter, Saturn and Mars are among the brightest compact
objects seen by \Planck\ HFI; the signal amplitude affects the data
handling in a number of ways.  Moving solar system objects are flagged
and removed from the TOI in the standard processing
pipeline.  Specialized processing for the planet data is required,
with two major differences from the nominal processing (see
\cite{planck2013-p03} for details).

\begin{table}[tmb]                 % table* is a two-column table.  Drop the * for one column.
\begingroup
\newdimen\tblskip \tblskip=5pt
\caption{Observation seasons of the planets observed by \Planck: date
  range and position in Galactic coordinates.}                          % Caption goes here.
\label{table:PlanetSeasons}                            % Label goes here.
\nointerlineskip
\vskip -3mm
\footnotesize
\setbox\tablebox=\vbox{
   \newdimen\digitwidth 
   \setbox0=\hbox{\rm 0} 
   \digitwidth=\wd0 
   \catcode`*=\active 
   \def*{\kern\digitwidth}
   \newdimen\signwidth 
   \setbox0=\hbox{-} 
   \signwidth=\wd0 
   \catcode`!=\active 
   \def!{\kern\signwidth}
\halign{\hbox to 2cm{#\leaderfil}\tabskip 0.5em&\hfil#\hfil&#\hfil\tabskip 1em&\hfil#\hfil\tabskip 1em&\hfil#\hfil\tabskip 0pt\cr                            % Template goes here.
\noalign{\doubleline}
                                    % Table headings go here.
\omit\hfil Planet\hfil&\omit\hfil Season\hfil&\omit\hfil Date Range\hfil&
\omit\hfil$l$\hfil&\omit\hfil$b$\hfil\cr
\omit&&&\omit\hfil[$\deg$]\hfil&\omit\hfil [$\deg$]\hfil\cr
\noalign{\vskip 3pt\hrule\vskip 5pt}
Mars & 1 &2009 22 Oct -- 29 Oct&204.3&*30.6\cr
\omit         & 2 &2010 10 Apr -- 15 Apr&203.3&*31.5\cr
 \omit        & 3 &2011 20 Dec -- 25 Dec&251.3&*60.5\cr
\noalign{\vskip 1pt\hrule\vskip 3pt}
Jupiter & 1 &2009 25 Oct -- 29 Oct&*33.6&$-$40.2\cr
 \omit          & 2 &2010 03 Jul -- 09 Jul&102.4&$-$61.4\cr
\omit & 3 &2010 06 Dec -- 12 Dec&*83.8&$-$61.0\cr
\omit & 4 &2011 03 Aug -- 09 Aug&156.4&$-$43.0\cr
\omit & 5 &2012 10 Jan -- 13 Jan&147.5&$-$49.2\cr
\noalign{\vskip 1pt\hrule\vskip 3pt}
Saturn & 1 &2010 04 Jan -- 08 Jan&286.0&*62.2\cr
\omit & 2 &2010 11 Jun -- 17 Jun& 271.6&*62.5\cr
\omit & 3 &2011 18 Jan -- 22 Jan& 310.3&*58.2\cr
\omit & 4 &2011 29 Jun -- 05 Jul& 298.4&*60.9\cr
\noalign{\vskip 1pt\hrule\vskip 3pt}
Uranus & 1 &2009 05 Dec -- 10 Dec& *81.3&$-$60.1\cr
\omit & 2 &2010 30 Jun -- 05 Jul&  *97.3&$-$60.9\cr
\omit & 3 &2010 10 Dec -- 15 Dec&*89.2&$-$60.8\cr
\omit & 4 &2011 05 Jul -- 10 Jul&105.4&$-$60.6\cr
\omit & 5 &2011 22 Dec -- 26 Dec&*97.4&$-$60.9\cr
\noalign{\vskip 1pt\hrule\vskip 3pt}
Neptune & 1 &2009 31 Oct -- 05 Nov&*39.9&$-$44.5\cr
\omit & 2 &2010 17 May -- 22 May&*45.0&$-$48.1\cr
\omit & 3 &2010 03 Nov -- 07 Nov&*42.0&$-$46.1\cr
\omit & 4 &2011 20 May -- 25 May&*47.4&$-$49.6\cr
 \omit& 5 &2011 16 Nov -- 20 Nov&*44.3&$-$47.7\\\cr
\noalign{\vskip 3pt\hrule\vskip 5pt}
}
}
\endPlancktable                    % ends one-column \halign
%\endPlancktablewide                 % ends two-column \halign
\endgroup
\end{table}                        % table* is a two-column table.  Drop the * for one column.

While the pickup from the $^4$He-JT cooler is removed from the data as
usual, pointing periods containing very bright sources such as the
planets cannot be used to reliably estimate the line amplitude.  Instead,
during the planet observations, these are extrapolated from 
  neighbouring pointing periods.  

To better detect glitches near the extremely bright planet crossings,
an estimate of the planet signal is subtracted from the data prior to
glitch detection.  Glitch template subtraction is performed on the
signal-subtracted timeline in the same way as during nominal
observations.

In order to remove the (quasi-stationary) astrophysical background
from the planetary data, a bilinear interpolation of the
frequency averaged map is subtracted.  The full mission
map is used for the five-season Jupiter data, while the nominal survey
sky maps are used in the processing of the other planetary data.

The planets are oblate spheroids, and appear as ellipses slightly
extended along the direction of the ecliptic.  The \Planck\ planet range
and \Planck\ sub-latitude calculated by JPL Horizons are used in combination
with the polar and equatorial radii of the planets reported by Horizons to
compute the angular size and ellipticity of each planet.
During HFI observations, the mean angular radii of
Jupiter, Saturn and Mars are 20\parcs44, 8\parcs542, and 4\parcs111 ,
respectively.  The ratio of the equatorial to polar radii are 1.069,
1.109 and 1.006, respectively. 

\begin{figure}[!ht]

\centerline{\includegraphics[width=1.0\columnwidth]{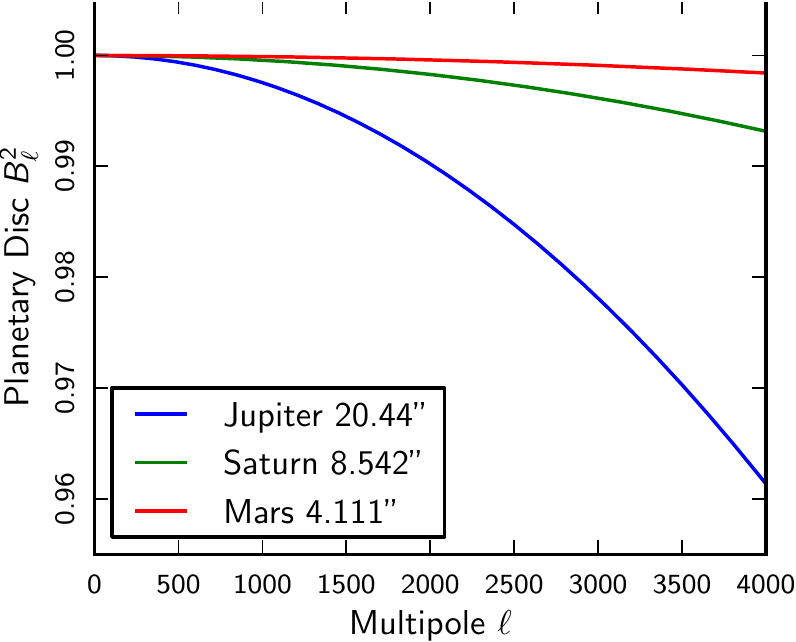}}
\caption{\label{fig:planet_disksize_bl} Window functions of the
  planetary discs of Jupiter, Saturn, and Mars, equivalent to the bias
  in the inferred effective beam window function if the beam is
  reconstructed from observations of one of these planets alone. The
  labels show the corresponding angular radius of each disc.}
\end{figure}

The finite planetary disc size increases the apparent size of the
scanning beam and biases the inferred effective beam window function.
The filtering in multipole space of a circular disc of angular radius
$R$ can be written as $B_{{\rm disc}}(\ell) = 2 J_1 (\ell R) / (\ell R)$,
where $J_1 (x) $ is the Bessel function of order 1.
Figure~\ref{fig:planet_disksize_bl} shows the $B_{{\rm disc}}^2(\ell)$ for
the three planetary discs. In practice, the effects of the disc size
are mitigated by merging observations from the three brightest
planets.  The effects of the large Jupiter disc size are greatest
where the spatial gradient of the beam is greatest, between the -3 to
$-10$\,dB contours of the beam.  By excluding the Jupiter observations
above $-10$\,dB, the disc size smearing is reduced, and setting a
$-20$\,dB threshold results in a bias in the window function below
$10^{-3}$ at all multipoles.

\Planck\ observes Saturn at a range of ring inclination angles:
6\pdeg03, 2\pdeg45, 12\pdeg6  and 9\pdeg4 for seasons 1, 2, 3 and 4,
respectively.  While emission from the solid angle of the ring
increases the effective planetary disc area, the brightness temperature
of the ring tends to be much less than the planetary disc temperature
for the \Planck\ bands. For example \cite{weiland2010} fit a single
90\,GHz ring brightness temperature 14\,\% that of the Saturn
disc.  In our beam reconstruction the average of the first three
\Planck\ observations of Saturn is used, and even in the extreme limit where
the ring brightness temperature is the same and only Saturn data
are used, the multipole space correction is $2\times10^{-3}$ at
$\ell=3000$; this correction is ignored.

The elliptical shape of the planetary disc gives a further bias in the
inferred window function, depending on whether the long axis of the
beam is aligned with or perpendicular to the long axis of the planet.
In this case the planetary disc is approximated as an elliptical
Gaussian with $\sigma = 0.5 R$.  The worst case is a $10^{-3}$ effect
in the case of 100\,GHz beams measured on Jupiter at $\ell=3000$
(where the 100\,GHz window function is vanishingly small).  At 143 and
217\,GHz, the $10^{-3}$ level is reached only at $\ell=4000$.  The
effect is negligible in the range of \Planck's sensitivity.  With
Saturn and Mars, the ellipticity effects are $< 10^{-4}$ and $<
10^{-6}$, respectively, at all multipoles; the ellipticity of the
planetary discs has a negligible effect on the estimate of the
scanning beam.

\subsection{Main beam model} % (fold)
\label{sub:tests_of_the_main_beam_model}

Two representations are used to describe the (two dimensional) main beam;  a
Gauss-Hermite basis (described in Appendix~\ref{sec:gausshermite}
following \cite{huffenberger2010} and \cite{planck2011-1.5}) and a
B-spline basis (Appendix~\ref{sec:griddedbeamdescription}).

The Gauss-Hermite (GH) and B-spline bases have very different
characteristics. The GH representation uses a relatively small number
of parameters and, in practice, amounts to a perturbative expansion
about a Gaussian shape. The B-spline is quite general, using many more
degrees of freedom to fit the data on a defined grid, with the spline
controlling the behaviour in between. The bias and correlation
structure of the noise of these two representations are characterized
using Monte Carlo simulations of the planetary data which include all
the details of the beam-processing pipeline used on the data (these
simulations are described more completely in
Sect.~\ref{sec:errors}).  In each case, the parameters of the
representation are derived directly from the time-ordered data,
without recourse to a pixelized map.

\begin{figure*}[!ht]
\centerline{\includegraphics[width=1.0\textwidth]{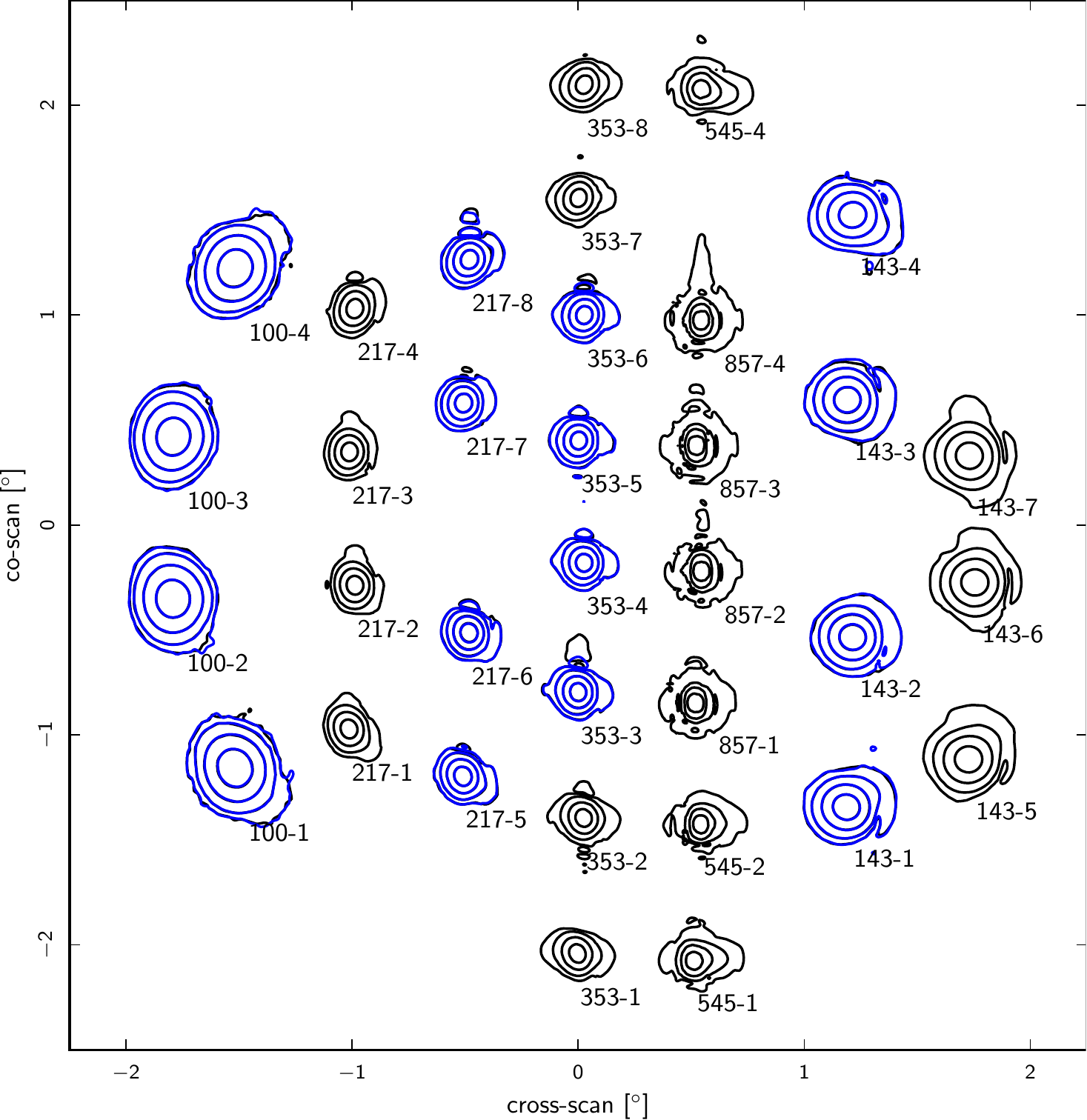}}
\caption{\label{fig:FocalPlanePlot} B-spline scanning beam profiles
  reconstructed from Mars, Saturn, and Jupiter seasons 1, 2 and 3
  data for near sidelobe studies.  The beams are 
  plotted in logarithmic contours of $-3$, $-10$, $-20$ and $-30$\,dB from the
  peak. PSB pairs are indicated with the $a$ bolometer in black and
  the $b$ bolometer in blue.}
\end{figure*}

\begin{figure*}[!ht]
\centerline{\includegraphics[width=1.0\textwidth]{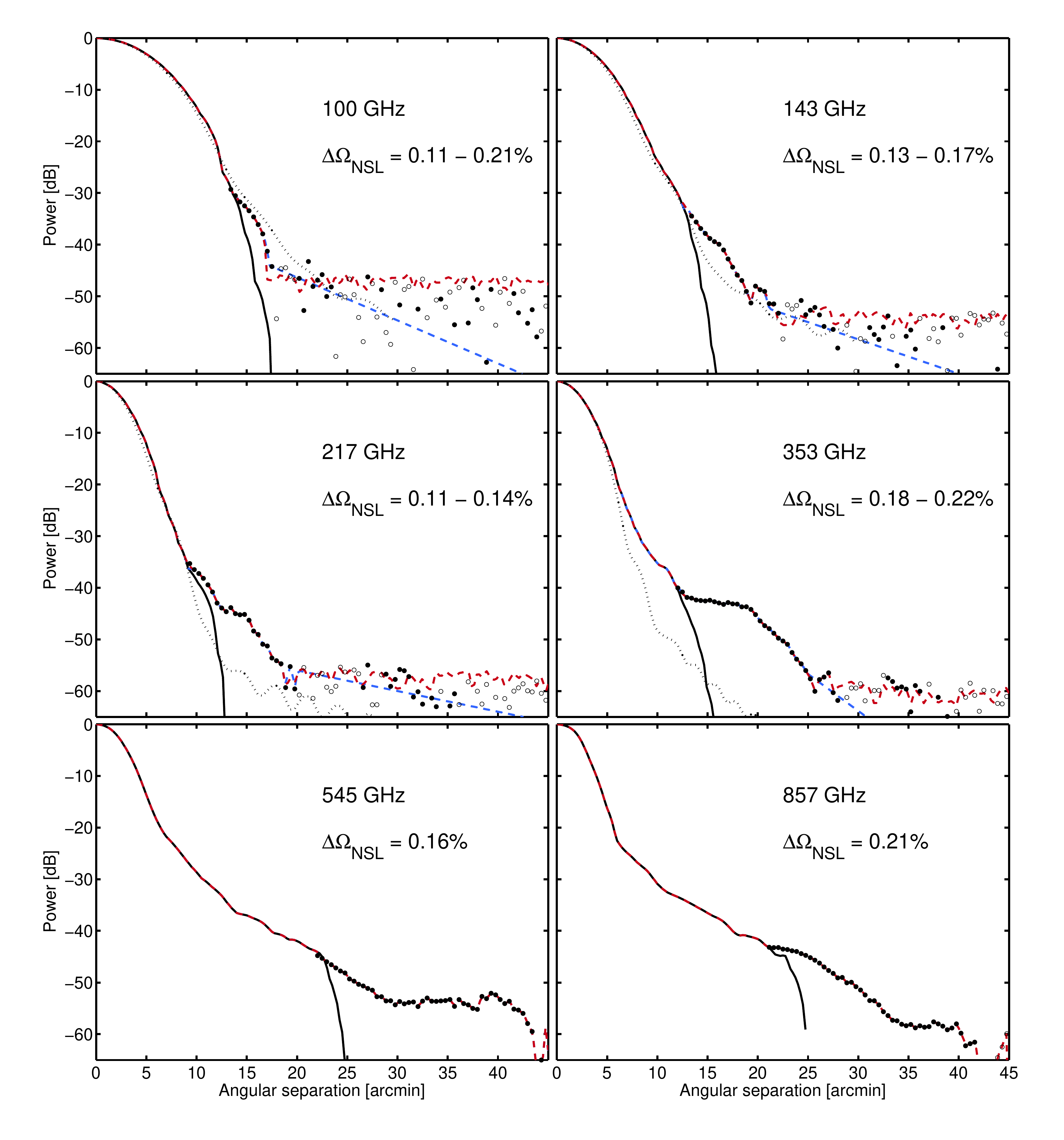}}
	\caption{The azimuthally- and band-averaged main beam profiles
          (black solid curve) derived from the B-spline representation
          of the first and second Mars observations compared to that
          derived from a combination of Mars, Jupiter and Saturn
          observations (filled and open markers represent positive and
          negative data respectively). The red dashed line is defined
          as the joint envelope of the main beam and near sidelobe
          dataset, the integral of which represents the maximal solid
          angle that is compatible with these data.  A nominal near
          lobe model, provided as a reasonable extrapolation of the
          data below the noise floor, is shown as the blue dashed
          line. The fractional increase in solid angle, relative to
          the Mars-alone derived beam profile, is displayed in each
          panel.  The black dotted line shows the {\tt GRASP}
          physical  
          optics model averaged over a subset of detectors that have
          been simulated (100--353\,GHz).  The data show a clear excess
          in power over the model at 143, 217 and 353\,GHz that is
          consistent with a spectrum of surface errors on scales
          between 2 and 12\,cm, with an RMS of order
          10\,$\mu$m. Table~\ref{table:ScanningBeamSolidAngleErrorBudget}
          contains an estimate of the fraction of the solid angle in
          the near sidelobes that is not captured in the B-spline
          representation.  For clarity, the figure extends only to
          45\arcm.  In all cases the solid angle is derived from the
          profile extending out to 5\deg. Due to the high
          signal-to-noise of the Jupiter data ($-40$ to $-55$\,dB,
          depending on the frequency), and the rapidly falling
          response of the beam, the solid angle estimates are
          insensitive to the limit of
          integration.\label{fig:azimuthalBeams}}
\end{figure*}

Figure~\ref{fig:FocalPlanePlot} shows the B-spline scanning beams for
the entire HFI focal plane, as reconstructed from the Jupiter and
Saturn data. Figure~\ref{fig:azimuthalBeams} shows the radially
binned, frequency averaged beam profile for the HFI channels,
comparing the B-spline representation of the Mars data (solid black
lines) with the combined Mars, Jupiter and Saturn data (filled and open points,
and the blue dashed line).  The B-spline maps are apodized with a
Gaussian at a radius beyond the signal-to-noise floor of the Mars
data: 13\parcm4, 13\parcm0, 11\parcm4, 12\parcm9, 17\parcm8, 17\parcm8
on average at 100, 143, 217, 353, 545,and 857\,GHz respectively.

The azimuthally averaged beam window function, $B_\ell^2$, from each
of these models is compared to the known input beam model. At 100, 143
and 217\,GHz the two methods perform comparably,
with the B-spline having slightly smaller bias and variance; at the
higher frequencies the B-spline performs demonstrably better,
especially at 545 and 857\,GHz, as expected due to the highly
non-Gaussian shape of these multi-mode detectors.

\subsection{Near sidelobes}
\label{sec:nearsidelobes}

While the HFI beams are Gaussian at the $-25$\,dB level,
non-trivial structure in the beam is captured in the data at lower
power.  There are two distinct components to the near lobe response: a
discrete pattern of secondary lobes evident at frequencies of 217\,GHz
and above; and a diffuse shoulder consistent with phase errors in the
aperture plane.

The \Planck\ reflectors suffer from print-through of the honeycomb
structure that supports the carbon-fiber face sheets. The size of the
deformation has been measured during thermal testing to be less than
$20\,\mu$m \citep{tauber2010b}. While small in amplitude, the strict
periodicity of this pattern results in a correspondingly periodic
structure in the near lobes, seen clearly in
Fig.~\ref{fig:JupiterStack},which is slightly larger than predicted
based 
on the pre-launch measurements. A simple grating equation describes
the angular positions of the resulting contributions to the near
sidelobes,
\begin{equation}
    \sin \theta _n = \frac{n \lambda}{Yd}.
    \label{eq.grating}
\end{equation}
where $\theta _n$ is the angular position of the order $n$ lobe from
the central beam peak, $\lambda$ is the wavelength of the radiation,
$d$ is the grating periodicity and $Y$ is a factor that
describes the position of each reflector along the optical
path, with $Y=1.00$ for the primary reflector and $Y=1.80$ for the
secondary reflector.  Three possible periodicities (19.6\,mm, 30\,mm,
52\,mm) in the honeycomb array dominate the \Planck\ dimpling pattern
for the 857\,GHz detectors, though only those for the 52\,mm periodicity can be
seen for the 545\,GHz and 353\,GHz detectors. For the highest
frequency detectors, only the weaker lobes due to the 19.6\,mm and
30\,mm periodicities are seen outside the 40\arcm\ main beam model, but
they contribute at most $(0.050\pm 0.008)$\,\%\ to the integrated beam
solid angle.  A forthcoming publication \citep{oxborrow2013} will
present an in-depth study of the mirror surface.

The second component is a beam shoulder becoming significant near the $-30$\,dB
contour, and extending to 2--4 FWHM
from the beam centre. This shoulder is consistent with scattering from
random surface errors on the primary and secondary reflectors, and is
reasonably well-described by a spectrum of surface errors with
correlation lengths ranging from 2 to 12\,cm, with an RMS of order
10\,$\mu$m \citep{Ruze1966}.  The contribution of each of these
components is included in the radially binned profiles shown in
Fig.~\ref{fig:azimuthalBeams}. 

\begin{figure*}[!ht]
\centerline{\includegraphics[width=1.0\textwidth]{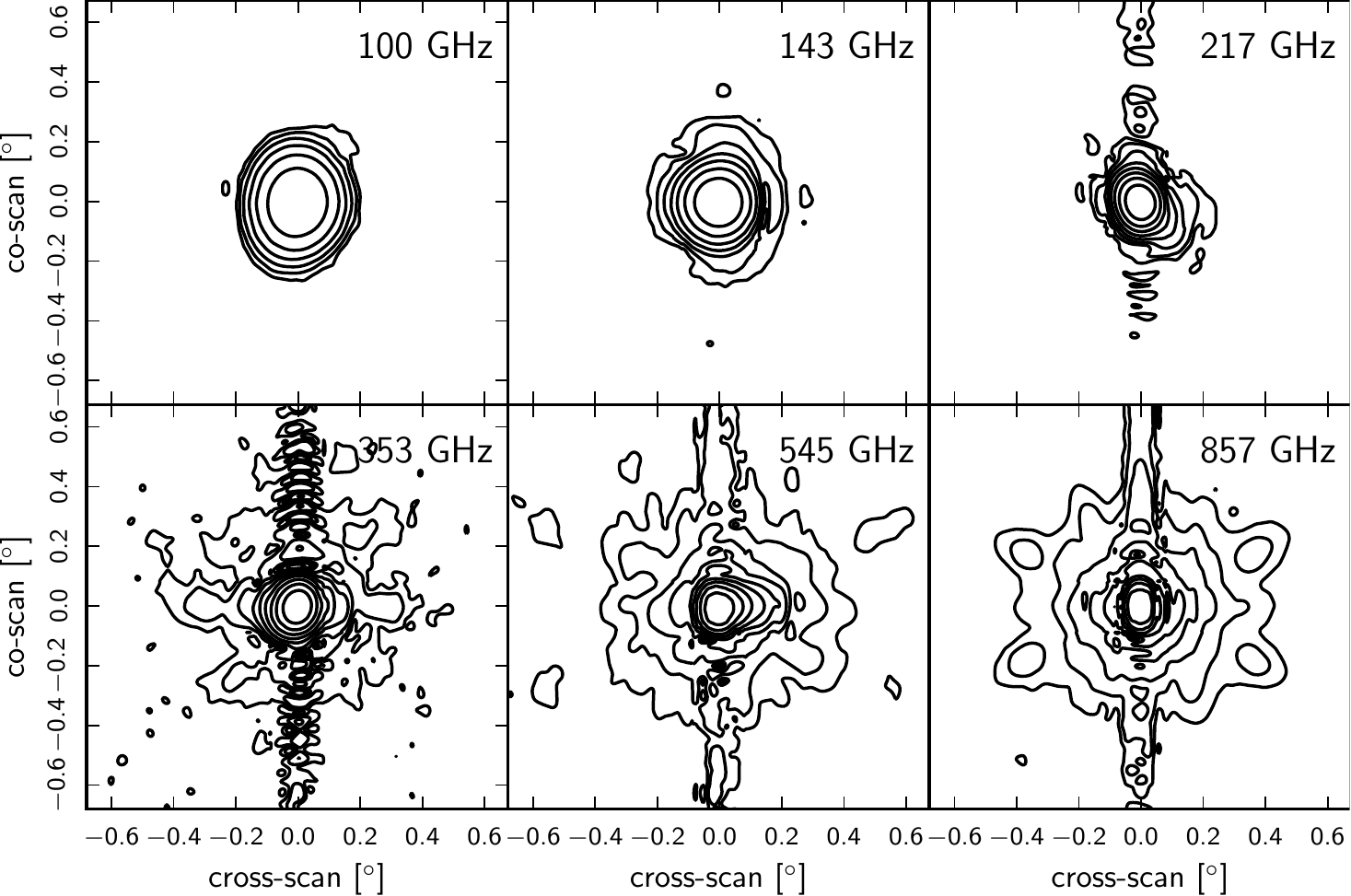}}
\caption{One scanning  beam at each HFI frequency (100-3b, 143-6,
  217-1, 353-7, 545-1, and 857-3).  Contours are in dB from the peak in 
steps of $-5$\,dB.  The lowest contours are set at $-30$\,dB, $-35$\,dB, $-40$\,dB, $-45$\,dB,
$-45$\,dB, $-45$\,dB at 100\,GHz, 143\,GHz, 217\,GHz, 353\,GHz, 545\,GHz, and 857\,GHz, respectively.}
\label{fig:BeamNearSidelobeExamples} 
\end{figure*}

\begin{figure*}[!ht]
\centerline{\includegraphics[width=1.0\textwidth]{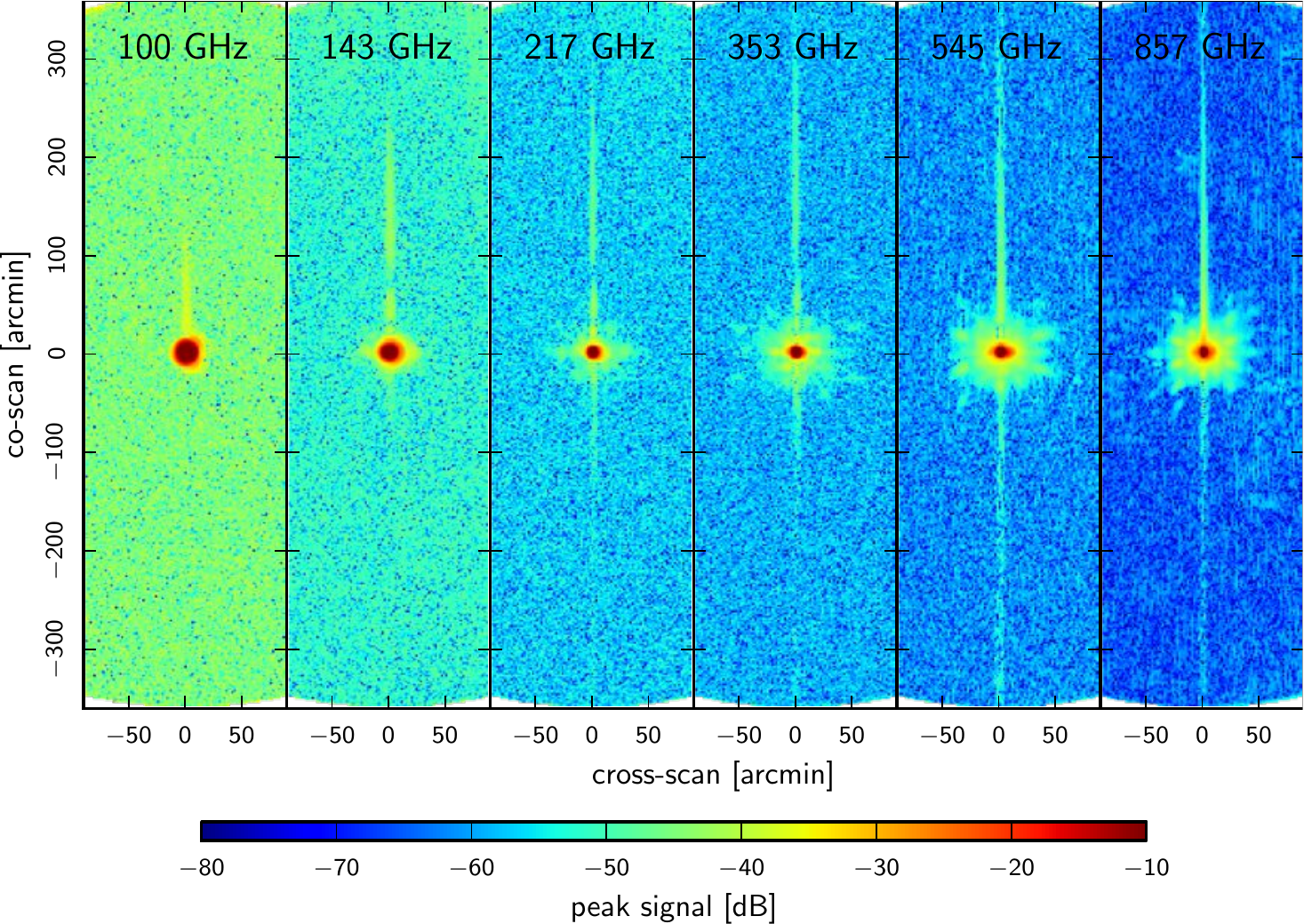}}
\caption{\label{fig:JupiterStack} Gridded data from all five seasons
  of Jupiter.  The colour scale shows the absolute value of the peak signal.}
\end{figure*}

While the B-spline parameterization captures both the main beam and
near sidelobe structure, the extended features must be separately
included in the Gauss-Hermite beam representation.  
Figure~\ref{fig:BeamNearSidelobeExamples} shows contour
plots of a B-spline beam using Mars, Jupiter and Saturn data at each
frequency.

The B-spline representation of the scanning beam used to
compute the window function includes only that portion of the near
sidelobe structure that falls within the signal-to-noise of the Mars
data; the Jupiter and Saturn data provide an estimate of the beam
solid angle that is neglected in the scanning beam product.  The near
sidelobe solid angle, and the resulting window function error, are
sensitive to the details of the analysis, including the sky
subtraction, offset removal and masking of in-scan ringing at the
part-per-million level.  Although a comprehensive study of these
effects in the Jupiter and Saturn data are underway, a conservative,
and model independent upper limit is obtained by taking the envelope
of the noise floor to define the maximum solid angle allowed by the
data (the red dashed line in Fig.~\ref{fig:azimuthalBeams}). A
reasonable estimate of the true solid angle in the near lobes can be
obtained by extrapolating the data below the noise floor (the blue
dashed line in Fig.~\ref{fig:azimuthalBeams}).  By either measure, the
grating lobes and diffuse shoulder account for a small fraction of the
total beam solid angle; for the 100, 143 and 217\,GHz channels this
contribution represents less than 0.15\,\% of the total solid angle (see
Table~\ref{table:ScanningBeamSolidAngleErrorBudget}).  The
amplitude of the impact on the window function is estimated by comparing the
Legendre transform of the maximal envelope of the Jupiter and Saturn
data with that of the nominal Mars derived scanning beam.  The result
is shown as the family of green curves in
Fig.~\ref{fig:allsystematics}.  Because the Monte Carlo ensembles that
are used to derive the error envelope neglect this near sidelobe
structure in the beam that is input to the simulations, the window
function error amplitudes have been scaled to accomodate the upper
limit defined by the noise floor of the Jupiter and Saturn data.

\subsection{Residual time response}
\label{sec:jupiterresidual}

The \Planck\ spacecraft spin rate is constant to 0.1\,\%, making the
time response of the electronics and the bolometer degenerate with
the angular response of the optical system. 

The B-spline beam model extends $\pm $20\arcm\ from the centre of the
beam.  An error in the time response on fast timescales will thus be
exactly compensated by the scanning beam. However, errors in the
time response beyond the limit of the scanning beam
reconstruction will not be accounted for, and will bias the
recovered beam window function.

To look for residual long tails due to incomplete deconvolution of the
time response, all five seasons of Jupiter observations
are binned into a 2-D grid of 2\arcm\ pixel size extending 6\deg\  
from the planet (Fig.~\ref{fig:JupiterStack}).  These are
background-subtracted using the \Planck\ 
maps and stacked by fitting a Gaussian to estimate 
the peak amplitude and centroid of each observation.  The data are
binned as a function of pointing relative to the planet centre.

While a static nonlinearity correction is included in the TOI
processing, partial ADC saturation and dynamical nonlinearity 
bias the normalization of these tails by underestimating the Jupiter peak
signal.  An estimate of the nonlinear correction is derived by fitting a
gridded map of all three Mars observations to a map of all five
Jupiter observations for each detector.  A signal reduction at the
peak of Jupiter is ruled out at the 1\,\% level at 100 and 143\,GHz, but
detected at higher frequencies. Relative to Mars, the peak Jupiter signal is
reduced by nonlinearity on average by 3$\pm$3\,\%, 12$\pm$3\,\%,
12$\pm$4\,\% and 65$\pm$20\,\% at 217, 353, 545 and 857\,GHz, respectively.
The tail normalizations are scaled by these factors.

\begin{figure*}[!ht]
\centerline{\includegraphics[width=1.0\textwidth]{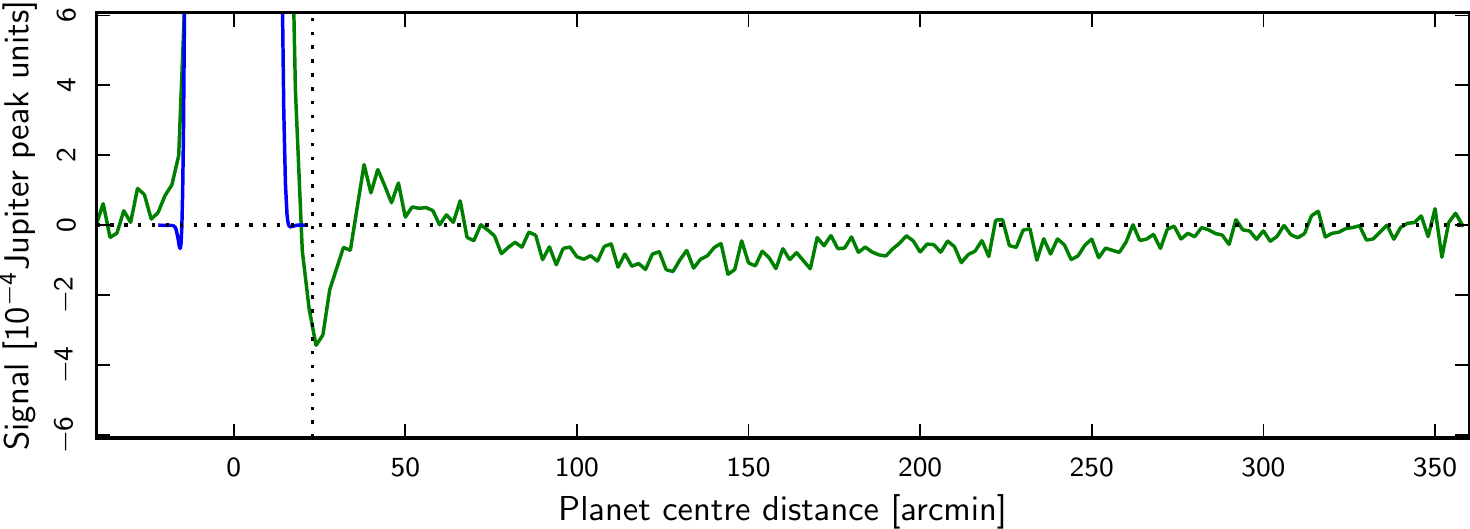}}
\caption{\label{fig:BeamTail1DExample} A slice through stacked Jupiter
  data for bolometer 143-6, illustrating residual long time response
  after deconvolution.  The vertical dotted line shows the extent of
  the scanning beam map (plotted in blue).}
\end{figure*}

An example of the residual tail is shown in
Fig.~\ref{fig:BeamTail1DExample}.  As well as a tail following the
planet due to imperfect deconvolution of the time response, there is
also  a tail preceding the planet crossing; this is due to the lowpass
filter applied in the Fourier domain.  The residual beam tails have
amplitudes typically at the level of $10^{-4}$ of the peak but extend
several degrees from the centre of the beam.  The response beyond
6\deg\ on the sky is consistent with noise for all
detectors.

The signal to noise of the tail measurement greater than 6\deg\ from Jupiter
sets a model-independent limit on the knowledge of the time response at signal
frequencies from 0.016\,Hz to 0.1\,Hz at $<10^{-4}$.

These stacked Jupiter data are integrated to determine the expected
bias in total beam solid angle from this remaining uncorrected time
response (see Table~\ref{table:ScanningBeamSolidAngleErrorBudget}).
The mean integral values are typically a few times $10^{-4}$ of the
main beam solid angle, an order of magnitude lower than the error in
the beam solid angle due to noise and other systematics.

The residual scanning beam tails can also bias the effective beam
window function.  The spherical harmonic transform of the
residual that is not included in the model for the main scanning
beam is computed,
$B_{\ell m}^{\rm{full}} = B_{\ell m}^{\rm{main}} + B_{\ell m}^{\rm{tail}} $, where
$B_{\ell m}^{\rm{main}}$ is the multipole space representation of the main
scanning beam model and $ B_{\ell m}^{\rm{tail}} $ is the multipole space
representation of the long tail model. In all cases, the $m$=0
(symmetric) mode of the ratio of $B_{\ell m}^{\rm{full}}$ to $B_{\ell
  m}^{\rm{main}}$ dominates higher order modes by at least a factor of 1000, meaning
that the coupling to the scan strategy is negligible and the bias in
the effective beam window function can be approximated by  
\begin{equation}
\delta W_{\ell} = \sum_m \frac{|B_{\ell m}^{\rm{full}}|^2}{|B_{\ell
    m}^{\rm{main}}|^2} - 1 ~.
\end{equation}
The main effect on the effective beam window function is that at low
$\ell$, the bias $\delta W_{\ell} $ approaches a value of twice the
fractional contribution to the total solid angle.  When the window
function is normalized to unity at the dipole frequency, the effect is
a roughly constant bias in the window function at a level of a few
$\times 10^{-4}$ for $\ell > 100$.  The contribution of the
residual tail to the window function is neglected in the error budget.

\begin{table*}[tmb]                 % table* is a two-column table.  Drop the * for one column.
\begingroup
\newdimen\tblskip \tblskip=5pt
\caption{Scanning beam solid angle ($\Omega_{{\rm SB}}$) error budget,
  showing the bias and fractional error due to the residual time
  response ($\Delta\Omega_\tau$), near sidelobes
  ($\Delta\Omega_{{\rm NSL}}$) and solid angle colour correction
  ($\Delta\Omega_{{\rm CC}}$). The Monte Carlo error
  ($\Delta\Omega_{{\rm MC}}$)
  includes noise and pointing uncertainty. The colour correction is
  the upper limit in solid angle change due to colour correction from
  a planet spectrum source (roughly $\nu^2$) to \textit{IRAS}-convention
  ($\nu^{-1}$).  The near sidelobe contribution of the nominal
  (maximal) near sidelobe envelopes is shown for the 100 -- 353\,GHz
  bands; there is no appreciable difference between the two in the
  sub-mm bands, due to the extremely high signal-to-noise of these
  data.}
%Caption goes here.
\label{table:ScanningBeamSolidAngleErrorBudget}                            
% Label goes here.
\nointerlineskip
\vskip -3mm
\footnotesize
\setbox\tablebox=\vbox{
   \newdimen\digitwidth 
   \setbox0=\hbox{\rm 0} 
   \digitwidth=\wd0 
   \catcode`*=\active 
   \def*{\kern\digitwidth}
   \newdimen\signwidth 
   \setbox0=\hbox{-} 
   \signwidth=\wd0 
   \catcode`!=\active 
   \def!{\kern\signwidth}
\halign{\hbox to 2cm{#\leaderfil}\tabskip 1em&\hfil# \tabskip 2em&\hfil# \tabskip 2em&\hfil# \tabskip 2em&\hfil# \tabskip 2em&\hfil# \tabskip 0pt\cr                            % Template goes here.
\noalign{\doubleline}
                                    % Table headings go here.
\omit\hfil Band\hfil&\omit\hfil $\Omega_{{\rm SB}}$ \hfil&\omit\hfil
$\Delta\Omega_\tau$ \hfil& \omit\hfil $\Delta\Omega_{{\rm NSL}}$ \hfil
& \omit\hfil $\Delta\Omega_{{\rm CC}}$ \hfil &  \omit\hfil
$\Delta\Omega_{{\rm MC}}$ \hfil\cr
\omit\hfil [GHz]\hfil&\omit\hfil [arcmin$^2$]\hfil&&&&\cr
\noalign{\vskip 3pt\hrule\vskip 4pt}
                                    % Body of table goes here.
100&104.2&$-$0.03*$\pm$*0.04\,\%&*0.11*(0.21)\,\% & $<0.3$\,\%&0.53\,\%\cr
143&*58.4&$-$0.03*$\pm$*0.01\,\%&*0.13*(0.17)\,\% & $<0.3$\,\%&0.14\,\%\cr
217&*26.9&$-$0.03*$\pm$*0.01\,\%&*0.11*(0.13)\,\% & $<0.3$\,\%&0.11\,\%\cr
353&*25.1&$-$0.002$\pm$*0.01\,\%&*0.18*(0.22)\,\% & $<0.5$\,\%&0.10\,\%\cr
545&*25.4&*0.04*$\pm$*0.01\,\%&*0.16******\% & $<2.0$\,\%&0.13\,\%\cr
857&*23.0&*0.09*$\pm$*0.01\,\%&*0.21******\% & $<1.0$\,\%&0.15\,\%\cr
\noalign{\vskip 3pt\hrule\vskip 5pt}
}
}
%\endPlancktable                    % ends one-column \halign
\endPlancktablewide                 % ends two-column \halign
\endgroup
\end{table*}                        % table* is a two-column table.  Drop the * for one column.

\subsection{Far sidelobes}
\label{sec:FSL}
\begin{figure}[!ht]
\centerline{\includegraphics[width=1.0\columnwidth]{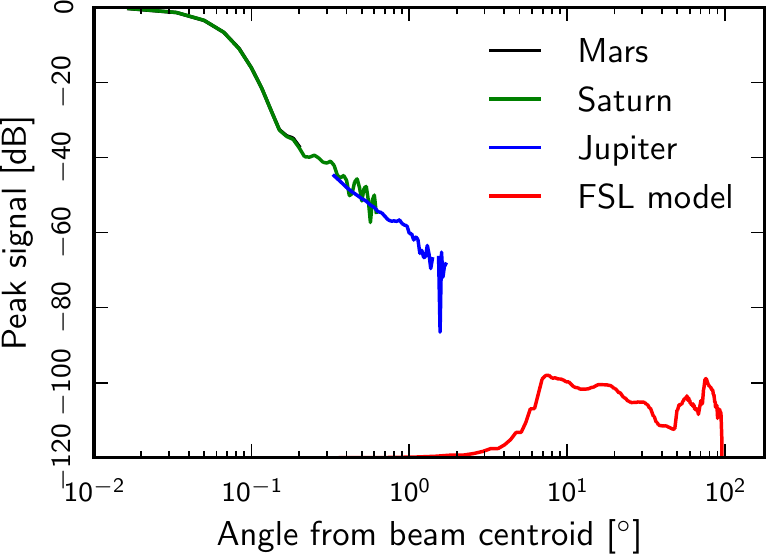}}
\caption{\label{fig:FSLprofile} Azimuthally averaged profiles of
  measured beams of channel 353-1 compared to the
  azimuthal average of the far sidelobe physical optics model.}
\end{figure}

The far sidelobes (FSL) are defined as the
response of the instrument at angles more than $5^{\circ}$ from the
main beam centroid.  \cite{tauber2010b} describes the pre-launch
measurements and predictions of the far sidelobe response using physical
optics models.  Figure~\ref{fig:FSLprofile} compares the measured
beam profile of detector 353-1 with the FSL physical optics model.
The way the FSL are treated in the dipole calibration 
and in the scanning beam model affects the effective beam window
function, and care is needed to check whether the off-axis
response could bias the window function at $\ell > 40$ (angular scales
5\deg) relative to
$\ell < 40$.  To the extent that the physical optics simulations
correctly predict the far sidelobe response (which is shown to be
roughly the case in the survey difference maps), they produce effects
negligible in the effective beam window function of HFI.
Appendix~\ref{sec:FSLappendix} presents the details of this
calculation. 

As a check of the quality of the physical optics model of the far
  sidelobes, \cite{planck2013-pip88}  
attempt a template fit of the physical optics model to the survey
difference maps in combination with a  zodiacal light model.  The
template fits are presented in Table~4 of 
\cite{planck2013-pip88}.      The FSL signature is clearly detected
at 857\,GHz at a level much 
higher than predicted. As these channels are multi-moded \citep{maffei2010}, the
differences are not that surprising; it
is very difficult to perform the calculations necessary for the prediction.
In addition, the specifications for the horn fabrication were
quite demanding, and small variations, though still within the
mechanical tolerances, could give large variations in the amount
of spillover.

For the lower frequency, single-moded channels, there is no clear
detection of primary 
(PR) spillover (i.e., pickup from close to the spacecraft spin axis; see
Appendix~\ref{sec:FSLappendix}). While
the significant negative values of the best fit template amplitude may indicate 
some low-level, large-scale systematic, there seems to be
nothing with the distinctive signature of PR spillover at frequencies
between 100 and 353\,GHz. These values indicate that
the PR spillover values in Table~2 of \cite{tauber2010b} may
be  overestimated.

For the direct contribution of the secondary (SR) spillover, the situation is similar at 
   353\,GHz, but at 217 and 143\,GHz there is a 3\,$\sigma$ detection at about 
   the level expected, while at 100\,GHz the value is about 2.5
   times higher than expected, though the signal-to-noise level of the detection is 
   less than 2\,$\sigma$. The baffle contribution to the SR spillover
   appears higher than predicted, though
   \cite{planck2013-pip88} note that the baffle spillover is difficult
   to model and the diffuse signal is easily contaminated by other
   residuals in the survey difference. 

\section{Effective beams}
\label{sec:effectivebeams}

Unlike \textit{WMAP} \citep{jarosik2010}, for large portions of the sky the scan strategy of \Planck\ does not azimuthally symmetrize the effect of the beams on the CMB map.  Treating the beams as azimuthally symmetric leads to a flawed power spectrum reconstruction.  To remedy this, the effective beam takes into account the coupling between the azimuthal asymmetry of the beam and the uneven distribution of scanning angles across the sky.

The effective beam is computed for each HFI frequency scanning beam and scan
history in real space using the  {\tt FEBeCoP} \citep{mitra2010}  method, as
 in \Planck's early release \citep{planck2011-1.7}.  A
companion paper describes the details of the application of {\tt FEBeCoP} to
\Planck\ data \citep{febecop2}. 

{\tt FEBeCoP} calculates the effective beam at a position
in the sky by computing the real space average of the scanning beam over all
crossings angles of that sky position.   The observed temperature sky
$\widetilde{\mathbf{T}}$ is a convolution of the true sky $\mathbf{T}$
and the effective beam $\mathcal{B}$, 
\begin{equation}
\widetilde{\mathbf{T}} \ = \ \Omega_{\rm{pix}} \, \mathcal{B} \otimes \mathbf{T},
\end{equation}
where $\Omega_{\rm{pix}}$ is the solid angle of a pixel, and the effective beam can be written in terms of the pointing matrix $A_{ti}$ and the scanning beam $P(\hat{r}_j,\hat{p}_t)$ as
\begin{equation}
\mathcal{B}_{ij} \ = \ \frac{\sum_t A_{ti} \, P(\hat{\mathbf{r}}_j, \hat{\mathbf{p}}_t)}{\sum_t A_{ti}} \, ,
\label{eq:EBT}
\end{equation}
where $t$ is the time-ordered data sample index and $i$ is the pixel 
index. $A_{ti}$ is $1$ if the pointing direction falls in pixel number
$i$, else it is $0$, $\mathbf{p}_t$ represents the pointing
direction of the time-ordered data sample, and
$\hat{\mathbf{r}}_j$ is the centre of pixel number $j$, where the
scanning  beam $P(\hat{\mathbf{r}}_j, \hat{\mathbf{p}}_t)$ is being
evaluated (if the pointing direction falls within the cut-off
radius).   

\begin{figure}[!ht]
\centerline{\includegraphics[width=1.0\columnwidth]{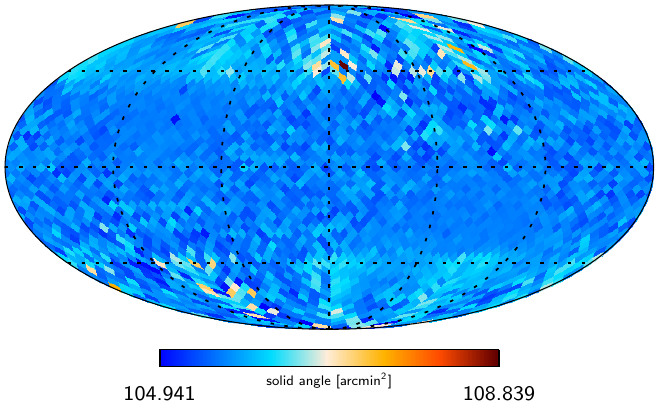}}
\centerline{\includegraphics[width=1.0\columnwidth]{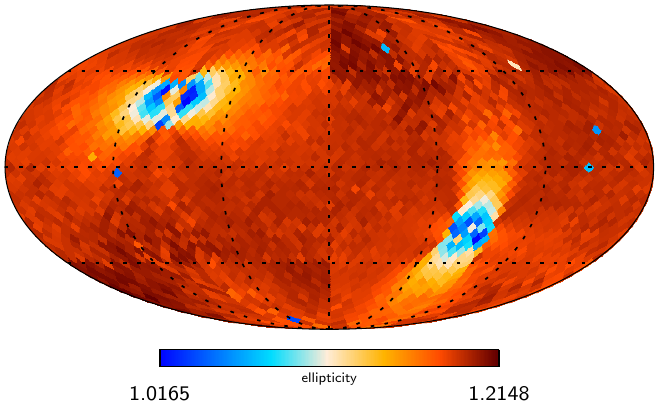}}
\caption{\label{fig:febecop_skydistribution} The effective beam solid
  angle (upper panel) and the best-fit Gaussian
  ellipticity (lower panel) of the 100\,GHz effective beam across the
  sky in Galactic coordinates.}
\end{figure}

The sky variation of the effective beam solid angle and the ellipticity of the best-fit
Gaussian to the effective beam at {\tt HEALPix}
$N_{\rm side}=16$ pixel centres are shown for 100\,GHz in 
Fig.~\ref{fig:febecop_skydistribution}.  The effective beam is less
elliptical near the ecliptic poles, where more scanning angles
symmetrize the beam.  

The mean and RMS variation of the effective beam
solid angle across the sky for each HFI map are presented in Table~\ref{table:BeamSolidAngle}. 

\subsection{Effective beam window functions}

The multipole space window function of one (or two) observed map(s) is defined, in the
absence of instrumental noise and other systematics, as the
ratio of the ensemble averaged auto- or cross-power spectrum of the map(s) to the true theoretical
sky power spectrum
\begin{equation}
	W^\text{\rm eff}_\ell \ = \ \langle C^\text{\rm obs}_\ell \rangle /
	C^\text{\rm sky}_\ell \, .
	\label{eq:effwl_def}
\end{equation}
It must account for
the azimuthal asymmetry of the scan history and the beam profile. This is done
in the HFI data processing pipeline
using the harmonic space method {\tt Quickbeam} (described in
Appendix~\ref{sec:quickbeam}), which allows a quick determination of
the nominal effective beam window  
functions {\em and} of their Monte Carlo based error eigenmodes
(Section~\ref{sec:erroreigenmodes}) for all auto- and cross- 
spectra pairs of HFI detectors, the error determination being
computationally intractable with {\tt FEBeCop}.

In the {\tt FEBeCoP} approach, many (approximately 1000) random 
realisations of the CMB sky are generated starting from a
given fiducial power spectrum $C^\text{\rm in}_\ell$.  
For each beam model, 
the maps are convolved with the pre-computed effective beams, and the pseudo-power spectra 
${\tilde C^\text{\rm    conv}}_\ell$ of the resulting maps are
computed and corrected by the mode coupling kernel matrix $M$ \citep{Hivon2002} for a
given sky mask:
$C_\ell^{\rm{obs}} = M_{\ell \ell'}^{-1} {\tilde C}^\text{\rm
  conv}_{\ell'} \, .$
The Monte Carlo average of kernel-corrected power spectra compared to the
input power spectrum then gives the effective beam window function
(Eq.~\ref{eq:effwl_def}, with $C^\text{\rm in}_\ell$ replacing $C^\text{\rm sky}_\ell$).

In addition, another harmonic space method ({\tt FICSBell};
Appendix~\ref{sec:FICSBell}) was also tested and all three methods give consistent
results for the nominal window functions
at 100--353\,GHz (see Fig.~\ref{fig:EffBeamDifference}).

%-------------------------------------------------------------------------
\begin{figure}[!ht]
\begin{center}
\includegraphics[width=1.0\columnwidth]{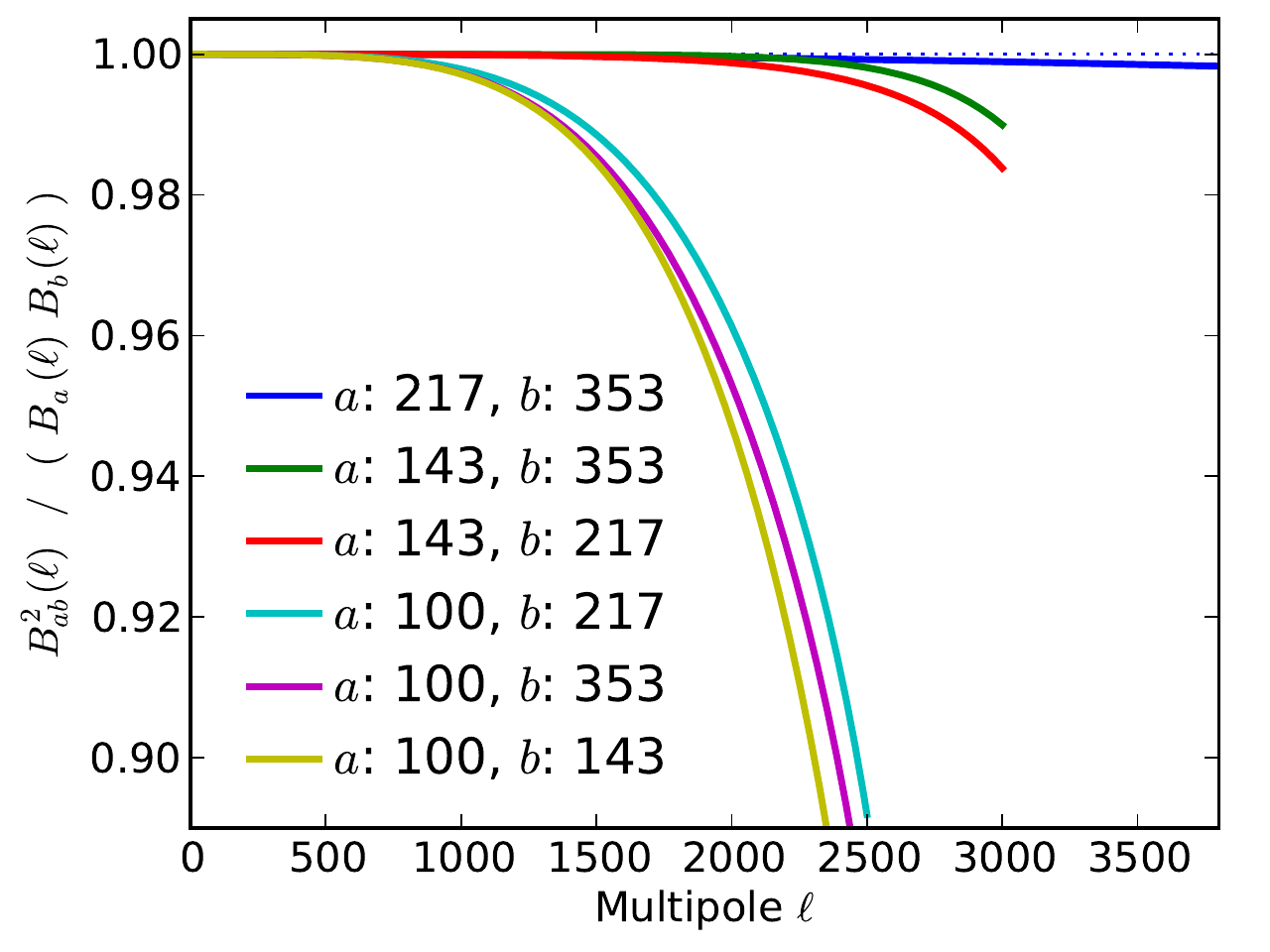}
\caption{Ratios of the cross-beam window functions $B_{ab}^2(\ell)$ to the product of
the respective auto-beam window functions $B_a(\ell) B_b(\ell)$ for maps at 100,
143, 217 and 353\,GHz, illustrating Eq.~\ref{eq:crossbeam}.
\label{fig:Bl_Ratios}%
}%
\end{center}
\end{figure}
%-------------------------------------------------------------------------
For two different maps obtained with different detectors or combination of
detectors $X$ and $Y$, because of the optical beam
non-circularity and \Planck's scanning strategy, the cross-beam window function
is not the geometrical mean of the respective auto-beam window
functions, i.e., 
\begin{equation}
W^{XY}(\ell) \ne \left[W^{XX}(\ell) W^{YY}(\ell)\right]^{1/2}
\quad\text{if}\ X \ne Y, 
\label{eq:crossbeam}
\end{equation}
as illustrated in Fig.~\ref{fig:Bl_Ratios}, while of course $W^{XY} = W^{YX}$
for any $X$ and $Y$.%

\begin{figure}[!ht]
\centerline{\includegraphics[width=1.0\columnwidth]{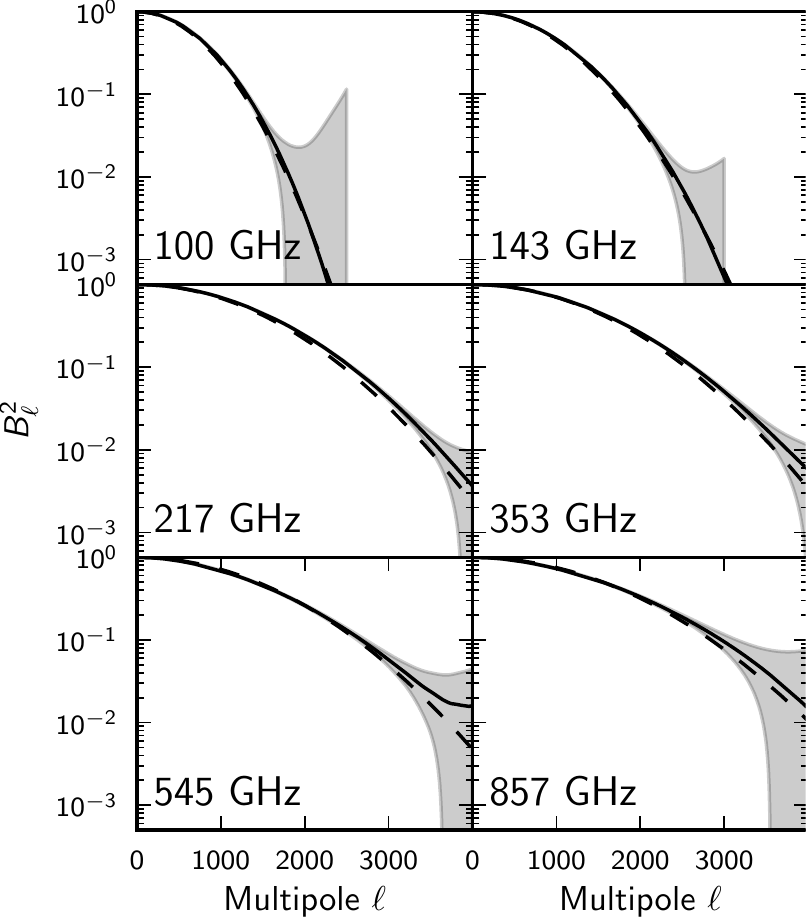}}
\caption{\label{fig:FebecopWindowFunctions} Effective beam window
  functions (solid lines) for each HFI frequency.  The shaded region
  shows the $\pm 1 \sigma$ error envelope.  Dashed lines show
  the effective beam window function for Gaussian beams with
  FWHM  9\parcm65, 7\parcm25, 4\parcm99, 4\parcm82, 4\parcm68, and 4\parcm33 
  for 100, 143, 217, 353, 545 and 857\,GHz, respectively.}
\end{figure}
The effective beam window functions for the 2013 maps are shown in
Fig.~\ref{fig:FebecopWindowFunctions}.  Variations in the effective
beam window function from place to place on the sky are significant;
the published window functions have been appropriately weighted for
the analysis of the nominal mission on the full sky.  Analyses
requiring effective beam data for more restricted data ranges or for
particular regions of the sky should refer to the specialized tools
provided in \cite{planck2013-p28}.

\begin{table*}[tmb]                 % table* is a two-column table.  Drop the * for one column.
\begingroup
\newdimen\tblskip \tblskip=5pt
\caption{Mean values of effective beam parameters for each HFI
  frequency. The error in the solid angle $\sigma_\Omega$ comes from
  the scanning beam error budget.  The spatial variation $\Delta \Omega$ is the RMS
  variation of the solid angle across the sky.  Two values are
  reported for the FWHM: the first is the FWHM of the
  Gaussian whose solid angle is equivalent to that of the effective
  beams; the second in parenthesis is the mean best-fit
  Gaussian. $\Omega_1$ and $\Omega_2$ are the solid angles contained
  within a ring with radius 1 and 2\,FWHM respectively
  (used for aperture photometry as described in Appendix A of \cite{planck2013-p05}).  The
  ellipticity $\epsilon$ is the ratio of the major to minor axis of
  the best fit Gaussian, averaged over the full sky. $\Delta \epsilon$
  is the RMS variability over the sky of the ellipticity.}  % Caption goes here.
\label{table:BeamSolidAngle}                            % Label goes here.
%\nointerlineskip
\vskip -3mm
\footnotesize
\setbox\tablebox=\vbox{
   \newdimen\digitwidth 
   \setbox0=\hbox{\rm 0} 
   \digitwidth=\wd0 
   \catcode`*=\active 
   \def*{\kern\digitwidth}
   \newdimen\signwidth 
   \setbox0=\hbox{+} 
   \signwidth=\wd0 
   \catcode`!=\active 
   \def!{\kern\signwidth}
\halign{\hbox to 2cm{#\leaderfil}\tabskip 1em&\hfil#\hfil \tabskip 1em&\hfil#\hfil \tabskip 1em &\hfil#\hfil \tabskip 1em&\hfil#\hfil \tabskip 1em &\hfil#\hfil \tabskip 1em &\hfil#\hfil \tabskip 1em &\hfil#\hfil \tabskip 1em &\hfil#\hfil \tabskip 0pt\cr                            % Template goes here.
\noalign{\doubleline}
                                    % Table headings go here.
\omit\hfil Band\hfil&$\Omega$&$\sigma_\Omega$&$\Delta \Omega$&FWHM&$\Omega_1$&$\Omega_2$&$\epsilon$&$\Delta \epsilon$\cr
\omit\hfil [GHz]\hfil&[arcmin$^2$]&[arcmin$^2$]&[arcmin$^2$]&[arcmin]&[arcmin$^2$]&[arcmin$^2$]&&\cr
\noalign{\vskip 3pt\hrule\vskip 5pt}
                                    % Body of table goes here.
100&105.78&0.55&0.31&9.66 (9.65)&100.83&105.78&1.186&0.023\cr
143&*59.95&0.08&0.25&7.27 (7.25)&*56.81&*59.95&1.036&0.009\cr
217&*28.45&0.03&0.27&5.01 (4.99)&*26.44&*28.43&1.177&0.030\cr
353&*26.71&0.02&0.25&4.86 (4.82)&*24.83&*26.65&1.147&0.028\cr
545&*26.53&0.03&0.34&4.84 (4.68)&*24.29&*26.30&1.161&0.036\cr
857&*24.24&0.03&0.19&4.63 (4.33)&*22.65&*23.99&1.393&0.076\cr
\noalign{\vskip 3pt\hrule\vskip 5pt}
}
}
%\endPlancktable                    % ends one-column \halign
\endPlancktablewide                 % ends two-column \halign
\endgroup
\end{table*}                        % table* is a two-column table.  Drop the * for one column.

\section{Uncertainties and robustness}
\label{sec:errors}
Ensembles of simulated planetary observations are used to propagate
noise and other systematic effects in the time-ordered data into
errors in the beam reconstruction. These simulations are also used to
estimate bias in the reconstruction of the beam by comparing the
ensemble average beam with the input beam.

To assess the completeness of the statistical and systematic noise
model, the consistency of the beam reconstruction derived from
different planetary observations is measured against corresponding
Monte Carlo ensembles.  The MC-derived beam statistics, including both
the bias and the correlation structure of the errors, have been found
to be somewhat sensitive to the near lobe structure that is included
in the beam used as an input to the simulations. An investigation of this
effect is ongoing.  

In addition, several sources of systematic error that can potentially
impact the beam determination, but are not included in the current
simulation pipeline, have been investigated. The most significant of
these are the beam colour correction and distortion due to ADC
nonlinearity.  These errors are estimated separately and found to be
small in comparison to the error bars estimated in the eigenmode
analysis described in Sect.~\ref{sec:totalerrors}.

\subsection{Simulation of planet observations}
\label{sec:simulatedplanets}
The bias and uncertainty in the scanning beams are determined using
ensembles of simulated planet observations.  The simulation pipeline
uses the LevelS \Planck\ simulation code \citep{reinecke2006} to
generate simulations of the first two observations of Mars and the
first three observations of Jupiter and Saturn for each bolometer.
This pipeline is used to generate 100 simulations for each bolometer
at 353\,GHz, 545\,GHz and 857\,GHz, 200 at 143\,GHz and 217\,GHz, and 400
at 100\,GHz. Each simulation is reconstructed into a beam model using
the identical procedure and software as for the real data.

The components of the simulations are as follows.
\begin{enumerate}
\item The scanning beams used as input to the simulations are
  the same Mars beams used to calculate the effective beam window function.  As shown in
  Appendix~\ref{sec:gausshermite}, the reconstruction bias depends
  more on the beam representation than the exact input beam used in a
  simulation.
\item The \Planck\ spacecraft pointing and the 
  Horizons ephemerides are used to calculate the pointing relative to the planets
  for the simulation. An additional random 2\parcs5\  
  RMS per sample pointing error is added in both the in-scan and cross-scan
  directions with a power spectral density given by the pointing error estimate,
  consistent with the estimated error in the spacecraft pointing (see
  section 4, and figure 7 of \cite{planck2013-p03}).  Error in the beam
  centroid determination is not simulated.
\item Detector noise is generated in the timeline by a random
  realization of Gaussian noise with a power spectral density as
  reported in \cite{planck2013-p03}.
\end{enumerate}

For a small number of the simulations, cosmic ray glitches are added
to the simulation with the energy spectrum and rate measured in the
data \citep{planck2013-p03e}, and the deglitching procedure from the
standard pipeline is applied to detect samples contaminated by glitch
transients and to subtract the long tails \citep{planck2013-p03}.

Each simulation in the ensemble provides an estimate of the scanning
beam, which is then used as input to {\tt Quickbeam}
(Appendix~\ref{sec:quickbeam}) to estimate an effective beam window
function 
(EBWF).  The resulting ensemble of EBWFs are then used to compute the
bias and errors on the EBWF derived from the data; the mean of the
ensemble provides a measure of the reconstruction bias, and the
distribution of the ensemble gives the uncertainty.  The bulk of the
RMS of the ensemble can be captured with a small number of
eigenmodes (Sect.~\ref{sec:erroreigenmodes}).

This procedure allows for the direct determination of the EBWF and
associated errors for each of the detector cross-correlations input to
the angular power spectrum likelihood, described in \cite{planck2013-p08}.

\subsection{Absolute consistency: comparison of systematics against simulations} 
\subsubsection{Seasonal consistency}
\label{sec:seasonalconsistency}
\begin{figure}[htbp]
\begin{center}
        \includegraphics[width=1\columnwidth]{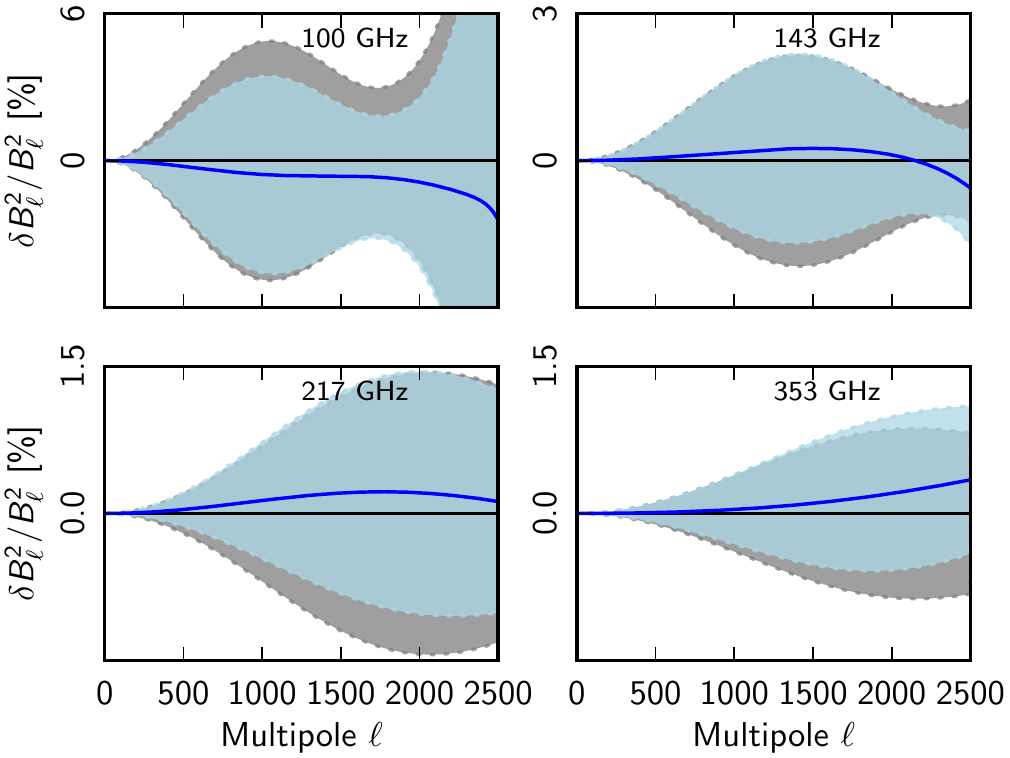}
	\caption{\label{fig:Mars1-Mars2-Bl2-all}The seasonal
          consistency of the beams at four HFI frequencies.Thick blue lines show the difference
            of the Mars season 1 and Mars season 2 window functions
            normalized to the 
            band-average full window function. The grey shaded band
            is the $\pm$1\,$\sigma$ error for Mars 1 and the blue
            shaded band is the $\pm$1\,$\sigma$ error for Mars 2.}
\end{center}
\end{figure}

One important test of the consistency of the scanning beam measurement
is the stability over time, as measured in different seasons and on
different objects over the course of the mission. The difference of
the scanning beam window function between the first two observations
of Mars is shown in Fig.~\ref{fig:Mars1-Mars2-Bl2-all}.  The
residuals are well within the estimated uncertainty that is derived
from the simulation ensemble for 100--353\,GHz. Due to the sparse
cross-scan sampling, the B-spline beam representation does not
converge for a single planetary observation; for this test the
Gauss-Hermite (GH) formalism is used. 

\subsubsection{Time variability of planet flux density}
Mars is known to have diurnal variability at  HFI frequencies due to
the non-uniformity of the surface 
albedo\footnote{\url{http://www.lesia.obspm.fr/perso/emmanuel-lellouch/mars/}}.
A single detector scans Mars on time scales of a few hours, during
which the brightness temperature of Mars can vary by several percent.
Additionally, the Mars planetary disc area varies by several percent
during an observation as it moves relative to \Planck.  To assess the
impact of Mars variability on the scanning beam determination, GH
representations of the beam are derived for each bolometer, with and
without a model for the diurnal variability of Mars.  Using $\chi^2$
tests for goodness of fit, these results are indistinguishable.

\subsubsection{Beam colour correction}
The scanning beams are measured using planets, whose spectral energy
distribution (SED) roughly follows the Rayleigh-Jeans spectrum.
However, the beam window functions from these measurements are used to
correct the angular power spectra of CMB anisotropies, which is
characterized by a very different spectrum, one that is falling in
power as a function of frequency relative to a Rayleigh-Jeans spectrum
across the HFI bands.  

Physical optics simulations, using the {\tt
  GRASP}\footnote{TICRA, \url{http://www.ticra.com}} software, are used 
to investigate this 
effect.  For each HFI channel from 100--353\,GHz, monochromatic
simulations are generated at five frequencies across each band using
the pre-launch telescope model \citep{maffei2010,tauber2010b}. The
solid angle of these simulated beams varies across the band, due to
diffraction and focusing effects.  For 100--217\,GHz, the
beam size comes to a minimum near the band centre, while at 353\,GHz
the beam size \textit{rises} with frequency.

No attempt is made to colour-correct the planet-derived window
functions, because a telescope solution has not yet been determined
that agrees with the measured solid angles.  Spot checks at 143 and
353\,GHz with a defocused telescope model improve agreement between
the data and the model, but does not change the trend of solid angle
with frequency across the band.  The investigation of these effects
suggests that numerical models can bound the uncertainty, but cannot
reliably predict a bias. Rayleigh-Jeans-weighted average and CMB
anisotropy-weighted average beams are produced and used to compute an
effective beam window function (blue curves in Fig.~\ref{fig:allsystematics}).
The largest bias results at 353\,GHz.  For the frequencies 100, 143,
and 217\,GHz, however, the band average 
effect is less than 0.1\,\% across the entire multipole range used in
the CMB likelihood.

The beam solid angle also varies as a function of source SED; the
{\tt GRASP} simulations constrain the size of the beam solid angle colour
correction from a $\nu^2$ source SED (like the planets) to the \textit{IRAS}
convention $\nu^{-1}$. At 545 and 857\,GHz, the {\tt GRASP} models of
\cite{murphy2010} provide a measure of the effec,t which is found to be
significant at 353\,GHz, but not at the other frequencies, at
the level of a few tenths of a percent (Table~\ref{table:ScanningBeamSolidAngleErrorBudget}).

\subsubsection{ADC nonlinearity}

Nonlinearity in the ADC in the HFI readout
electronics mainly manifests itself as an apparent gain drift
\citep{planck2013-p03,planck2013-p03f}, however,  
the measured beam shapes are also biased by the effect. 

Correcting the ADC nonlinearity relies on a model that predicts which
ADC codes contribute to each data sample.  The 
presence of large signal gradients, such as a planet, or pickup from
the $^4$He-JT cooler, make modeling difficult.  A model for
correcting every detector is still under development.

\begin{figure}[!ht]
\centerline{\includegraphics[width=1.0\columnwidth]{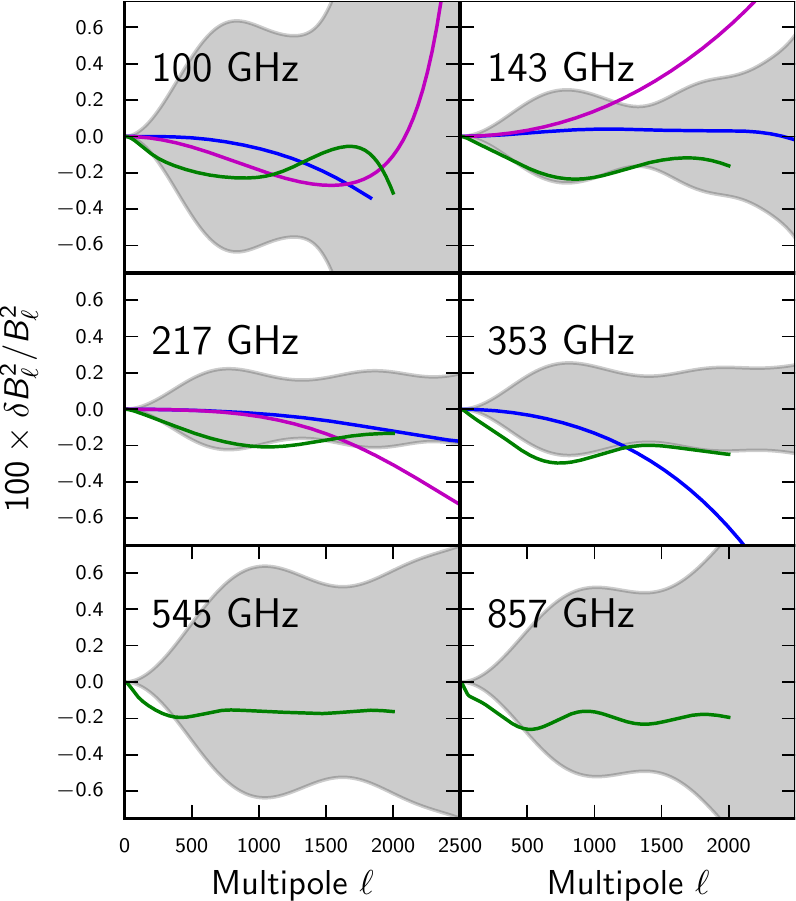}}
\caption{\label{fig:allsystematics} An estimate of known biases in the
  effective beam window function compared to the RMS statistical error
  including the additional factor 2.7  (grey shaded region) for each
  HFI frequency band.  Green is the bias 
  due to near sidelobes, blue is the colour-correction bias, and
  magenta shows an upper limit on the effect of ADC nonlinearity.}
\end{figure}

A model of the behaviour of the ADCs is used to apply nonlinearity to
simulated Mars observations.  The B-spline scanning beams
reconstructed from these simulations predict biases that are typically
under 2\,\% at $\ell \le 2000$, less than 
the RMS error (see the magenta curves in Fig.~\ref{fig:allsystematics} for simulated bias of
the 100 through 353\,GHz channels).

Using the brighter planets in future \Planck\ main beam models will
tend to reduce the effect, as the higher signal amplitudes sample a
broader range of the ADC, tending to average down the resulting bias
\citep{firas1993}.

\subsubsection{Pointing errors}
While the simulated planet observations account for random pointing
errors, pointing drifts remain at the level of
a few arcseconds per pointing period in
the co-scan and cross-scan directions
common to all detectors \citep{planck2013-p01,planck2013-p03}. These
have the effect of widening the beam.  Even in
the worst case, if remaining errors are
Gaussian-distributed with a 2\parcs5 residual, the window function
$B_\ell^2$ is 0.2\,\% biased at multipole $\ell = 3000$.  Since this is
a very small effect relative to the other beam errors, it is considered negligible.

\subsubsection{Glitches}
The HFI data are affected by transient signals due to interactions
with high energy particles.  The planet simulation tools allow for the
addition of a random population of glitches to simulated Mars observations with a
realistic rate and energy spectrum \citep{planck2013-p03e}, testing the
performance of the deglitching algorithm near planet 
signals and the effects of undetected low energy glitches in the
channel with the highest glitch rate at each frequency band.

The RMS of the noise increases by at most 20\,\% in the bolometer
with the highest glitch rate, but more typically less than 10\,\%, and
the reconstruction bias changes by a negligible amount.  The effects
are small enough that they are not included in the error budget.

\subsection{Relative consistency: window function ratios vs. spectral ratios}

The CMB sky itself allows an additional, and statistically
powerful, check on the relative consistency (but not the absolute
accuracy) of the effective beam window functions within a frequency
band.  See Fig.~33, and the associated discussion in
Sect.~7.3 of \cite{planck2013-p03}, showing the self-consistency of
the window functions at a level better than 0.4\,\%, across the full
range of multipoles employed in cosmological parameter estimation.

\section{Total error budget} 
\label{sec:totalerrors}
\label{sec:erroreigenmodes}
The distribution of beam solid angles reconstructed from simulated
planet reconstructions provides the statistical error budget for the
beam solid angles (Table~\ref{table:ScanningBeamSolidAngleErrorBudget}).  Other systematic
effects are small in comparison.

The uncertainty in the effective beam window function is determined
with the distribution of window functions computed from the simulated
beams. These errors are highly correlated in multipole space, and we
find that they can be fully described by a small number of eigenmodes
and their covariance matrix (Appendix~\ref{sec:erroreigenmodes_1pair}
and \ref{sec:erroreigenmodes_npairs}), which can be retrieved, together
with the nominal beam window functions, from \cite{planck2013-p28} and
the \Planck\ Legacy
Archive\footnote{\url{http://www.sciops.esa.int/index.php?project=planck&page=Planck_Legacy_Archive}}. These
simulations also allow us to determine the bias induced by the beam
reconstruction pipeline, and correct the final beam window functions
from it (Appendix~\ref{sec:MCbias}).

Studies have been performed to probe the impact of
assumptions regarding the near sidelobe response of the beam.  These
studies suggest the potential that the Monte Carlo ensembles used to
derive the error eigenmodes may not capture aspects of the bias and
the correlation structure of the errors.  A visual summary of the
total error budget, including the impact of the near sidelobe
response, is provided in Fig.~\ref{fig:allsystematics}.  For the
initial data release a conservative scaling of the MC-derived beam
errors, a factor of 2.7 in power, is applied in order to accommodate
the maximum extent of the bias that is allowable, given the
signal-to-noise ratio of the Jupiter and Saturn data. 

\section{Conclusion}

A combination of Jupiter and Saturn observations, survey difference
maps, and checkout and performance verification (CPV) phase data is
used to derive the bolometer/readout electronics time response
transfer function.  This transfer function is deconvolved from the HFI
time-ordered data.  Residuals can be detected as long tails, stretching
up to 6\deg\ from compact sources, but are at a level less than
$10^{-4}$ of the beam peak, and so are insignificant for the scanning beams.

The effective beams and window functions are estimated for all HFI
detectors using a scanning beam derived from a combination of the
first two seasons of Mars observations.  The effective beam products
account for the partial symmetrization of the scanning beam which 
results from the scan strategy.
The {\tt FEBeCoP} algorithm produces effective beams in real
space, while the {\tt Quickbeam} method is used to produce the
effective beam window functions and errors in harmonic space.

The final error budget for each effective beam window function (both
auto- and cross-beams) is well-described by five eigenmodes for each
beam and a correlation matrix of eigenvalues.  The error parameters
are estimated with an ensemble of simulated Mars observations.  The
simulations are based on random realizations of noise and pointing
errors.

No significant time variation of the scanning beam or the time
response is found.  Cross-power window functions are also produced
that describe the beam filtering effects in cross-correlations between
independent HFI maps.

The high signal-to-noise ratio of the Jupiter and Saturn data limit the
contribution of the near sidelobe response (between 30\arcm\ and
5\deg\ of the beam centroid) to the total beam solid angle to less
than 0.2\,\% at 100 to 217\,GHz.  This portion of the beam, which is
beyond the signal-to-noise of the Mars data, is not fully represented
in the beam simulations.  The Monte Carlo derived error amplitudes are
scaled by a factor of 2.7 to account for the contribution of the near
sidelobe response to the beam window function.

Far sidelobes contribute negligibly to the window function, but may
potentially impact the primary calibration and result in pickup from
bright Galactic sources.  Improved physical optics models for the far
sidelobes will be forthcoming for future releases.

The impact of nonlinearity in the ADC, the
sensitivity of beam shape to the spectral intensity of the source, and
residual cosmic ray transients are found to be insignificant sources
of systematic error.

Knowledge of the beam window functions allows \Planck\ HFI to
extrapolate the dipole calibration over more than three orders of
magnitude in angular scale.  For the current data release, beam errors
are subdominant to noise, foreground marginalization and sample
variance in cosmological parameter estimation.  Future \Planck\ data
releases will fully exploit all of the available planetary data and
create wide-field beam maps, allowing an even more precise, and
accurate, measurement of the main beams and near sidelobes.

%________________________________________________________________

\begin{acknowledgements}

The development of \Planck\ has been supported by: ESA; CNES and
CNRS/INSU-IN2P3-INP (France); ASI, CNR, and INAF (Italy); NASA and DoE
(USA); STFC and UKSA (UK); CSIC, MICINN, JA and RES (Spain); Tekes,
AoF and CSC (Finland); DLR and MPG (Germany); CSA (Canada); DTU Space
(Denmark); SER/SSO (Switzerland); RCN (Norway); SFI (Ireland);
FCT/MCTES (Portugal); and PRACE (EU). A description of the Planck  
Collaboration and a list of its members, including the technical or
scientific activities in which they have been involved, can be found
at
\url{http://www.sciops.esa.int/index.php?project=planck&page=Planck_Collaboration}. 
\end{acknowledgements}

% we load the A&A style for the biblio (''aa.bst'' file)
\bibliographystyle{aa}

% we load the bilbiographic file containing the entries
\bibliography{../../../../Repositories/BibTeX/Planck_bib,P03c_extra}
%\bibliography{Planck_bib,P03c_extra}
%\bibliography{../Repositories/BibTeX/Planck_bib,P03c_extra}

\appendix
\section{Electronics model}
\label{sec:electronicsmodel}
This Appendix derives the effect of the HFI electronics filtering on 
the TOI.
If the input signal (power) on a bolometer is 
\begin{equation}\label{bol_in}
s_0(t)=e^{i\omega t},
\end{equation}
the bolometer physical impedance can be written as:
\begin{equation}\label{bol_out}
s(t)=e^{i\omega t}F(\omega),
\end{equation}
where $\omega$ is the angular frequency of the signal and $F(\omega)$
is the complex intrinsic bolometer transfer function.  For HFI the
bolometer transfer function is modelled as the sum of 4 single pole low pass filters, 
\begin{equation}\label{bol_tf}
F(\omega) = \sum_{i=1,4} \frac{a_i}{1 + i\omega\tau_i}.
\end{equation}
The modulation of the signal is done with a square wave, written here as a composition of sine waves of decreasing amplitude,
\begin{equation}\label{sigmod}
s'(t)=e^{i\omega t}F(\omega)\sum_{k=0}^{\infty} \frac{e^{i\omega_r(2k+1)t}-e^{-i\omega_r(2k+1)t}}{2i(2k+1)},
\end{equation}
where the Euler relation $\sin x=(e^{ix}-e^{-ix})/2i$ is used, and $\omega_r$ is the angular frequency of the square wave.
The modulation frequency is $f_{\rm{mod}} = \omega_r/2\pi$ and was set to 
$f_{\rm{mod}} = 90.18759$\,Hz in flight.  
This signal is then filtered by the complex electronic transfer
function $H(\omega)$. Setting 
$$\omega_k^+=\omega+(2k+1)\omega_r$$
and
$$\omega_k^-=\omega-(2k+1)\omega_r$$
gives 
\begin{equation}\label{sigele}
\Sigma(t)=\sum_{k=0}^\infty\frac{F(\omega)}{2i(2k+1)}\left[H(\omega_k^+)e^{i\omega_k^+t}-H(\omega_k^-)e^{i\omega_k^-t}\right].
\end{equation}
This signal is then sampled at high frequency ($2 f_{\rm{mod}} N_{\rm{samp}}$). 
$N_{\rm{samp}}$ is a parameter of the HFI electronics corresponding to the 
number of high frequency samples in each modulation semi-period.
In order to obtain an output signal sampled 
every $\pi/\omega_r$ seconds, the waveform is integrated on a
semi-period, as done in the HFI readout. To also include a time shift
$\Delta t$, the integral is calculated between 
$n\pi/\omega_r+\Delta t$ and $(n+1)\pi/\omega_r+\Delta t$
(with $T=2 \pi/\omega_r$ period of the modulation).
The time shift $\Delta t$ is encoded in the HFI electronics by the parameter $S_{\rm phase}$, with the relation  $\Delta t = S_{\rm phase}
/N_{\rm{samp}}/f_{\rm{mod}} $.

After integration, a sample of a bolometer at time $t_n$ can be written as
\begin{equation}\label{eqn:output}
Y(t_n) = (-1)^n F(\omega) H'(\omega) e^{i t_n \omega},
\end{equation}
where 
\begin{multline}\label{tfele}
H'(\omega) = \frac 12 \sum_{k=0}^\infty
e^{-i(\frac{\pi\omega}{2\omega_r}+\omega\Delta t)} \Bigg[
 \frac{H(\omega_k^+)e^{i\omega_k^+ \Delta t}}{(2k+1)\omega_k^+}
 \left(1-e^{\frac{i\omega_k^+\pi}{\omega_r}}\right) 
\\ - \frac{H(\omega_k^-)e^{i\omega_k^- \Delta t}}{(2k+1)\omega_k^-}  \left(1-e^{\frac{i\omega_k^-\pi}{\omega_r}}\right)
\Bigg].
\end{multline}

The output signal in Eq.~\ref{eqn:output} can be demodulated (thus removing the $(-1)^n$) and compared to the input signal in Eq.~\ref{bol_in}. The overall transfer function is composed of the bolometer transfer function and the effective electronics transfer function, $H'(\omega)$,
\begin{equation}
TF(\omega) = F(\omega) H'(\omega).
\end{equation}

The shape of  $H(\omega)$ is obtained by combining low and high-pass
filters with Sallen Key topologies \citep{sallen1955}, with their respective time
constants, and accounting also for the stray capacitance low pass
filter given by the bolometer impedance combined with the stray
capacitance of the cables.  The sequence of filters that define the electronic band-pass function
$H(\omega) = h_0*h_1*h_2*h_3*h_4*h_{5}$ are listed in
Table~\ref{table:readout_electronics_filters}, with parameters listed
in Table~\ref{table:LFER4pars}.

\begin{table*}[tmb]
\begingroup
\newdimen\tblskip \tblskip=5pt
\caption{HFI electronics filter sequence; here $s = i \omega$.}
\label{table:readout_electronics_filters}
\nointerlineskip
\vskip -3mm
\footnotesize
\setbox\tablebox=\vbox{
    \newdimen\digitwidth
    \setbox0=\hbox{\rm 0}
    \digitwidth=\wd0
    \catcode`*=\active
    \def*{\kern\digitwidth}
    \newdimen\signwidth
    \setbox0=\hbox{+}
    \signwidth=\wd0
    \catcode`!=\active
    \def!{\kern\signwidth}
\halign{\hbox to 1.5cm{#\leaderfil}\tabskip 1em&
         \vtop{\hsize
2.0in\hangafter=1\hangindent=1em\noindent\strut#\strut\par}\tabskip
2em&
         \vtop{\hsize 2.0in\noindent\strut#\strut\par}\tabskip 0pt&

         #\hfil\cr
\noalign{\doubleline}
\omit\hfil Filter\hfil&\omit\hfil Description\hfil&\omit\hfil
Parameters \hfil&\omit\hfil Function\hfil\cr
\noalign{\vskip 3pt\hrule\vskip 5pt}
0&Stray capacitance low pass filter&     $\tau_{\rm stray}= R_{\rm
bolo} C_{\rm stray}$ & $\displaystyle h_0 = \frac{1}{ 1.0+\tau_{\rm
stray} s}$\cr
\noalign{\vskip 5pt}
1&Low pass filter&                       $R_1=1$\,k$\Omega$ \\
                                          $C_1=100$\,nF &
$\displaystyle h_1 = \frac{2 +R_1 C_1 s}{2 (1 +R_1 C_1 s)}$\cr
\noalign{\vskip 5pt}
2&Sallen Key high pass filter &          $R_2=51$\,k$\Omega$\\
$C_2=1\,\mu$\,F&$\displaystyle h_2= \frac{(R_2 C_2 s)^2}{(1 +R_2 C_2
s)^2}$\cr
\noalign{\vskip 5pt}
3&Sign reverse with gain &\hglue 6mm\dots&         $\displaystyle h_3=-5.1$\cr
\noalign{\vskip 5pt}
4&Single pole low pass filter with gain &$R_4=10$\,k$\Omega$\\
                                          $C_4=10$\,nF&$\displaystyle
h_4= \frac{1.5}{1 +R_4 C_4 s}$\cr
\noalign{\vskip 5pt}
5&Single pole high pass filter coupled to a
Sallen Key low pass filter&              $R_9=18.7$\,k$\Omega$\\
                                          $R_{12}=37.4$\,k$\Omega$&
   $\displaystyle h_{5} = \frac{2*R_{12} R_9 R_{78} C_{18} s}{s^3 K_3
+ s^2 K_2+ s K_1 + R_{12} R_9}$\cr
\omit&&\omit                             $C=10.0$\,nF\hfil\cr
\omit&&\omit                             $R_{78}=510$\,k$\Omega$\hfil\cr
\omit&&\omit                             $C_{18}=1.0\,\mu$F\hfil\cr
\omit&&\omit                             $K_3 = R_9^2 R_{78} R_{12}^2
C^2 C_{18}$\hfil\cr
\omit&&\omit                             $K_2 = R_9 R_{12}^2 R_{78}
C^2+R_{9}^2 R_{12}^2 C^2$\hfil\cr
\omit&&\omit                               \quad\quad\quad $ + R_9
R_{12}^2 R_{78} C_{18} C$\cr
\omit&&\omit                             $K_1 =R_9 R_{12}^2 C+R_{12}
R_{78} R_9 C_{18}$\hfil\cr
\noalign{\vskip 5pt\hrule\vskip 3pt}}}
\endPlancktablewide
\endgroup
\end{table*}

\begin{table*}[tmb]                 % table* is a two-column table.  Drop the * for one column.
\begingroup
\newdimen\tblskip \tblskip=5pt
\caption{Parameters for LFER4 model that are deconvolved from the data.}                          % Caption goes here.
\label{table:LFER4pars}                            % Label goes here.
\nointerlineskip
\vskip -3mm
\footnotesize
\setbox\tablebox=\vbox{
   \newdimen\digitwidth 
   \setbox0=\hbox{\rm 0} 
   \digitwidth=\wd0 
   \catcode`*=\active 
   \def*{\kern\digitwidth}
   \newdimen\signwidth 
   \setbox0=\hbox{+} 
   \signwidth=\wd0 
   \catcode`!=\active 
   \def!{\kern\signwidth}
\halign{\hbox to 2cm{#\leaderfil}\tabskip 1em&\hfil#\hfil \tabskip 1em&\hfil#\hfil \tabskip 1em &\hfil#\hfil \tabskip 1em&\hfil#\hfil \tabskip 1em&\hfil#\hfil \tabskip 1em&\hfil#\hfil \tabskip 1em&\hfil#\hfil \tabskip 1em&\hfil#\hfil \tabskip 1em&\hfil#\hfil \tabskip 1em&\hfil#\hfil \tabskip 0pt\cr                            % Template goes here.
\noalign{\doubleline}
                                    % Table headings go here.
\omit Bolometer&$a_1$&$\tau_1$&$a_2$&$\tau_2$&$a_3$&$\tau_3$&$a_4$&$\tau_4$&$\tau_{\rm stray}$&$S_{\rm phase}$\cr
\omit&&[ms]&&[ms]&&[ms]&&[ms]&[ms]&\cr
\noalign{\vskip 3pt\hrule\vskip 5pt}
                                    % Body of table goes here.
100-1a&0.392&10.0*&0.534&20.9**&0.0656*&**51.3&0.00833& *572&1.59&0.00139\cr
100-1b&0.484&10.3*&0.463&19.2**&0.0451*&**71.4&0.00808& *594&1.49&0.00139\cr
100-2a&0.474&*6.84&0.421&13.6**&0.0942*&**37.6&0.0106*& *346&1.32&0.00125\cr
100-2b&0.126&*5.84&0.717&15.1**&0.142**&**35.1&0.0145*& *293&1.38&0.00125\cr
100-3a&0.744&*5.39&0.223&14.7**&0.0262*&**58.6&0.00636& *907&1.42&0.00125\cr
100-3b&0.608&*5.48&0.352&15.5**&0.0321*&**63.6&0.00821& *504&1.66&0.00125\cr
100-4a&0.411&*8.2*&0.514&17.8**&0.0581*&**57.9&0.0168*& *370&1.25&0.00125\cr
100-4b&0.687&11.3*&0.282&24.3**&0.0218*&**62.0&0.00875&*431&1.38&0.00139\cr
\omit\cr
143-1a&0.817&*4.47&0.144&12.1**&0.0293*&**38.7&0.0101*& *472&1.42&0.00125\cr
143-1b&0.49*&*4.72&0.333&15.6**&0.134**&**48.1&0.0435*& *270&1.49&0.00125\cr
143-2a&0.909&*4.7*&0.076&17.0**&0.00634&*100**&0.00871& *363&1.48&0.00125\cr
143-2b&0.912&*5.24&0.051&16.7**&0.0244*&**26.5&0.0123*& *295&1.46&0.00125\cr
143-3a&0.681&*4.19&0.273&*9.56*&0.0345*&**34.8&0.0115*& *317&1.45&0.00125\cr
143-3b&0.82*&*4.48&0.131&13.2**&0.0354*&**35.1&0.0133*& *283&1.61&0.00083\cr
143-4a&0.914&*5.69&0.072&18.9**&0.00602&**48.2&0.00756& *225&1.59&0.00125\cr
143-4b&0.428&*6.06&0.508&*6.06*&0.0554*&**22.7&0.00882&  **84&1.82&0.00125\cr
143-5&0.491&*6.64&0.397&*6.64*&0.0962*&**26.4&0.0156*& *336&2.02&0.00139\cr
143-6&0.518&*5.51&0.409&*5.51*&0.0614*&**26.6&0.0116*& *314&1.53&0.00111\cr
143-7&0.414&*5.43&0.562&*5.43*&0.0185*&**44.9&0.00545&*314&1.86&0.00139\cr
\omit\cr
217-5a&0.905&*6.69&0.080&21.6**&0.00585&**65.8&0.00986& *342&1.57&0.00111\cr
217-5b&0.925&*5.76&0.061&18.0**&0.00513&**65.6&0.0094*& *287&1.87&0.00125\cr
217-6a&0.844&*6.45&0.068&19.7**&0.0737*&**31.6&0.0147*& *297&1.54&0.00125\cr
217-6b&0.284&*6.23&0.666&*6.23*&0.0384*&**24**&0.0117*& *150&1.46&0.00111\cr
217-7a&0.343&*5.48&0.574&*5.48*&0.0717*&**23**&0.0107*& *320&1.52&0.00139\cr
217-7b&0.846&*5.07&0.127&14.40*&0.0131*&**47.9&0.0133*&*311&1.51&0.00139\cr
217-8a&0.496&*7.22&0.439&*7.22*&0.0521*&**32.5&0.0128*&*382&1.79&0.00111\cr
217-8b&0.512&*7.03&0.41*&*7.03*&0.0639*&**27.2&0.0139*&*232&1.73&0.00125\cr
217-1&0.014&*3.46&0.956&*3.46*&0.0271*&**23.3&0.00359&1980&1.59&0.00111\cr
217-2&0.978&*3.52&0.014&26.1**&0.00614&**42**&0.00194& *686&1.6*&0.00125\cr
217-3&0.932&*3.55&0.034&*3.55*&0.0292*&**32.4&0.00491& *279&1.74&0.00125\cr
217-4&0.658&*1.35&0.32*&*5.55*&0.0174*&**26.8&0.00424& *473&1.71&0.00111\cr
\omit\cr
353-3a&0.554&*7.04&0.36*&*7.04*&0.0699*&**30.5&0.0163*& *344&1.7*&0.00125\cr
353-3b&0.219&*2.68&0.671&*6.95*&0.0977*&**23.8&0.0119*& *289&1.57&0.00111\cr
353-4a&0.768&*4.73&0.198&*9.93*&0.0283*&**50.5&0.00628& *536&1.81&0.00125\cr
353-4b&0.684&*4.54&0.224&10.8**&0.0774*&**80**&0.0149*& *267&1.66&0.00111\cr
353-5a&0.767&*5.96&0.159&12.4**&0.0628*&**30.3&0.0109*& *357&1.56&0.00111\cr
353-5b&0.832&*6.19&0.126&11.1**&0.0324*&**35**&0.0096*& *397&1.66&0.00111\cr
353-6a&0.049&*1.76&0.855&**6.0**&0.0856*&**21.6&0.0105*& *222&1.99&0.00125\cr
353-6b&0.829&*5.61&0.127&*5.61*&0.0373*&**25.2&0.00696& *360&2.28&0.00111\cr
353-1&0.41*&*0.74&0.502&*4.22*&0.0811*&**17.7&0.0063*& *329&1.32&0.00097\cr
353-2&0.747&*3.09&0.225&*7.26*&0.0252*&**44.7&0.00267& *513&1.54&0.00097\cr
353-7&0.448&*0.9*&0.537&*4.1**&0.0122*&**27.3&0.00346& *433&1.78&0.00125\cr
353-8&0.718&*2.23&0.261&*6.08*&0.0165*&**38**&0.00408& *268&1.77&0.00111\cr
\omit\cr
545-1&0.991&*2.93&0.007&26.0**&0.00139&2600**&\dots&\dots&2.16&0.00111\cr
545-2&0.985&*2.77&0.013&24.0**&0.00246&2800**&\dots&\dots&1.87&0.00097\cr
545-4&0.972&*3.0*&0.028&25.0**&0.00078&2500**&\dots&\dots&2.22&0.00111\cr
\omit\cr
857-1&0.974&*3.38&0.023&25.0**&0.00349&2200**&\dots&\dots&1.76&0.00111\cr
857-2&0.84*&*1.48&0.158&*6.56*&0.00249&3200**&\dots&\dots&2.2*&0.00125\cr
857-3&0.36*&*0.04&0.627&*2.4**&0.0111*&**17**&0.002**&1900&1.52&0.00126\cr
857-4&0.278&*0.4*&0.719&*3.92*&0.00162&**90**&0.00152& *800&1.49&0.00056\\\cr
\noalign{\vskip 3pt\hrule\vskip 5pt}
}
}
%\endPlancktable                    % ends one-column \halign
\endPlancktablewide                 % ends two-column \halign
\endgroup
\end{table*}                        % table* is a two-column table.  Drop the * for one column.

\section{Gauss-Hermite beams}
\label{sec:gausshermite}
HFI's scanning beams are described by an elliptical Gaussian shape to
an accuracy of 2--5\,\% in solid angle.  A Gauss-Hermite representation of the beam uses
Hermite polynomials to provide higher-order corrections to a
Gaussian model \citep{huffenberger2010}.  There are fewer
degrees of freedom than in a gridded beam map, 
allowing higher signal-to-noise on each mode. However, because the order of
the decomposition is truncated, in practice the description is
ill suited to a description of features much beyond the extent of the
main beam.  Larger scale features of the beam, including a beam shoulder
\citep{Ruze1966} or the effect of the print-through of the
backing structure (grating lobes), must be included separately.

The initial two-dimensional Gaussian is parametrized as
\begin{equation}
        \label{eq:2dGaussian}
P(x_1,x_2) =  \frac{A}{|2\pi\mathbf{\Sigma}|^{1/2}}\exp\left[-\frac12\sum_{i,j=1}^{2}(x_i-{\bar x}_i)\Sigma^{-1}_{ij}(x_j-{\bar x}_j)\right], 
\end{equation}
where $A$ is an overall amplitude, $(x_1,x_2)$ are two-dimensional 
Cartesian coordinates (the pointing is projected to a flat sky), $({\bar x}_1,{\bar x}_2)
 $ are the coordinates of the beam center, and the correlation matrix is given by
\begin{equation}
        \mathbf{\Sigma} = \left( \begin{array}{cc}
        \sigma_1^2 & \rho\sigma_1\sigma_2 \\
        \rho\sigma_1\sigma_2 & \sigma_2^2 \end{array} \right)\;.
\end{equation}
Hence, the Gaussian model parameters $A, {\bar
  x}_1, {\bar x}_2, \sigma_1, \sigma_2, \rho$ are fitted. These can also be
expressed in terms of the ellipticity, $e$ (defined here as the ratio
between the major and minor axes), and rotation angle $\alpha$, of the
Gaussian ellipse.  

The coefficients to the Gauss-Hermite polynomials multiplying the
elliptical Gaussian are then fitted. The basis functions are defined as 
\begin{equation}
    \Phi_{n_1n_2}(\mathbf{x}) \propto H_{n_1}({x_1'})  H_{n_2}({x_2'}) 
    \exp\left(-\mathbf{x'\cdot x'}/2\right), 
\end{equation}
where $H_n(x)$ is the order-$n$ Hermite polynomial and the primed
coordinates $\mathbf{x'}$ rotates into a system aligned with the axes
of the Gaussian and scaled to the major and minor axes $\sigma_i$
(i.e., to the principle axes of the correlation matrix
$\mathbf{\Sigma}$).  Having determined the elliptical Gaussian
separately, the subsequent Gauss-Hermite polynomial fit is a
generalized least-squares procedure, solvable by the usual matrix
techniques, under the assumption of white noise. Because the full data
model includes further effects such as pointing error, glitch effects,
etc, a full error analysis requires a broad suite of simulations,
described in Sect.~\ref{sec:errors}.

\section{B-spline beams} 
\label{sec:griddedbeamdescription}
In this model of the scanning beam, a two-dimensional B-spline surface
is fit to the time-ordered planet data. A smoothing criterion is
applied during the fit to minimize the effects of high
spatial frequency variations due to noise.  This representation of the
beam is superior to a simple two-dimensional binning of the data in
its ability to capture large signal gradients.

B-splines are a linear combination of polynomials of degree $k$ and
order $k+1$. They are  defined by the location of their control points
(called \emph{knots}) of which there are 5 for 3rd degree polynomials.  

The $k$-degree B-spline built using the knots \{$\lambda_{i}, ...,
\lambda_{i+k+1}$\} \citep{deboor1972,cox1972} is given by
\begin{equation}
P_{i,1}(x) = \left\{
	\begin{array}{l}
	1, \mbox{ if } x \in \mbox{[} \lambda_{i}, \lambda_{i+1} \mbox{]}\\
	0, \mbox{ if } x \notin \mbox{[} \lambda_{i}, \lambda_{i+1} \mbox{]}
	\end{array} \right.
\end{equation}
with recursion relations
\begin{equation}
P_{i, l+1}(x) = \displaystyle{\frac{x - \lambda_{i}}{\lambda_{i+l} - \lambda_{i}}} P_{i,l}(x) + \displaystyle{\frac{\lambda_{i+l+1}-x}{\lambda_{i+l+1}-\lambda_{i+1}}} P_{i+1, l}(x).
\end{equation}
where the index $l$ runs from 1 through the B-spline degree $k$.  The
B-spline knots \{$\lambda_{i}, ..., z\lambda_{i+k+1}$\} are 
located on a regularly spaced grid in the detector coordinate
frame. At the edge of the reconstructed beam map area, 4 coincident
knots are added to avoid vanishing basis functions, allowing a unique
decomposition. 

In the solution to the B-spline coefficients $P_{i,l} (x)$, a smoothing
criterion is introduced as a constraint on the sum of the derivatives
of the beam at each knot, motivated by the assumption that the true
beam does not contain very high spatial frequencies  and prevents
noise from biasing the reconstruction at spatial frequencies smaller
than the smoothing scale. 

A smoothing criterion $\eta$ \citep{dierckx1993} is related to the the
sum of the order $k$ derivative of the beam model ($P^{k}$) evaluated
at the knot locations $\lambda_i$:
\begin{equation}
\eta = \displaystyle{\sum_{i=1}^{g}\left[P^{k}(\lambda_{i+})-P^{k}(\lambda_{i-})\right]^2}
\label{eq:smoothcrit}
\end{equation}
where $g$ is the total number of distinct knots and $\lambda_{i+}$ and
$\lambda_{i-}$ are the left and right derivative of the beam model
evaluated at the knot location. The smoothing criterion is introduced as an extra term in
the score function $\zeta$ in the  least-squares minimization of the
beam map with respect to the data, 
\begin{equation}
\zeta = \eta + p\times \delta.
\end{equation}
\noindent where $\delta$ is the usual squared difference between the data points
$y_r$ and the model $P(x_{r})$ at pointings $x_{r}$,
\begin{equation}
\delta = \displaystyle{\sum_{r=1}^{m}}\left[y_{r} - P(x_{r})\right]^2.
\end{equation}
and $p$ is a relative weighting factor for the smoothing
criterion. The knot spacing and the smoothing criterion weight $p$ are
determined separately for each frequency band based on physical optics
simulations and the coverage of the two Mars transits given by the
scanning strategy (see Table~\ref{table:knotspacingandsmoothcrit}).
Simulated planet observations  (see Sect.~\ref{sec:errors})   
show that the choice of B-spline knot spacing and smoothing parameters
do not significantly bias the beam reconstruction.

\begin{table}[tmb]                 % table* is a two-column table.  Drop the * for one column.
\begingroup
\newdimen\tblskip \tblskip=5pt
\caption{Summary of B-spline knot spacing and smoothing criterion
  weight $p$ used in the beam reconstruction.}                          % Caption goes here. 
\label{table:knotspacingandsmoothcrit}                            % Label goes here.
\nointerlineskip
\vskip -3mm
\footnotesize
\setbox\tablebox=\vbox{
   \newdimen\digitwidth 
   \setbox0=\hbox{\rm 0} 
   \digitwidth=\wd0 
   \catcode`*=\active 
   \def*{\kern\digitwidth}
   \newdimen\signwidth 
   \setbox0=\hbox{$0$} 
   \signwidth=\wd0 
   \catcode`!=\active 
   \def!{\kern\signwidth}
\halign{\hbox to 2cm{#\leaderfil}\tabskip 0em&\hfil#\hfil\tabskip 2em&\hfil#\hfil\tabskip 0pt\cr
  % Template goes here.
\noalign{\doubleline}                                    % Table headings go here.
\omit\hfil Band\hfil&Knot separation&$p$\cr
\omit\hfil [GHz]\hfil&[arcmin]&\cr
\noalign{\vskip 3pt\hrule\vskip 5pt}
% Body of table goes here.
100&1.5*& $10$!\cr
143&1.25& $10$!\cr
217&1.0*& $10$!\cr
353&1.0*& $10^4$\cr
545&0.75& $10^6$\cr
857&0.75& $10^6$\cr
\noalign{\vskip 3pt\hrule\vskip 5pt}
}
}
\endPlancktable                    % ends one-column \halign
%\endPlancktablewide                 % ends two-column \halign
\endgroup
\end{table}                        % table* is a two-column table.  Drop the * for one column.

\section{Far sidelobe effects on the effective beam window function}
\label{sec:FSLappendix}
The far sidelobes (FSL) are defined as the response of the instrument
at angles $> 5\deg$ from the main beam centroid. The FSL affect
  both 1. the gain calibration of the instrument with the dipole, and
  2. how the calibration is transfered to higher multipoles with the effective
beam window function.  Here we discuss the interplay of these effects.

The FSL can be separated into three main components (see Fig.~5 of \cite{tauber2010b}): 
\begin{enumerate}
\item \textit{Primary Reflector Spillover} (PR Spillover) is the response
  of the instrument to radiation from just above the primary mirror
  that reflects off the secondary mirror and arrives at the feed horns.
  This response is nearly aligned with the spin axis of the
  spacecraft, and therefore scans very little of the sky on 1 minute
  time-scales. 
\item \textit{Secondary Reflector Direct
  Spillover } (SR Direct Spillover, or SRD) is the response from
  directly above the secondary mirror.  The sidelobe peaks roughly
  10\deg\ from the main beam, and as such scans the sky in nearly the
  same way as the main beam as the satellite rotates.  The azimuthal
  extent is roughly 30\deg. 
\item \textit{Secondary Reflector Baffle Spillover} (SR Baffle
  Spillover, or SRB) is response from radiation
  reflecting off the baffles.  This is difficult to model, being diffuse
  radiation reflecting off the poorly known inner baffle surfaces.  It
  is spread over a large fraction of the sky.
\end{enumerate}

The HFI dipole calibration is performed assuming a delta-function
(pencil) beam \citep{planck2013-p03f}.  This leads to a bias in the
calibration described by the ratio of the dipole convolved with the
full-sky beam to the dipole convolved by a pencil beam,
$$
\tilde{g} = g \frac{P \otimes D}{P ^{\rm{pencil}} \otimes D}, 
$$
where $\tilde{g}$ is the estimated gain, $g$ is the true gain, $P$ is the true full-sky
beam, $P^{\rm{pencil}}$ is the assumed pencil beam, and $D$ is the
dipole.  The true full-sky beam is taken to consist of three portions, 
$$
P = P_{\rm{main}} +  P_{\rm{NSL}} + P_{\rm{FSL}},
$$
the main beam $P_{\rm{main}}$ (within 20\arcm\ of the centroid), the
near sidelobes $P_{\rm{NSL}}$ (between 20\arcm\ and 5\deg), and
the far sidelobes $P_{\rm{FSL}}$ (response further than 5\deg\ from
the centroid).   The quantity 
$$
P_{\rm{MNSL}} = P_{\rm{main}} +  P_{\rm{NSL}}
$$
is defined as the beam within 5\deg\ of the
centroid (the main lobe and near sidelobes). The bias in the calibration can be
rewritten as
\begin{equation} \label{eqn:FSLgainbias}
\tilde{g} = g \left(  1 - f_{\rm{FSL}} + \frac{P_{\rm{FSL}} \otimes D}{P^{\rm{pencil}} \otimes D}\right),
\end{equation}
where $f_{\rm{FSL}}$ is the integral of the far sidelobe response
relative to the full beam integral.   The first term, $1 -
f_{\rm{FSL}}$, is due to the loss in response of the main lobe and near sidelobes to
far sidelobes, while the second term represents the coupling of the
dipole into the sidelobes.

For each pointing period the convolution of
the pencil beam with the dipole is approximated by
\begin{equation} \label{eqn:pencilbmdipoleapprox}
P^{\rm{pencil}} \otimes D \simeq \sin \theta_{\rm{main}} d_{\rm{pencil}},
\end{equation}
where $\theta_{\rm{main}}$ is the angle between the main beam
centroid and the spin axis, and $d_{\rm{pencil}}$ is the dipole
amplitude that would be observed with a pencil beam.  The
components of the far sidelobes enter into
Equation~\ref{eqn:FSLgainbias} differently, and assuming the majority
of the response to the dipole in each component comes from one
direction on the sky, can be approximated as
\begin{align*}
P_{\rm{FSL}} \otimes D \simeq &  \epsilon_{\rm{PR}} \sin \theta_{\rm{PR}}
 d_{\rm{pencil}} \\
&  + \epsilon_{\rm{SRD}} \sin \theta_{\rm{SRD}} d_{\rm{pencil}} \\
& + \epsilon_{\rm{SRB}} \sin \theta_{\rm{SRB}} d_{\rm{pencil}},
\end{align*}
where $\epsilon_{\rm{PR}}$,
$\epsilon_{\rm{SRD}}$, and $\epsilon_{\rm{SRB}}$ are the fraction of
the total solid angle in the PR spillover, the SR direct spillover,
and the SR baffle spillover, respectively, and $\theta_{\rm{PR}}$ , 
$\theta_{\rm{SRD}}$ , $\theta_{\rm{SRB}}$ are the angles between the
spin axis and the peak of the PR spillover response, the SR direct
spillover response, and the SR baffle spillover response
respectively.  Eqn.~\ref{eqn:FSLgainbias} then simplifies to 
\begin{equation} \label{eqn:FSLgainbiassins}
\begin{aligned} 
\tilde{g} \simeq g \left( \right. &  1 - f_{\rm{FSL}} +
  \epsilon_{\rm{PR}} \frac{\sin \theta_{\rm{PR}}}{\sin
    \theta_{\rm{pencil}}}   \\
&\left. + \epsilon_{\rm{SRD}} \frac{\sin \theta_{\rm{SRD}}}{\sin
  \theta_{\rm{pencil}}} \right. \\
&\left. + \epsilon_{\rm{SRB}} \frac{\sin \theta_{\rm{SRB}}}{\sin \theta_{\rm{pencil}}} \right).
\end{aligned}
\end{equation}
From \cite{tauber2010b},  $\theta_{\rm{pencil}}  \simeq 85\deg$,
$\theta_{\rm{PR}}  \simeq 10\deg$, $\theta_{\rm{SRD}}  \simeq 75\deg$,
and $\theta_{\rm{SRB}}  \simeq 45\deg$ making the PR spillover and SR
baffle spillover negligible.  Equation~\ref{eqn:FSLgainbiassins} reduces further to
\begin{equation} \label{eqn:finalgainbias}
\tilde{g} \simeq g \left(  1 - f_{\rm{FSL}} + \epsilon_{\rm{SRD}}
  \frac{\sin \theta_{\rm{SRD}}}{\sin \theta_{\rm{pencil}}} \right).
\end{equation}
 SR direct spillover scans the dipole, but with a slightly different
 amplitude, since it is offset by 10\deg\ from the main lobe. The PR 
spillover does not modulate the dipole;  aligned with the spin axis,
the PR spillover contributes a nearly constant signal during each stable pointing
period.   The predicted values of the solid angle fractions are $f_{\rm{FSL}}= 3.3\times10^{-3}$ 
and $\epsilon_{\rm{SRD}} = 1.9\times10^{-3}$  at 100\,GHz \citep{tauber2010b}, making
the measured gain $\tilde{g} \simeq 0.9985 g$.

However, the effect of the gain bias on the angular power
spectrum is further reduced by corrections to the effective beam
window function due to the FSL.   Not including the FSL in the beam
model  tends to bias the window 
function at very low multipoles relative to high multipoles.
Considering the measured sky signal at small scales 
$\tilde{T}_{\rm{sky}}$ as compared to the true sky $T_{\rm{sky}}$ one has
$$
\tilde{T}_{\rm{sky}} = \frac{S}{\tilde{g}} = \frac{P
  \otimes  T _{\rm{sky}}}{\left(  1 - f_{\rm{FSL}} + \frac{P_{\rm{FSL}} \otimes
      D}{P^{\rm{pencil}} \otimes D}\right)}, 
$$
where $S$ is the signal measured by the bolometer. Solving for
the true sky signal gives
$$
P_{\rm{MNSL}}  \otimes T _{\rm{sky}} =
\left( 1 - f_{\rm{FSL}} + \frac{P_{\rm{FSL}} \otimes D}{P^{\rm{pencil}} \otimes
    D}\right) \tilde{T}_{\rm{sky}} - P_{\rm{FSL}} \otimes T _{\rm{sky}}. 
$$
Considering that $P_{\rm{MNSL}} \otimes T _{\rm{sky}}  \simeq (1 -
f_{\rm{FSL}} ) T _{\rm{sky}}$ , 
$$
T_{\rm{sky}} \simeq \left( 1 + \frac{P_{\rm{FSL}} \otimes
    D}{P^{\rm{pencil}} \otimes D} \frac{1}{1-f_{\rm{FSL}}}\right) \tilde{T}_{\rm{sky}} -
\frac{P_{\rm{FSL}} \otimes T_{\rm{sky}} }{1 -f_{\rm{FSL}} }.
$$
Now also considering that $P_{\rm{MNSL}} \otimes D \simeq (1 -
f_{\rm{FSL}}) P^{\rm{pencil}} \otimes D$, the true
sky signal can be written as
$$
T_{\rm{sky}} \simeq \tilde{T}_{\rm{sky}}  \left( 1 +
  \frac{P_{\rm{FSL}} \otimes D}{P_{\rm{MNSL}} \otimes D}  -
  \frac{P_{\rm{FSL}} \otimes T_{\rm{sky}}}{P_{\rm{MNSL}} \otimes
    T_{\rm{sky}}} \frac{T_{\rm{sky}}}{\tilde{T}_{\rm{sky}} }\right) ,
$$
or
\begin{equation} \label{eqn:FSLEBWNnorm}
T_{\rm{sky}} = \left( 1 + \phi_D - \phi_{\rm{sky}}
\right) \tilde{T}_{\rm{sky}},
\end{equation}
where the second-order correction term $T_{\rm{sky}} /
\tilde{T}_{\rm{sky}} \simeq 1$ is dropped, and  
the following factors are defined:
$$
\phi_D = \frac{P_{\rm{FSL}} \otimes D}{P^{\rm{pencil}} \otimes D} ;
$$
and
$$
\phi_{\rm{sky}} = \frac{P_{\rm{FSL}} \otimes T_{\rm{sky}}  }{P^{\rm{pencil}} \otimes T_{\rm{sky}} }.
$$ 
Here $\phi_{\rm{sky}}$ represents the relative pickup of anisotropy in the
sidelobe beam as compared to the main beam and near sidelobes.  The
response of the FSL to CMB anisotropy is negligible, but Galactic
response may not be. The PR
spillover contributes only a constant value per pointing
period, because it is not modulated with the scan.  Only the SR direct 
spillover enters into the formula.  Since the SR direct spillover
is nearly aligned with the main beam, Galactic signal is not picked up
in the CMB anisotropy except close to the Galactic plane.  So for
foreground-clean regions of the sky, $\phi_{\rm{sky}} \ll f_{\rm{FSL}}$.

The quantity $\phi_D$ is the bias due to the ratio of dipole response in the far sidelobes
to dipole response in the main beam.  Again the PR spillover contributes only an offset
per pointing period, which is removed by the {\tt polkapix} destriping
algorithm \citep{tristram2011,planck2013-p03f}, so 
we are left with the SR direct spillover and SR baffle spillover.
Applying the same approximation as in
Eq.~\ref{eqn:pencilbmdipoleapprox} gives
% removed SRB terms as pointed out by referee
$$
\phi_D \simeq \epsilon_{\rm{SRD}} \frac{\sin \theta_{\rm{SRD}}}{\sin
  \theta_{\rm{pencil}}},
$$
and Eq.~\ref{eqn:FSLEBWNnorm} becomes 
\begin{equation}
T_{\rm{sky}} \simeq \left(  1 + \epsilon_{\rm{SRD}} \frac{\sin \theta_{\rm{SRD}}}{\sin
  \theta_{\rm{pencil}}} \right) \tilde{T}_{\rm{sky}}.
\end{equation}
This result implies that FSL bias of the effective beam window
function tends to cancel the gain bias (Eq.~\ref{eqn:finalgainbias}),
and the response to CMB anisotropy is unaffected to first order.   

Simulations of the sky scanned with the far sidelobe physical optics model
\citep{planck2013-pip88,planck2013-p03} confirm this.

\section{Harmonic space computation of the effective beam window
  function}
Two harmonic space techniques ({\tt FICSBell} and {\tt Quickbeam}),
developed independently but closely related in 
their formalism, have been used to compute the effective beam window functions.
They provide a valuable cross-check of the pixel-based results obtained with
{\tt FEBeCoP} (Fig.~\ref{fig:FlowOfBeams}) and their low computational requirements
allow fast calculation of the window function errors
through the processing of Monte Carlo simulations of planet observations.
\label{sec:harmonicspace_effbeams}
\subsection{FICSBell}
\label{sec:FICSBell}
{\tt FICSBell} is a harmonic space method for computing the effective beam
window function directly from the scanning beam and the scan
history. The steps of this method are as follows.
\begin{enumerate}
\item The statistics of the orientation of each detector $d$ within each
  map pixel $p$
is computed first, and only once for a given observation
  campaign:
\begin{equation}
	w^d_s({\bf r}_p) = \sum_j e^{i s \psi_j},
\end{equation}
where $\psi_j$ is the orientation of the detector with respect to the local
meridian during the measurement $j$ occurring in 
the direction ${\bf r}_p$. Note that the $s=0$ moment is simply the
hit count map. The orientation hit moments are computed up to
  degree $s=4$. At the same time,
  the first two moments of the distribution of samples within each
  pixel (i.e., the centre of mass and  moments of inertia) are computed and stored.%
\item The scanning beam map of each detector $d$ is transformed into
  spherical harmonics:
\begin{equation}
	b^d_{\ell s} = \int d{\bf r}  B_d({\bf r}) Y_{\ell s}({\bf r}),
\end{equation}
where $B_d(\bf{r})$ is the beam map centered on the North pole, and the
$Y_{\ell s}(\bf{r})$ are the spherical harmonic basis functions.
Higher $s$ indices describe higher degrees of departure from azimuthal symmetry
and, for HFI beams, the coefficients $b^d_{\ell s}$ are decreasing functions of $s$ at
most multipoles considered. It also appears that, for $\ell<3000$, the coefficients with
$|s| > 4$ account for much less than 1\,\% of the beam
solid angle.  Spot checks where window functions are computed with
$|s|\le 6$ show a difference of less than $10^{-4}$ for
$\ell<2000$ at 100\,GHz and for $\ell<3000$ at 143 and 217\,GHz. For these reasons, only modes with $|s| \le 4$ are
considered in the present analysis.  \cite{armitage-caplan2009} reached a similar conclusion in their
deconvolution of LFI beams.
\item  For a given CMB sky realization $t$, described by its spherical harmonics coefficients
$a_{\ell m} = \int d{\bf r}  t({\bf r}) Y_{\ell m}({\bf r})$, the $b^d_{\ell s}$ coefficients computed above are
used to generate $s$-spin weighted maps,  
\begin{equation}
	m^d_{s}({\bf r}) = \sum_{\ell m} b^d_{\ell s}\  a_{\ell m}\ {}_sY_{\ell m}({\bf r}),
\end{equation}
as well as the first and second 
derivatives, using standard {\tt HEALPix} tools. 
\item The spin-weighted maps and orientation hit moments of the same order $s$ are
combined for all detectors involved, to provide an ``observed'' map
\begin{equation}
	m({\bf r}) = \left(\sum_d \sum_s  w^d_s({\bf r}) m^d_s({\bf r})\right) / \sum_d   w^d_0({\bf r}).
\end{equation}
 Similarly
the local spatial derivatives are combined with the location hit moments to
describe the effect of the non-ideal sampling of each pixel (see Appendix~\ref{sec:pixelization}). In this
combination, the respective number of hits of each detector in each pixel is
considered, as well as the detector weighting (generally proportional to the inverse
noise variance).
\item The power spectrum of this map can then be computed, and compared to the
input CMB power spectrum to estimate the effective beam window function over 
the whole sky, or over a given region of the sky. 
\end{enumerate}
Monte Carlo (MC) simulations in which the sky realizations 
are changed can be performed by repeating steps 3, 4 and 5. The impact of beam
model uncertainties can be studied by including step 2 in the MC simulations.

\subsection{Quickbeam}
\label{sec:quickbeam}
By decomposing the scanning beam into harmonic coefficients $B_{\ell
  m}$, each TOI sample can be modelled as (neglecting the contribution from instrumental noise, which is independent of beam asymmetry)
\begin{equation}
%T_i = \sum_{\ell ms} D^{l}_{-m s} (\phi_i, \theta_i, \alpha_i) b_{ls}
%(-1)^{m) T_{\ell m} + n_i,
T_i = \sum_{\ell ms} e^{-i s \alpha_i} B_{\ell s} \tilde{T}_{\ell m} {}_s
Y_{\ell m}(\theta_i, \phi_i),
\label{eqn:tod_beam}
\end{equation}
where the TOI samples are indexed by $i$, and $\tilde{T}_{\ell m}$ is the underlying sky signal. 
The spin spherical harmonic ${}_s Y_{\ell m}$ rotates the scanning beam to the pointing location $(\theta, \phi)$, while the $e^{-i s \alpha_i}$ factor gives it the correct orientation.
Equation~\eqref{eqn:tod_beam} may be evaluated using techniques developed
for convolution in \cite{wandelt2001} and
\citet{prezeau2010}, although manipulating  a TOI-sized object is necessarily slow.

On the small angular scales comparable to the size of the beam, it is a good approximation to assume that the procedure of mapmaking from TOI samples is essentially a process of binning:
\begin{equation}
T(p) = \sum_{i \in p} T_i / H(p),
\label{eqn:map_beam_full}
\end{equation}
where $H(p)$ is the total number of hits in pixel $\hat{n}$.

%Quickbeam is based on the discussion around Eqs. 6, 7 of \cite{Hanson:2010gu}. 
Given a normalized, rescaled harmonic transform of the beam
$B_{\ell m}$, sky multipoles $\tilde{T}_{\ell m}$ and a scan history object $w(\hatn, s)$ given by
\begin{equation}
w(\hatn, s) = \sum_{j \in p} e^{i s \alpha_j} / H(\hatn),
\end{equation}
where the sum is over all hits $j$ of pixel $p$ at location $\hatn_p$, and $\alpha_j$ is the scan angle for observation $j$.
The harmonic transform of this scan-strategy object is given by
\begin{equation}
\wslm{s}{L}{M} = \int d^2 \hatn {}_s Y_{LM}^*(\hatn) w(\hatn, s).
\end{equation}
The beam-convolved observation is then given by
\begin{equation}
\tilde{T}(\hatn) = \sum_{s\ell m} w( \hatn, -s ) B_{\ell s} T_{\ell m} {}_s
Y_{\ell m}(\hatn).
\end{equation}
Taking the ensemble average of the pseudo-$C_\ell$ power spectrum of these
$T_{lm}$ gives \citep{Hanson2010,hinshaw2007} %Eq. B3 of \cite{Hanson:2010gu}:
\begin{multline}
\tilde{C}_{L}^{{\rm TT}} = \sum_{S S'} \sum_{\ell_1 \ell_2} \frac{(2\ell_1+1)(2\ell_2+1)}{4\pi}
{}_{(-s -s')}{\cal W}_{\ell_1} B_{\ell_2 S} B_{\ell_2 S'}^* C^{{\rm TT}}_{\ell_2}
\\ \times
 \threej{\ell_1}{\ell_2}{L}{s}{-s}{0} \threej{\ell_1}{\ell_2}{L}{s'}{-s'}{0} 
\end{multline}
where
\begin{equation}
{}_{(s s')}{\cal W}_{L} = \frac{1}{2L+1} \sum_{M} {}_{S} w_{LM} {}_{S'} w_{LM}^*
\end{equation}
is a cross-power spectrum of scan history objects. Note that the
$w(n,s)$  used here can also incorporate a position dependent
weighting to optimize the pseudo-$C_\ell$ estimate, such as
inverse-noise or a mask, the equations are unchanged.
Writing the pseudo-$C_\ell$ in position space (following \cite{dvorkin2008})
with Wigner-d matrices gives  
\begin{multline}
\tilde{C}_{L}^{{\rm TT}} = \frac{1}{8\pi} \sum_{S S'} \int_{-1}^{1} dz \ d^{L}_{00}(z)
\\ \times
\left[\sum_{\ell_1} d^{\ell_1}_{-s -s'}(z) {}_{(-s -s')}{\cal W}_{\ell_1} (2\ell_1+1) \right] 
\\ \times
\left[ \sum_{\ell_2} d^{\ell_2}_{s s'}(z) B_{\ell_2 S} B_{\ell_2 S'}^*
  C^{{\rm TT}}_{\ell_2}(2\ell_2+1) \right].
\end{multline}
This integral can be implemented exactly using Gauss-Legendre
quadrature, at a cost of $\clo(\ell_{\rm max}^2 s_{\rm max}^2)$. For
simplicity,  the equations here are written for the auto-spectrum
of a single detector, but the generalization to a map made by adding
several detectors with different weightings is straightforward. The
cost to compute all of the necessary terms exactly in that case
becomes $\clo(\ell_{\rm max}^2 s_{\rm max}^2 N_{\rm det}^2)$.

On the flat sky, beam convolution is multiplication in Fourier space
by a beam rotated onto the scan direction. Multiple hits with
different scan directions are incorporated by averaging (as the scan
history objects above encapsulate).

A scan strategy which is fairly smooth across the sky is nearly
equivalent to  observing many independent flat-sky patches at high
$L$. There is a fairly good approximation to the beam convolved 
pseudo-power spectrum which is essentially a flat-sky approximation. In the
limit that $L \gg \ell_1$, with $C_{\ell_2}$ and $B_{\ell_2}$ is a slowly-varying
function in $\ell_2$, and using the equality
\begin{equation}
\sum_{\ell_2} (2 \ell_2+1) \threej{\ell_1}{\ell_2}{L}{s}{-s}{0}
\threej{\ell_1}{\ell_2}{L}{s'}{-s'}{0} = \delta_{ss'},
\end{equation}
 the pseudo-$C_\ell$ sum above can be approximated as
\begin{equation}
{\tilde{C}}_L^{TT} = C_L^{TT} \sum_{M} \left< \left| w(\hatn_p, M) \right|^2 \right>_p |B_{L M}|^2,
\end{equation}
where the average $\langle \rangle_p$ is taken over the full sky. It is
illustrative to consider two limits of this equation. Firstly, for a ``raster''
scan strategy in which each pixel is observed with the same direction:
\begin{equation}
\left< \left| w(\hatn, M) \right|^2 \right>_p = 1,
\end{equation}
and the predicted transfer function is just the power spectrum of the
beam. Secondly, for a ``best-case'' scan strategy, in which each pixel is
observed many times with many different orientation angles, 
\begin{equation}
\left< \left| w(\hatn, M) \right|^2 \right>_p = \delta_{M0},
\end{equation}
and the transfer function is the azimuthally symmetric part of
the beam. Note that this is for a full-sky observation; in the
presence of a mask, the average above produces an $f_{{\rm sky}}$ factor, as
expected but neglects the coupling between $L$ multipoles (which
can be calculated with the more complete equations above).

\begin{figure}[!ht]
\centerline{\includegraphics[width=1.0\columnwidth]{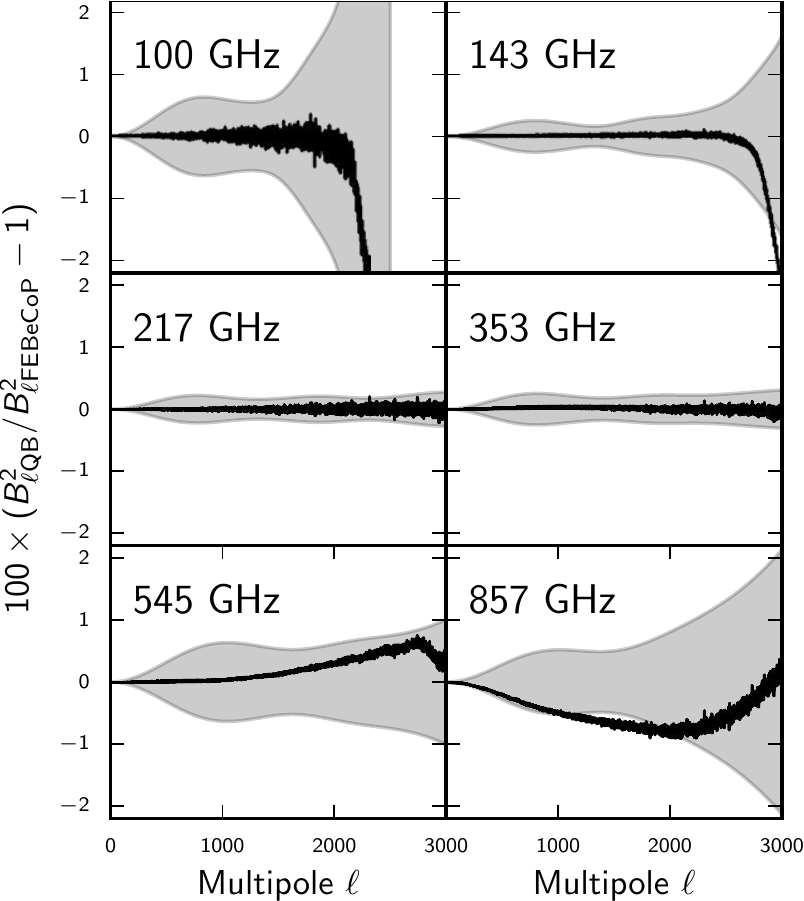}}
\caption{\label{fig:EffBeamDifference} Difference between effective beam
  window functions computed with a real space method ({\tt FEBeCoP}) and a
  harmonic space method ({\tt Quickbeam}).  The shaded region denotes the
  RMS error at each mutipole.}
\end{figure}

\section{Pixelization artefacts}
\label{sec:pixelization}

\begin{figure}[!ht]
\centerline{\includegraphics[width=\columnwidth]{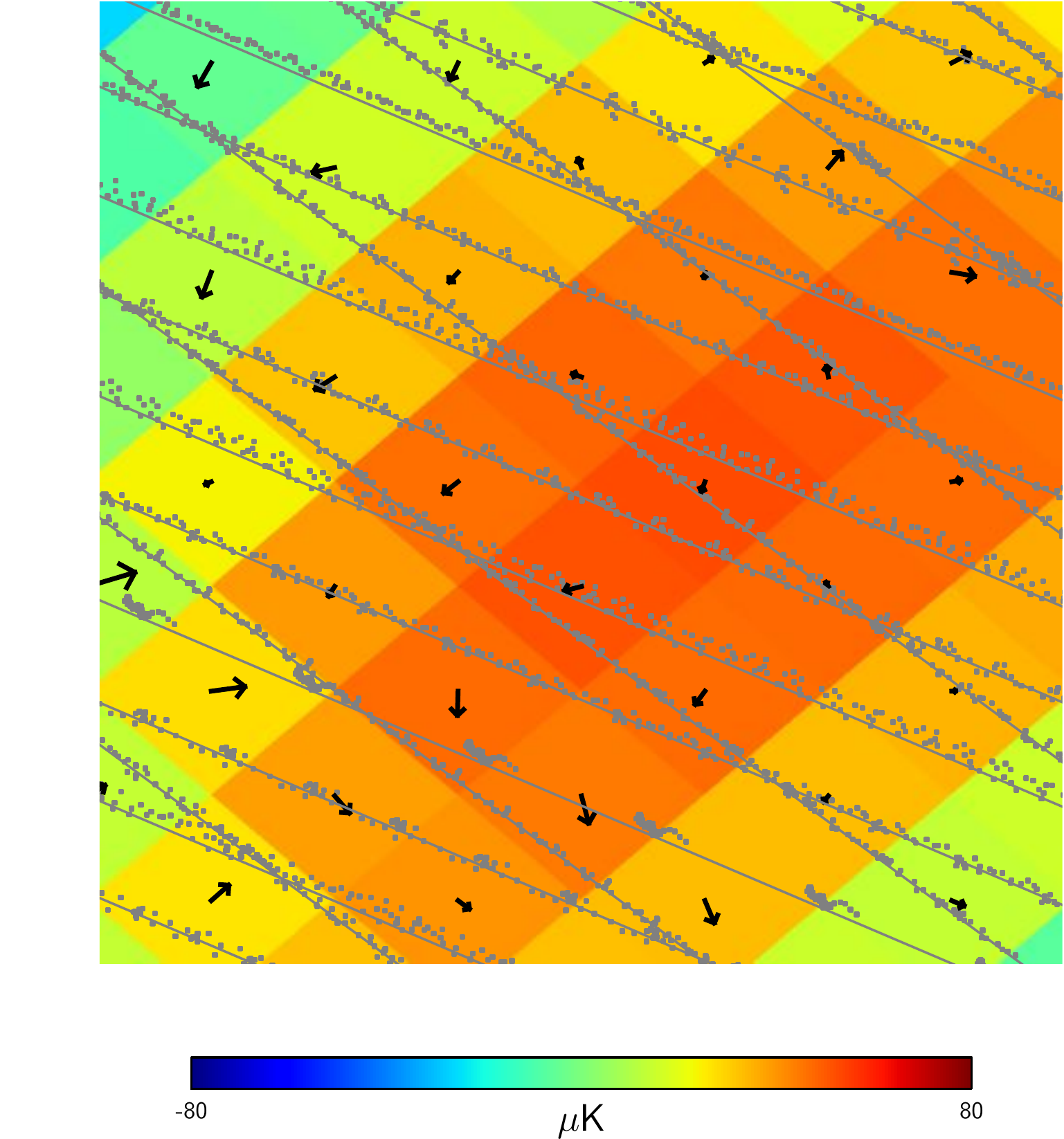}}
\label{fig:pixcoverage}
\caption{Illustration of TOI samples near the Galactic plane (grey
  dots), over-plotted on a simulated CMB realization which has been
  convolved with a Gaussian 7\arcm\ FWHM beam and pixelized at
  $(N_{\rm side} = 2048)$. Associated scanning rings (grey lines) as
  well as centres of mass for the hit distribution (black arrows) are
  also plotted.} 
\end{figure}

\Planck HFI maps are produced at {\tt HEALPix} resolution 11 $(N_{\rm side} = 2048)$, corresponding to pixels with a typical dimension of 1\parcm7. %sqrt(4 * pi / 12 / 2048^2) * 180. * 60. / pi
With the resolution comparable to the spacing between scanning
rings \citep{planck2011-1.1} there is an uneven distribution of
hits within pixels, introducing a complication in the analysis and
interpretation of the \Planck\ maps. 
A sample of the \Planck\ distribution of sample hits within pixels is
illustrated in Fig.~\ref{fig:pixcoverage}. 

The three effective beam codes may also be used to simulate the effect
of pixelization on the observed sky: {\tt LevelS/TotalConvoler/Conviqt}
(\citep{reinecke2006,wandelt2001,prezeau2010}));  {\tt FEBeCoP}
(\cite{mitra2010}; and {\tt FICSBell} (Appendix~\ref{sec:harmonicspace_effbeams}). 

For the measurement of CMB fluctuations, the effects of pixelization
may be studied analytically. On the small scales relevant to
pixelization, the observed CMB is smooth, both due to physical damping
and the convolution of the instrumental beam. Taylor expanding
the CMB temperature about a pixel centre to second order, the typical
gradient amplitude is given by
\begin{equation}
\langle |\nabla T |^2 \rangle = \frac{1}{4\pi} \sum_{\ell}
\ell(\ell+1)(2\ell+1) C_\ell^{T} W_\ell \approx 1\times10^9 \mu{\rm K}^2 / {\rm rad}^2.
\end{equation}
Here the approximate value is calculated for a $\Lambda$CDM
cosmology with a 7\arcm\ FWHM Gaussian beam.  The typical
curvature of the observed temperature, on the other hand is given by
\begin{equation}
\langle |\nabla^2 T |^2 \rangle = \frac{1}{4\pi} \sum_{\ell} [\ell(\ell+1)]^2(2\ell+1) C_\ell^{T} W_\ell \approx 7\times10^{14} \mu K^2 / {\rm rad}^4.
\end{equation}
On the scales relevant to the maximum displacement from the centre of
a $1.7'$ pixel, the maximum displacement is of order 1\arcm 
(3$\times10^{-4} {\rm rad})$), and so the gradient term tends to
dominate, although the curvature term is still non-negligible.  For
each observation of a pixel, the displacement from the
pixel centre can be denoted as $d = d_{\theta} + i d_{\phi}$.  The average over all
hits within a pixel gives an overall deflection vector  for a pixel
center located at $\hat{n}$ denoted as $d(\hat{n})$.  This
represents the centre of mass of the hit distribution;
Fig.~\ref{fig:pixcoverage} shows these average deflections 
using black arrows.  The deflection field $d(\hat{n})$ may be
decomposed into spin-1 spherical harmonics as
\begin{equation}
d_{lm} = \int_{4\pi} {}_1 Y_{\ell m}^* d(\hat{n}).
\end{equation}
With a second-order Taylor expansion of the CMB temperature about each
pixel centre, it is then possible to calculate the average pseudo-$C_\ell$
power spectrum of the pixelized sky.  This is given by
\begin{multline}
C_\ell^{T} = [1-\ell(\ell+1)R^d] {C}_\ell^{T} W_\ell + \\ 
\frac{1}{2} \sum_{\ell_1 \ell_2} \frac{\ell_1(\ell_1+1)(2\ell_1+1)(2\ell_2+1)}{4\pi} \\
\times \threej{\ell_1}{\ell_2}{\ell}{1}{-1}{0}^2 C_{\ell_1}^{T} W_{\ell_1} \left[ C_{\ell_2}^{d+} + (-1)^{\ell + \ell_1 + \ell_2} C_{\ell_2}^{ d-} \right],
\label{eqn:clt_pixelized}
\end{multline}
where $R^{d} = \langle |d|^2 \rangle/2$ is half the mean-squared
deflection magnitude (averaged over hits within a pixel, as well as
over pixels), $C_\ell^{d+}$ is the sum of the gradient and curl power
spectra of $d_{\ell m}$, and $C_\ell^{d-}$ is the gradient spectrum minus the
curl spectrum. The $R^{d}$ term describes a smearing of the observed sky due to
pixelization. For uniform pixel coverage of $N_{\rm side}=2048$ pixels 
$\langle |d|^2 \rangle^{1/2} = (2 R^{d})^{1/2} = 0.725'$, while, for the hit distribution of \Planck\
frequency maps, $R^{d}$ is within 0.2\,\% of this value for CMB channels, 
and 0.4\,\% for all channels. This term is therefore accurately described by
the {\tt HEALPix} pixel window function, which is derived under the assumption of
uniform pixel coverage, and the resulting relative error on the beam window
function is at most $4\times 10^{-4}$ for $\ell \le 3000$.

The effect of pixelization is degenerate with that of gravitational
lensing of the CMB, with the difference that it: (1) acts on the
beam-convolved sky, rather than the actual sky; and (2) produces a
curl-mode deflection field as well as a gradient mode. This is
discussed further in \cite{planck2013-p12}, where the sub-pixel
deflection field constitutes a potential source of bias for the
measured \Planck\ lensing potential. Indeed,
Eq.~\ref{eqn:clt_pixelized} is just a slightly modified version of the
usual first order CMB lensing power spectrum \citep{Hu2000},
\cite{Lewis2006} to accommodate curl modes.

A useful approximation to Eq.~\ref{eqn:clt_pixelized} which is
derived in the unrealistic limit that the deflection vectors are
uncorrelated between pixels, but in practice gives a good description
of the power induced by the pixelization, is that the $d(\hat{n})$
couples the CMB gradient into a source of noise with an effective
level given by
\begin{equation}
\sigma^{N} \approx \sqrt{ R^T \frac{4\pi}{N_{\rm pix}} \langle | d(\hat{n}) |^2
\rangle }, % (\muKarcmin ),
\end{equation}
where the average is taken over all pixels and $R^T$ is half the mean-squared power in the CMB gradient:
\begin{equation}
R^{T} = \frac{1}{8\pi} \sum_{\ell} \ell(\ell+1)(2\ell+1) \tilde{C}_\ell^{T}.
\end{equation}
For frequency-combined maps, $\sqrt{ \langle | d(\hat{n}) |^2 \rangle  }$ is
typically on the order of 0\parcm1, and so the induced noise $\sigma^{N}$ is 
approximately $2 \muKarcmin$. This is small compared to the
instrumental contribution, although it does not disappear when taking
cross-spectra, depending on the coherence of the hit distributions of
the two maps in the cross-spectra.

\section{Beam window function error}
%
% about B(l) eigen-modes
% inserted in main LaTeX file with \input{beam_eigenmodes}
% to be merged with main file ... someday
%
\newcommand{\nmc}{\ensuremath{n_{\rm MC}}}
\newcommand{\nmodes}{\ensuremath{n_{\rm modes}}}
\newcommand{\Bmean}{\ensuremath{B_{\rm mean}}}
\newcommand{\Wmean}{\ensuremath{W_{\rm mean}}}
\newcommand{\lmax}{\ensuremath{\ell_{\rm max}}}
\newcommand{\lmin}{\ensuremath{\ell_{\rm min}}}
\newcommand{\bias}{\ensuremath{\varepsilon_{\rm bias}}}
\newcommand{\bsfudge}{\ensuremath{f_{s}}}

\newcommand{\matthreethree}[9]{\left(
\begin{array}{ccc}%
\!#1\! & \!#2\! & \!#3\!\!\\%
\!#4\! & \!#5\! & \!#6\!\!\\%  
\!#7\! & \!#8\! & \!#9\!\!%  
\end{array}%
\right)}

\subsection{Error eigenmodes}
\label{sec:erroreigenmodes_1pair}

%-------------------------------------------------------------------------
\begin{figure*}[!ht]
\begin{center}
\includegraphics[width=1.0\textwidth]{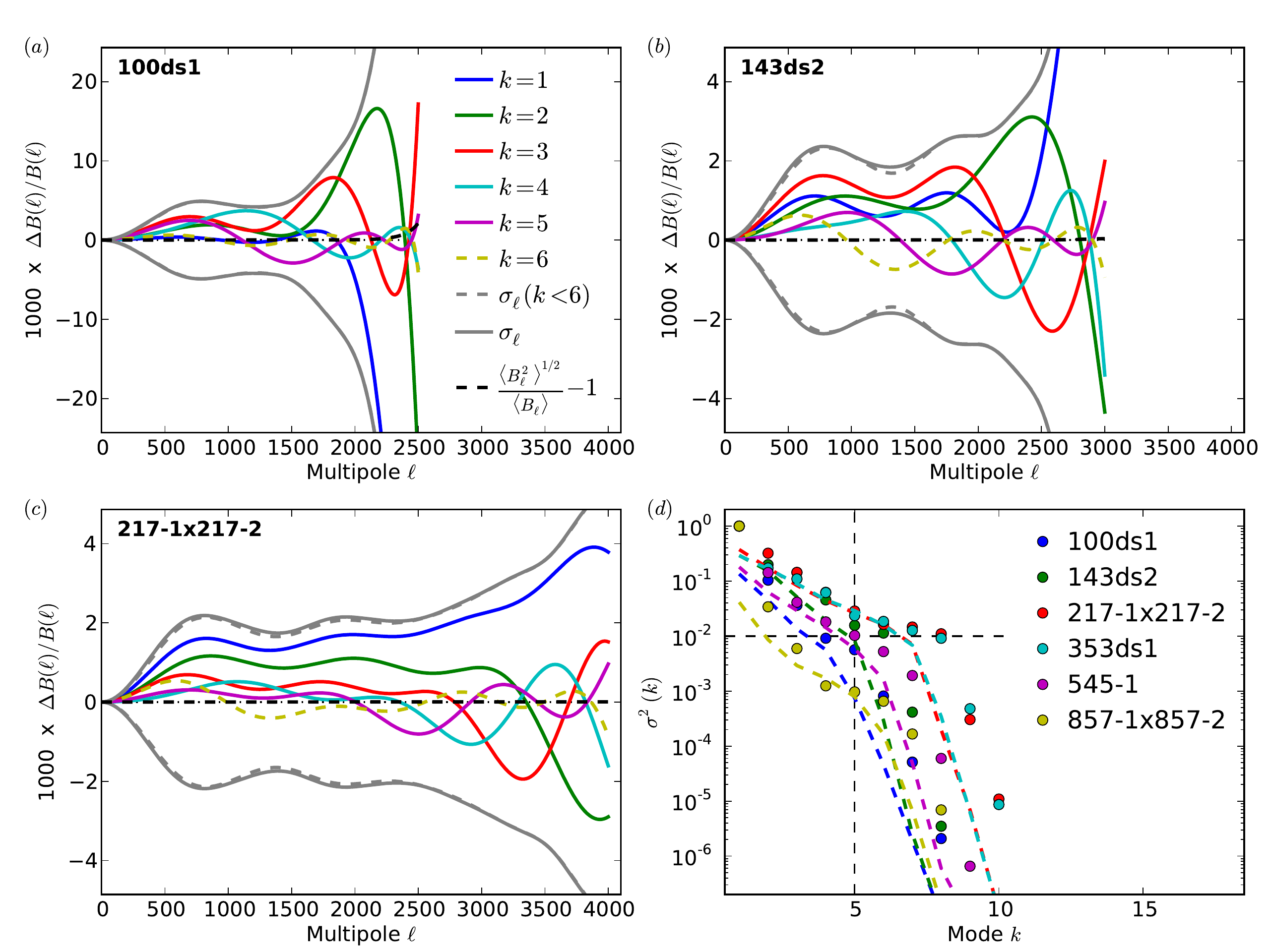}
\caption{Beam window function error modes. Panels $a$, $b$ and $c$ show the first six
eigenmodes defined in Eq.~\ref{def:bl_eigenmodes} for respectively the
effective auto-beam 100ds1 and 143ds2 and the effective cross-beam
217-1$\times$217-2. The first five modes used in the current beam error 
modeling are shown as solid colored curves, while the 6th mode (the first one to
be ignored) appears as yellow dashes. The grey dashes show the $\pm$1\,$\sigma$ envelope
obtained by adding the first five modes in quadrature, while the solid grey curve
is the $\pm$1\,$\sigma$ envelope estimated from the simulations (therefore including all
eigenmodes). The black dashes show the relative difference between
$\Wmean^{1/2}(\ell)$ and $\Bmean(\ell)$ defined in Eqs.~\ref{eq:wmean} and
\ref{eq:bmean}, respectively.
In panel $d$, for a selection of effective beams, the coloured symbols show
$(d_k/d_1)^2$ where $d_k$ is the $k$-th eigenvalue of the diagonal matrix ${\bf D}$ (Eq.~\ref{eq:bl_MC_SVD}), while the colored
dashes show the error made on the quadratic sum of the eigenvalues by truncating
it to $\nmodes$:
$1 - \sum_{k=1}^{\nmodes} d^2_k  / \sum_{k=1}^{\infty} d^2_k$. The vertical
dashes show the current $\nmodes=5$ value.
\label{fig:Bl_Emodes}%
}%
\end{center}
\end{figure*}
%-------------------------------------------------------------------------

%-------------------------------------------------------------------------
\begin{figure*}[!ht]
\begin{center}
\includegraphics[width=1.0\textwidth]{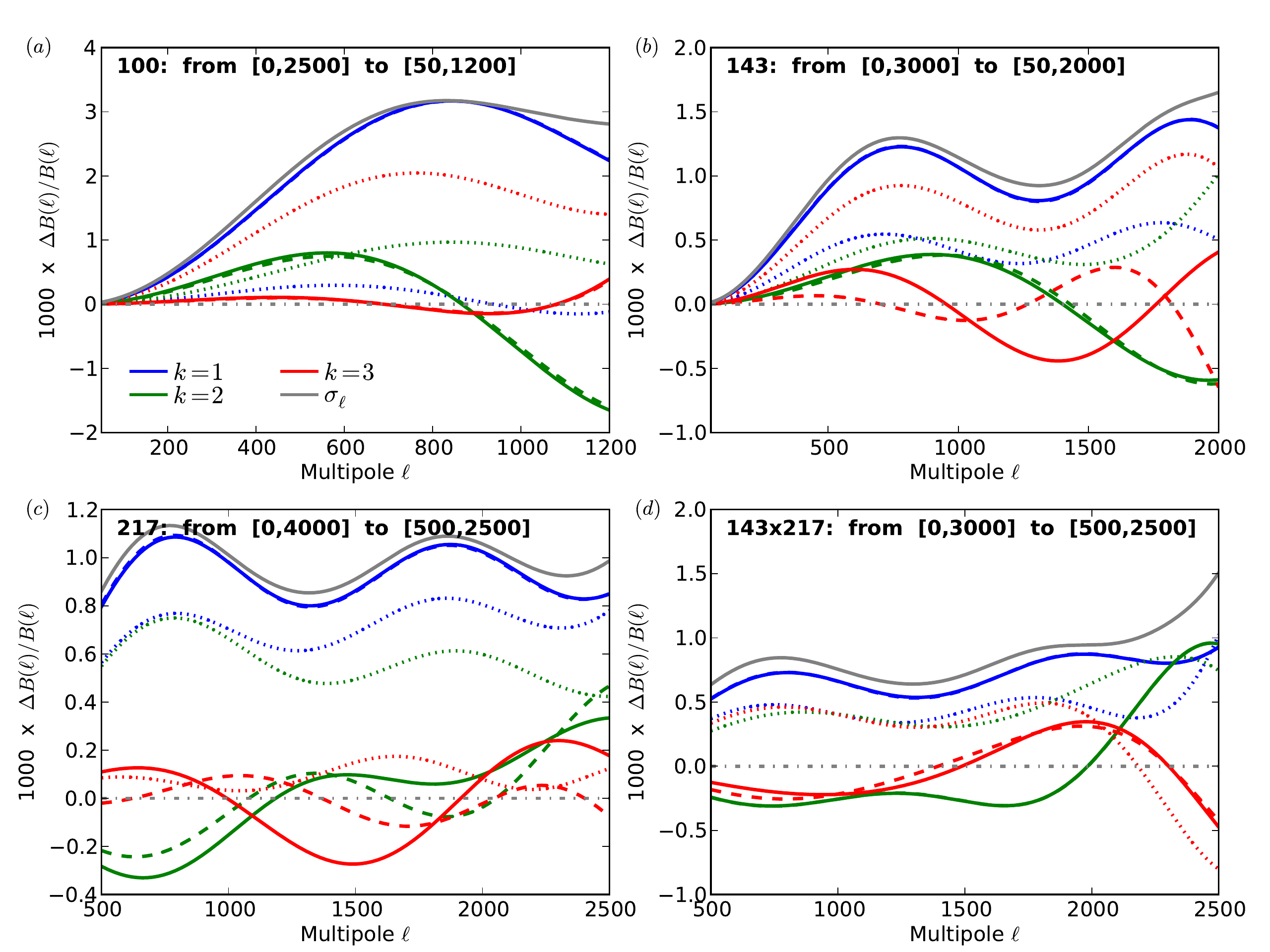}
\caption{Effect of $\ell$ truncation on beam error modes for frequency-averaged
beam window functions at 100, 143, 217\,GHz and 143$\times$217. For
clarity, only the three leading 
modes are shown, respectively, in blue, green and red, while the solid grey line
shows the 1\,$\sigma$ level, obtained by adding {\em all} modes in
quadrature.  
Dotted lines are the original eigenmodes computed on a wide
$\ell$-range. Solid lines are the eigenmodes computed 
directly from the MC simulations on the truncated $\ell$-range used for
cosmological analysis. Dashed lines show the  eigenmodes computed on
the truncated $\ell$-range with Eqs.~\ref{eq:shorter_svd} 
and \ref{eq:shorter_emodes}, starting
from the first five eigenmodes for the wide range. In all four cases considered,
the first leading mode on the truncated range, which dominates the error budget, is perfectly reconstructed out of the
information available, while the second leading mode is well estimated in all cases
except for 217\,GHz.
\label{fig:Bl_Em_cuts}%
}%
\end{center}
\end{figure*}
%-------------------------------------------------------------------------
%-------------------------------------------------------------------------
\begin{figure}[!ht]
\begin{center}
\includegraphics[width=1.0\columnwidth]{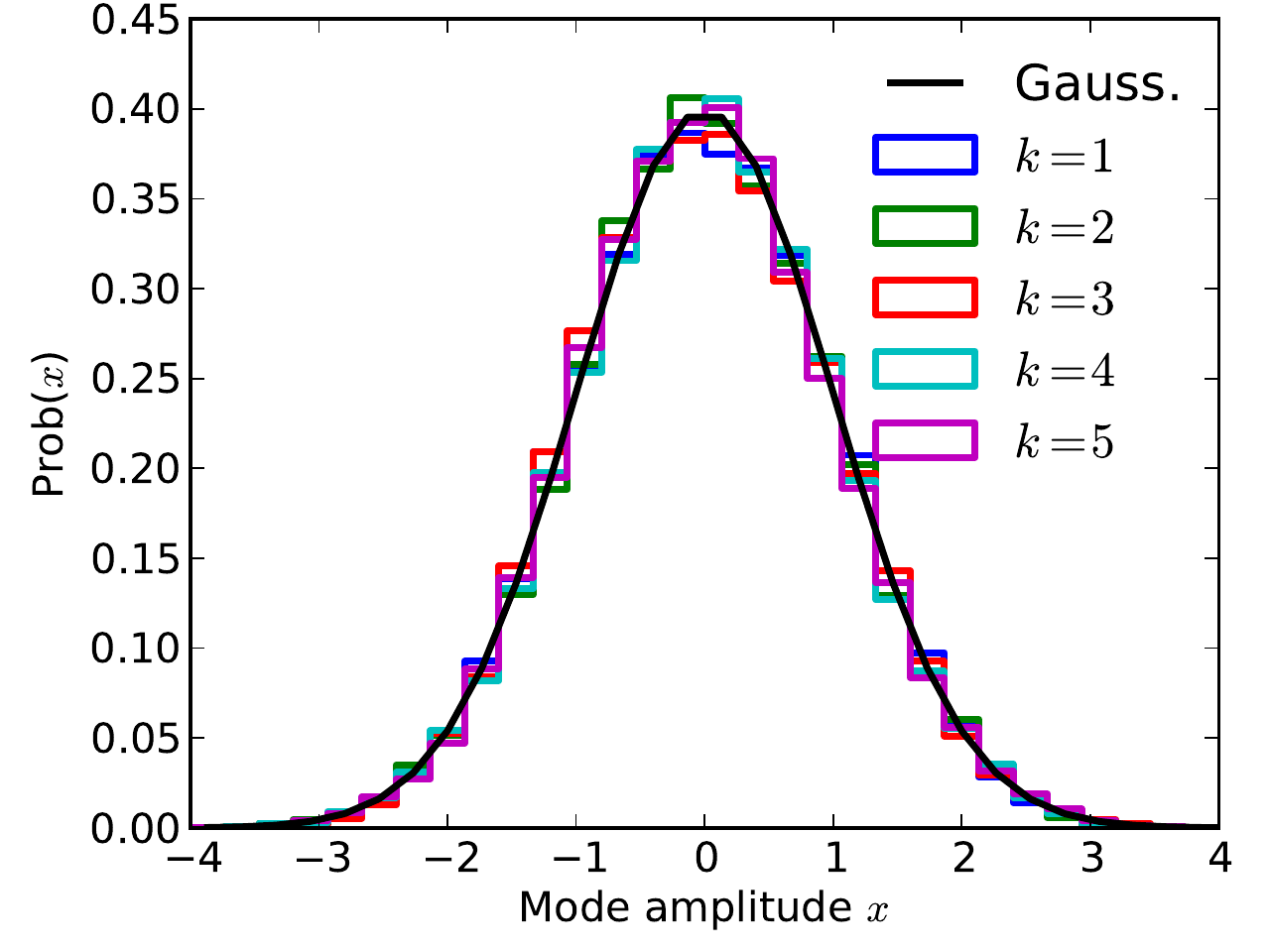}
\caption{Distribution of the eigenmodes determined from MC simulations, for all
HFI detector sets, for the first $\nmodes$=5 modes, compared to a Gaussian
distribution of zero mean and unit variance.
\label{fig:Bl_Em_stats}%
}%
\end{center}
\end{figure}
%-------------------------------------------------------------------------

Consider two individual detectors or detection units (weighted sum of
detectors) $X$ and $Y$ for which sky
maps are available. Here $X$ and $Y$ can be the same or different.
Putting aside the instrumental noise and other contingencies for the time being,
the cross- (or auto-) angular power spectrum measured of the observed maps
$C^{XY}_{\rm obs}(\ell)$ is on {\em average}
related to the real input signal $C^{XY}_{\rm sky}(\ell)$ through
\begin{equation}
	\langle C^{XY}_{\rm obs}(\ell)\rangle = C^{XY}_{\rm sky}(\ell) W^{XY}_{\rm eff,
true}(\ell) 
\end{equation}
where $W^{XY}_{\rm eff, true}$ is the effective beam window function. Note that
because of the optical beam non-circularity and \Planck\ scanning strategy,
\begin{equation}
W^{XY}(\ell) \ne \left[W^{XX}(\ell) W^{YY}(\ell)\right]^{1/2}
\quad\text{if}\ X \ne Y, 
%\label{eq:crossbeam}
\end{equation}
as illustrated in Fig.~\ref{fig:Bl_Ratios}, while $W^{XY} = W^{YX}$
for any $X$ and $Y$. It also appears that in the $\ell$ range of
interest, $W^{XY}(\ell) \ge 0$; therefore $W^{XY}
=\left(B^{XY}\right)^2$, analogous to what is usually done for simple 
(circular) beam models. In what follows, the $XY$ pair
superscript is dropped except when they are required for clarity.  In most cases,
scientific analyses will be conducted on a best guess $C_{\rm
  est}(\ell)$ of the sky power spectrum, in which the empirical
$C_{\rm obs}(\ell)$ is corrected from a nominal effective window
$W_{\rm eff, nom}(\ell)$
\begin{equation}
	C_{\rm est}(\ell) =C_{\rm obs}(\ell) / W_{\rm eff, nom}(\ell);
\end{equation}
therefore, on average,
\begin{eqnarray}
	\langle C_{\rm est}(\ell)\rangle 
	&=& C_{\rm sky}(\ell) W_{\rm eff, true}(\ell) / W_{\rm eff, nom}(\ell), \nonumber \\
	&=& C_{\rm sky}(\ell) \left(B_{\rm eff, true}(\ell) / B_{\rm eff, nom}(\ell)\right)^2.
\end{eqnarray}
The ratio $B_{\rm eff, true}(\ell) / B_{\rm eff, nom}(\ell)$ which determines the
uncertainties on the angular power spectrum originating from our beam knowledge
is studied using the planet transit MC simulations described
in Sect.~\ref{sec:errors}. The scanning beam map determined with the B-Spline code
described in Appendix~\ref{sec:griddedbeamdescription}
on each of these simulations is turned into an
effective beam window function $W_i(\ell)$ for $i=1,2,\dots, \nmc$ (where
$\nmc=100$ is the number of MC simulations) 
using the {\tt Quickbeam} formalism described in
Appendix~\ref{sec:quickbeam}.

Defining the means $\Bmean(\ell)$ and $\Wmean(\ell)$ as 
\begin{equation}
	\Bmean(\ell) = \sum_{i=1}^{\nmc} (W_i(\ell))^{1/2}  / \nmc,
\label{eq:bmean}
\end{equation}
and,
\begin{equation}
	\Wmean(\ell) = \sum_{i=1}^{\nmc} W_i(\ell)  / \nmc,
\label{eq:wmean}
\end{equation}
one can build the matrix of the deviations around the mean
\begin{equation}
	\Delta_i(\ell) = \bsfudge \ln\left(B_i(\ell)/\Bmean(\ell)\right),
	\label{eq:delta_beam}
\end{equation}
where the factor \bsfudge\ has been applied. As
discussed in Sect.~\ref{sec:totalerrors}, 
$\bsfudge=2.7$. This scaling factor is applied throughout the rest of
discussion and is included in the delivered products and plotted error
modes.\\ Sect.~\ref{sec:MCbias} contains a discussion of how the MC
average $\Bmean$ and nominal beam $B_{\rm eff,nom}$ are related and
focus here on the dispersion around the mean. \\ Since the relative
dispersion of the simulated $W_i(\ell)$ generally is very small (less
than 1\%), then $\Wmean(\ell) \simeq \Bmean(\ell)^2$ to a very good
approximation (as illustrated in Fig.~\ref{fig:Bl_Emodes}) and the
matrix $\Delta$ is well approximated by
\begin{eqnarray}
	\Delta_i(\ell) &\simeq& 1/2 \ \bsfudge
\ln\left[W_i(\ell)/\Wmean(\ell)\right], \nonumber \\
\ 			&\simeq& 1/2 \ \bsfudge \left[W_i(\ell)/\Wmean(\ell) - 1\right].
%%%%	\label{eq:delta_beam}
\end{eqnarray}
The quantity $B(\ell)$ was preferred over $W(\ell)$ in order to remain
consistent with the usual description of the beam in linear map space,
instead of quadratic space.

%%%%%
The statistical properties of the MC based beam window functions can be
summarized in the covariance of the deviations $\Delta$, defined as
\begin{equation}
	{\bf C} \equiv {\bf \Delta}^T \cdot {\bf \Delta} / (\nmc-1),
	\label{eq:bl_cov}	
\end{equation}
where ${\bf \Delta}$ is a matrix with $\nmc$ rows and $\lmax+1$ columns, and
${\bf C}$ is a square symmetric positive semi-definite matrix with $\lmax+1$ rows and columns.
It can be diagonalized into
\begin{eqnarray}
{\bf C}	&=&	{\bf V} \cdot {\bf D}^2 \cdot {\bf V}^T / (\nmc-1),\nonumber\\
	&=& {\bf E}^T \cdot {\bf E},
	\label{eq:bl_cov_diag}
\end{eqnarray}
where ${\bf D}$ is a diagonal matrix, with at most $\nmc$ positive eigenvalues, and the eigenmodes matrix
\begin{equation}
	{\bf E} \equiv {\bf D} \cdot {\bf V}^T/(\nmc-1)^{1/2}
	\label{def:bl_eigenmodes}
\end{equation}
of the beam uncertainty. 
Alternatively, one can perform a singular value
decomposition (SVD) of ${\bf \Delta}$, which reads
\begin{eqnarray}
	{\bf \Delta} &=& {\bf M} \cdot {\bf D} \cdot {\bf V}^T
	\label{eq:bl_MC_SVD}\\
		&=& {\bf M} \cdot {\bf E}\ (\nmc-1)^{1/2},
\end{eqnarray}
where ${\bf M}$ is an
orthogonal $\nmc\times\nmc$ matrix (i.e., 
${\bf M}^T \cdot {\bf M}= {\bf M} \cdot {\bf M}^T = {\bf I}_{\nmc}$), 
${\bf D}$ is a diagonal matrix with $\nmc$ non-negative eigenvalues of decreasing
amplitude, 
and ${\bf V}$ is a matrix with $\lmax+1$ rows whose $\nmc$ columns are
orthonormal vectors (i.e., ${\bf V}^T{\bf V} = {\bf I}_{\nmc}$), with ${\bf I}_{\nmc}$ being the
identity matrix of size $\nmc\times\nmc$.
The diagonalization of the covariance matrix ${\bf C}$
(Eq.~\ref{eq:bl_cov_diag}) has a numerical
complexity scaling like $\lmax^3$, while the SVD of ${\bf \Delta}$
(Eq.~\ref{eq:bl_MC_SVD}) scales like $\lmax\nmc^2$. Since $\nmc
\ll \lmax$ the latter approach was prefered because it is much faster, and it
naturally provides the mixing matrix ${\bf M}$. Equation~(\ref{eq:bl_MC_SVD}) has
a degeneracy on the sign of ${\bf M}$ and ${\bf V}$, which was lifted by
constraining the vectors of ${\bf V}$ (and therefore ${\bf E}$) to all be positive
at $\ell=200$.

It appears that most of the statistical content of ${\bf \Delta}$ or ${\bf C}$
is described by the first few modes $\nmodes$ with the largest eigenvalues, in
which case the ${\bf E}$ matrix is truncated to its first $\nmodes$ rows with
little loss of information. For HFI, $\nmodes=5$ is chosen, as illustrated in  Fig.~\ref{fig:Bl_Emodes}.\\
The statistics of the elements of the mixing matrix ${\bf M}$, and therefore of the MC measured
beam window function fluctuations, is shown in Fig.~\ref{fig:Bl_Em_stats} to be very
close to Gaussian, justifying the current analysis in terms of a 
covariance matrix.

The beam uncertainty model therefore is
\begin{eqnarray}
	B(\ell) &=& \Bmean(\ell) \exp\left({\bf g}^T.{\bf E}\right)_\ell\\
	     &=& \Bmean(\ell) \exp\left(\sum_{k=1}^{\nmodes} g_k e_k(\ell) \right),
\end{eqnarray}
where ${\bf g}$ is a $\nmodes$ element vector of independant Gaussian variates
of zero mean and unit variance and $e_k(\ell)$ is the $k$-th
row of ${\bf E}$. 

The SVD decomposition of the beam uncertainty was performed for the $\ell$ range
$[\lmin,\lmax]$ with $\lmin=0$ and $\lmax$ depending on the frequencies
of the two detectors involved in the beam considered.
Currently $\lmax=2500$ when the lowest frequency is 100\,GHz, $\lmax=3000$ when
that frequency is 143\,GHz, and $\lmax=4000$ at higher frequency.

If orthogonal error eigenmodes are needed for the range 
$[\lmin',\lmax']$, with $\lmin'\ge\lmin$ and $\lmax'\le\lmax$, the provided
${\bf E}$ must first be truncated to the new range to give the ${\bf E}_t$
matrix with $\nmodes$ rows and $\lmax'-\lmin'+1$ columns, and
then singular value decomposed into
\begin{equation}
	{\bf E}_t = {\bf R}' \cdot {\bf D}' \cdot {\bf V}^{'T},
	\label{eq:shorter_svd}
\end{equation}
where the new set of orthogonal eigenmodes is 
\begin{equation}
	{\bf E}' \equiv {\bf D}' \cdot {\bf V}^{'T}.
	\label{eq:shorter_emodes}
\end{equation}
This is illustrated on Fig.~\ref{fig:Bl_Em_cuts}, where the eigenmodes
are truncated to the frequency dependent $\ell$ ranges used in the \Planck\
$C(\ell)$ likelihood \citep{planck2013-p08}.

\subsection{Eigenmode covariance}
\label{sec:erroreigenmodes_npairs}
The previous approach, dedicated to the study of the uncertainty on the beam
window function associated with a single pair of detectors (or maps), can be
generalized to the simultaneous characterization of any number of pairs.
For instance, for the three disjoint pairs, $a=\{UV\}$, $b=\{XY\}$ and $c=\{ZT\}$, one writes

\begin{equation}
	\left({\bf \Delta}^a \ {\bf \Delta}^b \ {\bf \Delta}^c \right) = 
	\left({\bf M}^a \ {\bf M}^b \ {\bf M}^c \right).
	\matthreethree{{\bf E}^a}{0}{0}%
{0}{{\bf E}^b}{0}%
{0}{0}{{\bf E}^c}.
\end{equation}
and the covariance matrix reads

\begin{eqnarray}
{\bf C}^{abc} &=& \left({\bf \Delta}^a \ {\bf \Delta}^b \ {\bf \Delta}^c \right)^T.%
	\left({\bf \Delta}^a \ {\bf \Delta}^b \ {\bf \Delta}^c\right) \nonumber \\%
\	&=&\matthreethree{{\bf E}^a}{0}{0}%
{0}{{\bf E}^b}{0}%
{0}{0}{{\bf E}^c}^T.
	\matthreethree{%
{\bf I}}{{\bf M}^{aT}{\bf M}^b}{{\bf M}^{aT}{\bf M}^c}%
{{\bf M}^{bT}{\bf M}^a}{{\bf I}}{{\bf M}^{bT}{\bf M}^c}
{{\bf M}^{cT}{\bf M}^a}{{\bf M}^{cT}{\bf M}^b}{{\bf I}}.
	\matthreethree{{\bf E}^a}{0}{0}%
{0}{{\bf E}^b}{0}%
{0}{0}{{\bf E}^c}, \nonumber \\%
\	&=& {\bf E}^{a,b,cT} \cdot {\bf M}^{a,b,c} \cdot {\bf E}^{a,b,c},
\end{eqnarray}
where ${\bf M}^{a,b,c}$ is a square symmetric matrix with $3\nmodes$ rows, and
if one denotes ${\bf L}^{a,b,c}$ its Cholesky ``root,'' such that
${\bf M}^{a,b,c}={\bf L}^{a,b,c} \cdot {\bf L}^{a,b,cT}$, then
\begin{eqnarray}
	B^a(\ell) &=& \Bmean^a(\ell) \exp\left({\bf g}^T {\bf L}^{a,b,c} 
\left({\bf E}^a \; {\bf 0} \; {\bf 0} \right)\right)_\ell, \label{eq:corrbeammodel1}\\
	B^b(\ell) &=& \Bmean^b(\ell) \exp\left({\bf g}^T {\bf L}^{a,b,c} 
\left({\bf 0} \; {\bf E}^b \; {\bf 0} \right)\right)_\ell,\label{eq:corrbeammodel2}\\
	B^c(\ell) &=& \Bmean^c(\ell) \exp\left({\bf g}^T {\bf L}^{a,b,c} 
\left({\bf 0} \;  {\bf 0}  \;{\bf E}^c\right)\right)_\ell, \label{eq:corrbeammodel3}
\end{eqnarray}
where ${\bf g}$ is the $3 \nmodes$ element vector of independent Gaussian deviates
and is the same for Eqs.~(\ref{eq:corrbeammodel1}) to
(\ref{eq:corrbeammodel3}).

The \Planck-HFI Reduced Instrument Model (RIMO) available at 
\Planck\ Legacy
Archive\footnote{\url{http://www.sciops.esa.int/index.php?project=planck&page=Planck_Legacy_Archive}}
and described in \citet{planck2013-p28}
 %(PLA) 
contains the correlation matrix ${\bf M}^{a,b,c,d,\ldots}$,
where $a, b, c, d\ldots$ each are a different element of the set of pairs
that can be built out of the detection units available. So, for
$n_{\rm d}$ detection units, the number of pairs will be $n_{\rm d}
(n_{\rm d}+1)/2$ and the correlation matrix will be square and
symmetric, with the value ``1'' on
its diagonal and have $\nmodes n_{\rm d}(n_{\rm d}+1)/2$ rows and as many columns.
The $\nmodes$ relative to the same detector pair form adjacent rows and columns
in the matrix,
and the order of appearance of the pairs in the matrix is specified in the
header of the FITS extension containing the matrix.
The nominal $B_\ell$ and eigenmodes ${\bf E}$ are provided for each pair in a
different extension.

The results above were obtained in the basis of eigenmodes provided, if one
wants to obtain a beam correlated model on a different $\ell$-range, with the
eigenmodes ${\bf E}'$ defined in Eq.~(\ref{eq:shorter_emodes}), then the
covariance matrix becomes

\begin{equation}
{{\bf C}'}^{abc} = 
\matthreethree{{{\bf E}'}^a}{0}{0}%
{0}{{{\bf E}'}^b}{0}%
{0}{0}{{{\bf E}'}^c}^T \cdot 
{\bf M}'^{a,b,c}.
\matthreethree{{{\bf E}'}^a}{0}{0}%
{0}{{{\bf E}'}^b}{0}%
{0}{0}{{{\bf E}'}^c}, 
\end{equation}
with
\begin{equation}
{{\bf M}'}^{a,b,c} = 
\matthreethree{{{\bf R}'}^a}{0}{0}%
{0}{{{\bf R}'}^b}{0}%
{0}{0}{{{\bf R}'}^c}^T \cdot 
{\bf M}^{a,b,c}.
\matthreethree{{{\bf R}'}^a}{0}{0}%
{0}{{{\bf R}'}^b}{0}%
{0}{0}{{{\bf R}'}^c},
\end{equation}
where the ${\bf R}'$ matrices are obtained from the SVD in Eq.~(\ref{eq:shorter_svd}).

\subsection{Monte Carlo bias}
\label{sec:MCbias}
As discussed in Sect.~\ref{sub:tests_of_the_main_beam_model}, the Monte Carlo average
of the B-Spline reconstructed effective beam window function $\Bmean(\ell)$
can be different from the effective beam that would be obtained directly out of the
simulation input beam map $B_{\rm eff, in}(\ell)$, introduces a bias
\begin{equation}
	1 + \bias(\ell) \equiv \Bmean(\ell) \,/\, B_{\rm eff, in}(\ell),
	\label{eq:def_bias}
\end{equation}
which is interpreted as a consequence of the imperfect sampling of the beam
map by the planet, the effect of the instrumental noise and pointing uncertainty and
the smoothing feature of the B-spline functions.
% It is found that $\left|\bias(\ell{=}500)\right| \la 2\, 10^{-4}$ for all effective beams and that
% $\left|\bias(\ell{=}2000)\right| \la 2\, 10^{-3}$ at 100GHz, and $\la  10^{-3}$ at all other
% frequencies, 
It is found that $\left|\bias(\ell{=}500)\right| \le 2\times 10^{-4}$ and
$\left|\bias(\ell{=}1000)\right| < 5\times 10^{-4}$ for all effective beams,
making it smaller than the relative dispersion on the beam window function described
above.  It is assumed that the beam reconstruction
on the real data suffers from the same bias, and its beam window function
$B_{\rm eff, raw}(\ell)$ is corrected for this to provide the nominal beam
\begin{eqnarray}
	B_{\rm eff, nom}(\ell) &=&   B_{\rm eff, raw}(\ell) \,/\, \left[1 + \bias(\ell)\right], \\
   \ &=&   B_{\rm eff, raw}(\ell)\, B_{\rm eff, in}(\ell) \,/\, \Bmean(\ell).
\end{eqnarray}
In doing so, $\bias$ is assumed to be perfectly determined by the simulations, and no error is
associated with this correction.

\raggedright
\end{document}